%% file: aesin2.tex
\documentstyle[12pt,aaspp4]{article}
\catcode`\@=11 % This allows us to modify PLAIN macros.
\def\lsim{\mathrel{\mathpalette\@versim<}}
\def\gsim{\mathrel{\mathpalette\@versim>}}
\def\be{\begin{equation}}
\def\ee{\end{equation}}
\def\e{\epsilon}
\def\rtr{r_{tr}}
\def\mdot{\dot{m}}
\def\Mdot{\dot{M}}
\def\msun{M_{\odot}}
\def\@versim#1#2{\vcenter{\offinterlineskip
        \ialign{$\m@th#1\hfil##\hfil$\crcr#2\crcr\sim\crcr } }}
\catcode`\@=12 % at signs are no longer letters

\eqsecnum
\begin{document}
\date{}
\title{Advection-Dominated Accretion and the Spectral States of Black Hole
X-Ray Binaries:  Application to Nova Muscae 1991}
\author{Ann A. Esin, Jeffrey E. McClintock, Ramesh Narayan}
\affil{Harvard-Smithsonian Center for Astrophysics, Cambridge, MA 02138}

\begin{abstract}
Black hole X-ray binaries (BHXBs) are known to display five distinct
spectral states.  In order of increasing luminosity these are the
quiescent state, low state, intermediate state, high state and very
high state.  We present a self-consistent model of accretion flows
around black holes which unifies all of these states except the very
high state.  The model is an extension of the following paradigm which has
been applied successfully to the quiescent state.  The
accretion flow consists of two zones, an inner advection-dominated
accretion flow (ADAF) which extends from the black hole horizon to a
transition radius $\rtr$, and an outer thin accretion disk that is present 
beyond $r_{tr}$.  Above the disk is a hot corona which is a
continuation of the inner ADAF.  The model consistently treats the
dynamics of the accreting gas, the thermal balance of the ions and
electrons in the two-temperature ADAF and corona, and the radiation
processes that produce the observed spectrum.  

At low mass accretion rates, $\mdot\lsim0.01$ (in Eddington units),
the inner ADAF zone in the model radiates extremely inefficiently, and
the outer thin disk is restricted to large radii ($\rtr \sim 10^2-10^4$, in 
Schwarzschild units).  The luminosity therefore is low, and this 
configuration is identified with the quiescent state.  For $\mdot \gsim 0.01$ 
and up to a critical value $\mdot_{crit} \sim 0.08$, the radiative 
efficiency of the ADAF increases rapidly and the 
system becomes fairly luminous.  The spectrum is very hard and peaks 
around $100$ keV.  This is the low state.  The exact value of $\mdot_{crit}$
depends on the viscosity parameter $\alpha$ ($\mdot_{crit} \sim 1.3 \alpha^2$;
the paper assumes $\alpha = 0.25$).  For values of $\mdot > \mdot_{crit}$ 
and up to a second critical value about 10\% higher, the ADAF progressively 
shrinks in size, the transition radius decreases, and the X-ray spectrum 
changes continuously from hard to soft.  We identify this stage with the
intermediate state.  Finally, when $\mdot$ is sufficiently large,
the inner ADAF zone disappears altogether and the thin accretion disk 
extends down to the marginally stable orbit.  The spectrum is dominated by
an ultrasoft component with a weak hard tail. This is the high state.  
Model spectra calculated with this unified scenario agree well with 
observations of the quiescent, low, intermediate and high states.  Moreover, 
the model provides a natural explanation for the low state to high state 
transition in BHXBs.  We also make a tentative proposal for 
the very high state, but this aspect of the model is less secure.

A feature of the model is that it is essentially parameter-free.  We
test the model against observations of the soft X-ray transient Nova
Muscae during its 1991 outburst.  The model reproduces the observed
lightcurves and spectra surprisingly well, and makes a number of 
predictions which can be tested by observations of other BHXBs.

\end{abstract}


\section{Introduction}

Galactic X-ray sources are classified as black hole X-ray binaries (BHXB,
see van Paradijs \& McClintock 1995; Tanaka \& Lewin 1995; Tanaka \& 
Shibazaki 1996 and Liang 1997 for reviews) if the binary mass function is 
$\gsim 3 \msun$ 
(A0620-00, V404 Cyg, Nova Muscae 1991, GRO J1655-40, GS 2000-251, Nova 
Ophiuchi 1997) or if there is other less direct evidence that the accreting 
compact object is more massive than $3 \msun$ (e.g. Cyg X-1, LMC X-3,
GRO J0422+32).  In addition, those systems for which there is
no reliable mass estimate, but which exhibit spectral and temporal 
characteristics similar to the well established BHXBs, are often also
included in this category (e.g. GX 339-4, 1E1740.7-2942, GRS 1915+105).

Historically, five different spectral states of BHXBs have been identified,
based on the spectral shape and flux level in the 1--10 keV X-ray band (see 
van der Klis 1994; Nowak 1995; Tanaka \& Lewin 1995; Tanaka \& Shibazaki 
1996).  {\em Low/Hard state}: Systems that have a power-law spectrum with a 
photon index $\alpha_N \sim 1.5-1.9$ and an exponential cut-off around 100 
keV are referred to as being in the ``low'' or ``hard'' state.  The total 
X-ray luminosity in this state, computed for those systems where distance 
and mass estimates are available, is generally below $10\%$ of the Eddington 
luminosity (Nowak 1995).  {\em High/Soft state}: Other systems with spectra 
dominated by an ultrasoft blackbody-like component with a characteristic 
temperature $\sim 1$ keV (Tanaka \& Shibazaki 1996) and total luminosities 
exceeding the low state values are said to be in the ``high'' or ``soft'' 
state.  In addition to the thermal component a power-law tail is also seen 
in the spectra of these systems, although this hard component is much less 
luminous than in the low state.  The normalization of the power-law 
component is highly variable, as opposed to the soft blackbody component which 
is quite stable, and the photon index of the power-law is essentially fixed at 
$\alpha_N\sim 2.5$ (Tanaka \& Shibazaki 1996).  {\em Intermediate state}: 
Recently, a third state has been identified (Ebisawa et al. 1994, hereafter 
E94; Mendez \& van der Klis 1997; Belloni et al. 1996), referred to sometimes 
as the ``intermediate state''.  This state is seen during transitions between 
the low and high states.  As the name implies, the observed spectra in this 
state are intermediate in character between the low and high states.  {\em 
Very High state}: Occasionally, in some systems at very high luminosities, 
the non-thermal tail and blackbody component become comparable in flux, 
giving rise to the so called ``very high'' state (e.g. van der Klis 1994, 
Gilfanov et al. 1995).  The power-law component in this case has a photon 
index $\sim 2.5$ and does not show evidence of a cut off even out to a few
hundred keV.  {\em Quiescent/Off state}: Finally, in the last few years 
it has become possible to measure X-ray spectra of systems in the``off'' or 
``quiescent'' state.  These observations have been done on a subset of BHXBs, 
called soft X-ray transients (SXTs), which spend most of their lives in 
quiescence.  Quiescent spectra are distinctly non-blackbody, with photon 
indices somewhat softer than in the low state, and with flux levels lower by 
several  orders of magnitude than in the other four states (McClintock, 
Horne, \& Remillard 1995; Narayan, Barret, \& McClintock 1997, hereafter NBM; 
Narayan, Garcia \& McClintock 1997).

Within the limitations imposed by the uncertainties in the distance to each 
particular system and the mass of the accreting black hole, the five spectral 
states can be arranged in the following sequence of increasing total luminosity:
quiescent state, low state, intermediate state, high state, and very high 
state.  This order seems to be preserved when individual systems undergo 
state transitions and also when one binary is compared to another (provided 
the luminosity is expressed in Eddington units).  However not all BHXBs 
display all the states, and in fact most systems spend the bulk of their 
time in a single state, making only brief excursions to other 
states.  Cygnus X-1, for instance, is generally found in the low state, with 
occasional high state transitions, while LMC X-3 is nearly always in the high 
state (Tanaka \& Lewin 1995).  The transient sources (e.g. A0620-00, 
Nova Muscae, V404 Cyg) are the most interesting in this respect since they 
undergo dramatic outbursts during which their luminosities vary over many 
orders of magnitude.  Consequently, some transients cycle through all five 
states.  Nova Muscae 1991 (also known as GS 1124-683), for instance,  went 
very quickly from quiescence to the very high state and then decayed via the 
high, the intermediate and low states back to the quiescent state (E94).

During the past two decades a great deal of observational data has been 
accumulated on the behavior of BHXBs.  The temporal and spectral 
characteristics of these sources have been widely studied, but the precise 
nature of the different spectral states -- 
their relationships to one another and the mechanisms driving the state 
transitions -- remain very uncertain.  The thermal radiation seen in the high 
and very high states is generally modeled as modified blackbody emission
from a geometrically thin, optically-thick accretion disk extending down 
to the last stable orbit (Shakura \& Sunyaev 1973, Novikov \& Thorne 1973).  
The power-law component seen in most states is attributed to Compton 
upscattering of soft photons in an optically thin hot corona above the disk 
(e.g. Melia \& Misra 1993; Svensson \& Zdziarski 1994) or 
an optically thick Comptonizing cloud (Sunyaev \& Titarchuk 1980).  
However, no model has been able to explain the relationship between 
the thermal and the ``non-thermal'' spectral states and the cause of the 
transition from one to the other remains unknown.  The dramatic outbursts 
displayed by SXTs present an even greater challenge to existing models.  

Since SXT outbursts are similar in some respects to outbursts observed in 
dwarf novae, Mineshige \& Wheeler (1989) and Huang \& Wheeler (1989) 
proposed that the disk instability model developed to explain cataclysmic 
variables operates also in black hole systems.  This model has been fairly 
successful in reproducing the basic features of the observed light curves and 
appears to capture some of the underlying physics of the outbursts in BHXBs 
(see Cannizzo 1993 for a review).  However, Lasota, Narayan, \& Yi (1996)  
and Mineshige (1996) showed that the model is inconsistent 
with the observed  recurrence time between outbursts.  Moreover, since the 
model requires the thin disk to extend down to the last stable orbit, it 
cannot explain the X-ray emission observed in the spectra of quiescent SXTs
(Narayan, McClintock \& Yi 1996, hereafter NMY).  

During the last few years, two new accretion solutions, that go under the 
general name of advection-dominated accretion flows or ADAFs, have been 
extensively studied.  ADAFs are based on the usual $\alpha$-viscosity 
prescription, but differ from the standard thin accretion disk model in that 
a substantial fraction of the viscously dissipated energy is stored in the gas 
and advected to the central object with the accretion flow, rather than
being radiated.  The two branches of ADAFs are distinguished by the physical 
mechanism that causes energy advection.  When the mass accretion rate is very 
high, the flow is optically thick, so that most of the emitted photons are 
dragged into the black hole by the accreting gas.  This regime was first 
considered by Katz (1977) and Begelman (1978) and studied comprehensively by 
Abramowicz et al. (1988).  The second branch  (Narayan \& Yi 1994, 1995a,b; 
Abramowicz et al. 1995; Chen 1995; Chen et al. 1995) occurs when the accretion
rate is lower than a certain critical rate, $\mdot_{crit} \sim 1.3 \alpha^2$ 
in Eddington-scaled units (\S3.3, see also Rees et al. 1982).  In this case, 
the flow is optically thin and the gas is unable to cool efficiently.  The 
energy thus remains in the gas in the form of thermal energy of the particles 
and is once again advected into the black hole. In this paper we focus on the 
second, low-$\mdot$, optically thin branch of ADAFs.  This branch of solutions 
has been shown to be thermally and viscously stable (Abramowicz et al. 1995; 
Narayan \& Yi 1995b), except perhaps for a weak instability associated with 
certain short wavelength perturbations (Kato, Abramowicz, \& Chen 1996) 

Two key assumptions of the optically thin ADAF solutions are that (1) most 
of the viscous energy dissipated in the accretion process goes into 
heating the ions (Shapiro, Lightman \& Eardley 1976, hereafter SLE; Rees et 
al. 1982), and (2) energy transfer from ions to electrons is restricted to 
Coulomb collisions.  In the low density accreting gas Coulomb transfer is 
inefficient, so that the gas becomes a two-temperature plasma (SLE) where 
the ions are nearly virial and contain most of the viscously dissipated 
energy.  Because of the high ion temperature, pressure support is
important and the accreting flow assumes a nearly spherical configuration 
with sub-Keplerian rotation (Narayan \& Yi 1995a).  In contrast to the ions, 
the electrons  radiate 
efficiently via synchrotron and bremsstrahlung emission and inverse Compton 
scattering.  As a result the electrons are much cooler than the ions.

The optically thin ADAF solution has had significant success in explaining 
the quiescent states of several SXTs (see Narayan 1997 for a review).  NMY 
and NBM proposed that the accretion
flow in quiescent SXTs consists of two zones, an inner optically thin ADAF 
extending from the black hole horizon to a transition radius at $\sim 10^4$ 
Schwarzschild radii and an outer thin disk beyond this radius.  Using this 
model they explained the optical and X-ray spectra of two well-studied 
systems, V404 Cyg and A0620-00, while Hameury et al. (1997) found a good fit 
for GRO J1655-40.  As mentioned earlier, the observed spectra are inconsistent 
with any model based on a thin disk extending down to the last stable 
orbit. Further, Lasota et al. (1996b) showed that by limiting the thin disk to 
large radii the recurrence time between outbursts in the disk instability 
model is in better agreement with the observations.  Finally, Hameury et al. 
(1997) demonstrated by means of detailed numerical simulations that the 
model reproduces well the observed optical and X-ray lightcurves of 
GRO J1655-45 during its recent outburst.  In particular, the model naturally 
explains the 6-day X-ray delay observed by Orosz et al. (1997).  This delay is
hard to reconcile with the original version of the disk instability model 
as proposed by Mineshige \& Wheeler (1989).

In the NMY and NBM model of quiescent SXTs, the mass accretion rate is fairly 
low, of the order of $\sim 10^{-3}$ Eddington units, and the luminosity is 
extremely low (because of advection), of the order of $10^{-5}$ to $10^{-7}$ 
Eddington units.  Narayan (1996) extended the model to higher mass accretion 
rates and argued that many of the other spectral states of BHXBs are 
naturally explained with the same model.  The present paper develops these 
ideas further.  We present detailed self-consistent models of accretion flows 
around black holes for a wide range of mass accretion rate and compute 
realistic emission spectra of such flows.  In \S2 we review the basic 
equations governing ADAFs, and summarize the details of our model.   We focus 
mainly on various improvements introduced in this work, namely inclusion of 
the advection term in the electron energy equation whose importance has been 
emphasized by Nakamura et al. (1997), a more realistic treatment of the 
ADAF-disk interaction including X-ray reflection and iron fluorescent line
emission, and calculations of the equilibrium pair density.  In \S3 we explore 
the properties of the model and show that as we increase the mass accretion 
rate and track the transition radius consistently, the model spectra we 
calculate naturally go through a sequence of states which resemble the 
quiescent state, low state, intermediate state and high state.  We also 
propose a speculative model for the very high state.  In \S4 we focus on the 
outburst of a typical SXT, Nova Muscae 1991, and compare the results of our 
model with the X-ray data published by E94.  We conclude with a discussion 
in \S5 and point out some predictions of the model which can be tested with 
future observations.

\section{Modeling the Accretion Flow}

\subsection{Geometry of the Flow and Model Parameters}

We consider a binary system consisting of a Schwarzschild black hole of mass 
$M$ accreting gas from a companion star at a rate $\dot{M}=\mdot \Mdot_{Edd}$, 
where we define the Eddington accretion rate as $\Mdot_{Edd} = 10 L_{Edd}/c^2 
= 1.39\times 10^{18} (M/\msun)\, {\rm g\,s^{-1}}$, corresponding to a 
radiative efficiency of $0.1$.  We take the rotation axis 
of the accretion flow to be inclined at an angle $i$ to the line of sight.
In our model the accretion flow has two distinct zones.  From the outer 
radius of the flow, $r_{out}$ (hereafter all radii are in Schwarzschild units, 
$R_{Schw} = 2 G M/c^2$), to a transition radius, $\rtr$, the gas accretes via 
a standard thin disk with a hot corona above the disk.  For $r < \rtr$ the 
entire flow is assumed to be hot and quasispherical.  We model both the hot 
gas in the inner region and the gas in the corona above the outer thin disk 
as an advection-dominated accretion flow (ADAF).  We make no distinction 
between these two regions except for the fact that the mass accretion rate 
in the corona is a fraction of the total $\mdot$.  

Following NBM, we assume that $\mdot$ in the corona 
is inversely proportional to $r$, i.e.
\be
\label{mdotc}
\mdot_c = \mdot_{c,0} \left(\frac{\rtr}{r}\right), \ \ \ \ r \geq \rtr.
\ee
In general we take $\mdot_{c,0} = \mdot$ (\S\S3.1-3.3), except when the thin 
disk extends all the way to the last stable orbit, in which case we allow 
$\mdot_{c,0}$ to be less than $\mdot$ (see \S\S3.4,3.5).  The functional 
form of $\mdot_c$ in (\ref{mdotc}) is quite arbitrary, though it is not 
implausible.  Our physical picture of the corona is that it is a hot accretion 
flow above the thin disk which is fed by evaporation of mass from the 
surface of the disk (NBM).  Clearly, $\mdot_c$ must increase inward, and 
considering that the energy required for evaporation ultimately has to come 
from gravity (cf. the discussion of ``siphon flow'' in Meyer \& 
Meyer-Hofmeister 1994), a form such as (\ref{mdotc}) appears likely.  In 
principle, we could make $\mdot_c \propto r^{-\zeta}$ and treat $\zeta$ as 
a free parameter, but it is not clear that this would lead to any further 
insights into the physics of the corona.

The mass $M$ of the black hole, inclination $i$ of the binary, and the outer 
accretion disk radius $r_{out}$ are often constrained by observations.  Also
the transition radius, $\rtr$, can be estimated from the maximum width of the 
H$_{\alpha}$ emission line, as discussed in NMY and NBM.  However, there is 
in general no direct estimate of the mass accretion rate $\dot{m}$, so we 
consider it an adjustable parameter.

Our physical model of the ADAF is very similar to that described by NBM.   
We use the standard viscosity prescription (Frank, King, \& Raine 1992) 
parameterized through the viscosity coefficient $\alpha$, and assume that 
the hot gas and tangled magnetic fields in the ADAF are roughly in 
equipartition, so that $p_{gas} = [\beta/(1-\beta)] p_{mag} = \beta p_{tot}$.  
We take $\alpha$ and $\beta$ to be independent of radius.  

In the remainder of this section we summarize how we compute the 
physical properties of the ADAF.  Many of the details are described 
exhaustively in NBM, so we emphasize here only the modifications and 
enhancements that we have introduced in the present calculations.

\subsection{Flow Dynamics and Energy Balance}

Following NBM, we model the ADAF as a set of nested spherical shells truncated 
near the pole to mimic the flattening of the density profile.  The shells 
are uniformly spaced in $\log{(r)}$ and are centered at radii $r_j$.  
Within each shell the radial velocity $v_j$, angular velocity $\Omega_j$, 
isothermal sound speed  $c_{s,j}$, and viscous heating rate $Q_j^+$ are 
calculated using the global dynamical solutions described by Narayan, Kato 
\& Honma (1997) and Chen, Abramowicz \& Lasota (1997).  Each global solution 
is uniquely specified by the viscosity parameter $\alpha$, the ratio of the 
specific heats of the accreting gas $\gamma = (8-3 \beta)/(6-3 \beta)$ (see 
Esin 1997), and the advection parameter $f(r)$.  The last quantity is defined 
to be the ratio of the energy advected by the gas to the total energy 
dissipated in the ADAF and is in general a function of $r$, however, in 
calculating the dynamical solutions we replace it by a constant $f_{av}$ as 
described below.  The global solutions satisfy the basic conservation laws of 
accretion flows -- conservation of mass, radial momentum, angular momentum 
and energy -- and ensure the dynamical self-consistency of the models.

In previous applications of the ADAF model to BHXBs (NMY, NBM, Hameury et al. 
1997), the objective was to model only the quiescent state of the systems in 
which $\mdot$ is very low.  At such $\mdot$, the radiative efficiency is 
extremely low (cf. Narayan \& Yi 1995b), which implies that $f(r) \rightarrow 
1$ at each radius.  It was therefore possible in those studies to set $f(r)$ 
equal to unity without loss of self-consistency (NBM, NMY).  
The goals of the present paper are more ambitious.  Here we attempt to explain 
the entire range of observed BHXB spectral states, including quite 
luminous systems for which $f$ is significantly less than 1.  We therefore 
need a more sophisticated scheme where the global solution corresponds to
the actual value of $f$ as determined by the cooling. For this purpose,
we characterize the ADAF by the following average advection parameter, 
$f_{av}$,
\be
\label{fav}
f_{av} = \frac{\sum_j {Q^{adv}_j}}{\sum_j{Q^+_j}},
\ee
where $Q^+_j$ and $Q^{adv}_j$ are the viscous heating rate and the rate of 
energy advection in the $j^{\rm th}$ shell respectively, and the sums are 
computed over all spherical shells.  

In regions where the gas flow is a pure ADAF ($r<\rtr$), the
advected energy is simply the difference between the viscous heating and
total radiative cooling, 
\be
\label{adv1}
Q^{adv}_j = Q^+_j - Q^{-}_j, \ \ \ \ r < \rtr.  
\ee
The quantity $ Q^+_j$ is obtained from the global dynamical solution while 
$Q^{-}_j$ is known once the radiative processes and spectrum have been 
calculated
(\S2.3).  However, at radii where the hot flow is in the form of a corona 
above the thin disk, we need to take into account the energy required to
evaporate the gas from the thin disk into the corona.  The evaporation 
energy in shell $j$ is estimated to be the thermal 
energy of the gas evaporated locally, i.e.
\be
Q^{evap}_j = \left[\Mdot_c (r_j)-\Mdot_c (r_{j+1})\right] 
\left[\frac{3}{2} \frac{k T_i}{\mu_i m_u} + a(T_e) 
\frac{k T_e}{\mu_e m_u}\right], \ \ \ \ r \geq \rtr,
\ee
where $T_i$ and $T_e$ are the ion and electron temperatures in the ADAF, 
the coefficient $a(T_e)$ describes the thermal energy of the electrons
(Appendix A), $\mu_i = 1.23$ and $\mu_e = 1.14$ are the effective molecular 
weights of the ions and the electrons respectively, computed for a hydrogen
mass fraction of $75\%$ (cf. Narayan \& Yi 1995b), and we have neglected the 
thermal energy of the gas in the thin disk.   Where does the energy for 
evaporation come from?  The uncertainties in the physical processes 
responsible for evaporation make it difficult to answer this question.  
We simply assume that the energy is supplied entirely 
by the hot flow, perhaps via ion and electron conduction as
suggested by Meyer \& Meyer-Hofmeister (1994).  Under this assumption, the 
energy advected by the gas in the accretion flow is given by 
\be
\label{adv2}
Q^{adv}_j = Q^+_j - Q^{-}_j - Q^{evap}_j, \ \ \ \ r \geq \rtr.
\ee

Equations (\ref{fav})-(\ref{adv2}) give the average 
advection parameter $f_{av}$ of the flow.  Using this $f_{av}$, we compute 
the global solution using the method described in Narayan, Kato, \& Honma 
(1997).  As mentioned above, this gives the values of $v_j$, $\Omega_j$, 
$c_{s,j}$, and $Q_j^+$ in each shell $j$.  The radial velocity, combined 
with the equation of mass conservation, allows us to compute the mass density 
of the gas $\rho$ at each radius, as well as the ion and electron number 
densities, $n_i$ and $n_e$.

To solve for the ion and electron temperatures, $T_i$ and $T_e$, in the
two-temperature ADAF, we need two relations for each radial shell.  One 
constraint is provided by the equation for the local gas pressure,
\be
\label{pgas}
P_{gas} = \beta \rho c_s^2 = n_i k T_i + n_e k T_e,
\ee
while the second relation is given by the energy balance equation for 
the electrons,
\be
\label{eenergy}
Q^{e,adv}_j = Q^{ie}_j + \delta Q^+_j -Q^-_j.
\ee
Equation (\ref{eenergy}) states that the rate of energy advection by 
electrons (left-hand side) 
is equal to the rate of energy transfer from ions to electrons via Coulomb 
collisions, $Q^{ie}_j$, plus a fraction $\delta$ of the viscous dissipation,
minus the radiative cooling rate, $Q^-_j$.  In all our models we set $\delta 
= 10^{-3}\sim m_e/m_p$, which means that only $10^{-3}$ of the viscous 
heat goes directly into the electrons.  The exact value we choose for this 
constant is not important.  We find that the results presented in this paper
are practically unchanged 
for any value of $\delta \lsim 10^{-2}$. In our earlier work, we ignored 
$Q^{e,adv}_j$, assuming that it is small relative to the terms on 
the right.  However, Nakamura et al. (1997) have recently shown that 
electron energy advection can be important in some cases.  We therefore 
include this term and compute it self-consistently using the relations given 
in Appendix A.  We find that at high $\mdot$ the models are essentially 
unaffected by the electron advection term.  However, at low $\mdot$, this 
term acts as an extra energy source for the electrons, thus increasing the 
overall luminosity of such flows (see \S3.2).   Equations (\ref{pgas}) and 
(\ref{eenergy}) allow us to solve for $T_i$ and $T_e$ consistently in each 
radial shell.

\subsection{Radiative Transfer, Reflection, and the Iron Fluorescence Line}

The rate of cooling of the accreting gas is calculated using the iterative
scattering method described in detail in NBM.  Given the current guess of the
temperature and density profile, the synchrotron and bremsstrahlung emission 
from each shell is first computed.  Then, using the probability matrix 
elements $P^{aa}_{ij}$ (or $P^{da}_{ij}$), which give the probability that a 
photon  emitted in shell $i$ of the ADAF (or ring $i$ of the thin disk) is 
scattered by an electron in shell $j$ of the ADAF, the rate of cooling through 
Compton scattering is calculated (see NBM for details).  The iteration method 
ensures that multiple scattering within the ADAF, as well as the interaction 
between the ADAF and the outer disk are properly taken into account.  
Gravitational redshift, Doppler boosts and ray deflections are neglected.

The emission from the thin disk is calculated using the standard 
multicolor blackbody method (cf. Frank et al. 1992).  The effective 
temperature of the emission from ring $i$ is  determined by
\be
\sigma T^4_{eff,i} A_i= Q^+_{disk,i} + Q^{abs}_i,
\ee
where $A_i$ is the surface area of ring $i$ of the thin disk, $Q^+_{disk,i}$ 
is the viscous dissipation rate in the ring due to the local mass accretion 
rate in the disk, $\mdot_d = \mdot - \mdot_c$, and $Q^{abs}_i$ is the
rate at which energy is absorbed by the disk as a result of irradiation 
from the ADAF.  The rate of irradiation is computed using another matrix 
$P^{ad}_{ij}$ which gives the probability that a photon emitted in shell 
$i$ of the ADAF is incident on ring $j$ of the thin disk.

NBM made the simplifying assumption that all the radiation incident on the 
thin disk is absorbed and reprocessed into blackbody radiation, ignoring the 
fact that a significant fraction of the irradiating flux is scattered by the 
material near the disk surface and effectively reflected.  This is a valid 
approximation for modeling quiescent SXTs in which the disk is restricted to 
large radii so that the contribution of the disk to the observed spectrum is 
negligible.  However, in the present work we consider a range of transition 
radii and need to model the reflection more carefully.

Given the incident spectrum, we compute the spectrum of the reflected 
radiation using the angle-averaged 
Green's function for Compton reflection of monoenergetic photons from an 
optically thick cold disk (White, Lightman \& Zdziarski 1988;
Lightman \& White 1988).  The Green's function, $G (E,E_0)$, is defined 
as the probability that an incident photon with initial energy $E_0$ 
emerges from the cold medium with energy $E$.  Then, for the net irradiating 
spectrum $\sum_j {L_{j,i} (E_0)}$, where ${L_{j,i} (E_0)}$ is the specific 
luminosity incident on ring $i$ of the thin disk from ADAF shell $j$, the 
reflected spectrum is simply
\be
L^{\rm refl}_i (E) = \sum_j{\int_{E}^{\infty} {G (E,E_0)\,L_{j,i} (E_0)\, 
d E_0}}.
\ee
For photons with initial energies below $\sim 15$ keV, the value of the 
Green's function depends on the bound-free absorption opacity of the cold 
material, which is a function of the ionization structure.  Since the thin 
disk is relatively cold, its ionization state is largely determined by the 
irradiating X-ray flux, and in principle can be calculated self-consistently 
in our model.  However, the ionization does not have a strong effect on the 
dominant part of the reflected spectrum between 10 and 300 keV (Lightman \& 
White 1988), so for simplicity, we have used the absorption opacity for a 
completely neutral plasma with cosmic metal abundances (Morrison \& McCammon 
1983).  Having computed the reflected spectrum, the absorbed energy 
$Q^{abs}_i$ is simply the difference between the incident and reflected 
luminosities
\be
Q^{abs}_i = \int_0^{\infty}{[\sum_j{L_{j,i} (E)}-L^{\rm refl}_i (E)] d E}.
\ee
Both the reflected component and the standard thin disk blackbody emission
then become a part of the iterative scattering procedure.  This allows us to 
take into account Comptonization of the disk photons by the coronal electrons.

We also model the iron K$\alpha$ fluorescence line which is produced through 
photoelectric absorption of X-ray photons with energies above the K-shell 
absorption edge, $E_K$.   Such line features have been observed in the spectra 
of Seyfert I galaxies (e.g. Tanaka et al. 1995; Nandra \& Pounds 1994) and in 
the galactic BHXB Cyg X-1 (e.g. Ebisawa et al. 1996; Gierli$\acute{\rm n}$ski
et al. 1997).  As in our calculations of the reflection component, we assume 
that the disk contains mainly neutral material with a cosmic abundance of 
$3.3\times10^{-5}$ iron atoms per hydrogen atom (Morrison \& McCammon 1983).  
We compute the flux in the iron line using the function ${\cal G}(E_0, \cos 
\theta_0)$ given by George \& Fabian (1991), which they defined as the 
fraction of ADAF photons of initial energy $E_0$, incident on the thin disk at 
an angle $\theta_0$, that give rise to iron fluorescence photons able to escape 
the disk.  The resultant luminosity in the iron line from ring $i$ of the thin 
disk is then
\be
L^{\rm Fe}_i = \sum_j {\int_{E_K}^{\infty} {L_{j,i}(E_0)\, 
{\cal G}(E_0, \langle \cos{\theta_0} \rangle_{j,i})\, d E_0}},
\ee
where $\langle \cos \theta_0 \rangle_{j,i}$ is averaged over all photons from 
ADAF shell $j$ incident on ring $i$ of the thin disk, and the K-shell 
absorption edge is $E_K = 7.1\,{\rm keV}$ for neutral iron atoms (George \& 
Fabian 1991).  In our model both $L_{j,i}(E_0)$ and 
$\langle \cos{\theta_0} \rangle_{j,i}$ are completely specified for every pair 
of indices  $j$ and $i$, so that the integral is easily computed.  We do not 
model the K$\alpha$ line profile in detail, but simply take it to be a 
Gaussian centered at 6.4~keV with a fractional Doppler width equal to 
$\sin{i}/(2 r)^{1/2}$, where $i$ is the disk inclination.  

\subsection{Pair Production}

The electrons in the inner parts of the ADAF reach relativistic temperatures 
of $10^9-10^{10}\,{\rm K}$ making pair production possible there, as a result 
of particle-particle, particle-photon, and photon-photon interactions.
Kusunose \& Mineshige (1996) and  Bj$\ddot{\rm o}$rnsson et al. (1996) 
concluded that the effect of pair production on the structure of ADAFs is 
small as long as $\alpha < 1$.  Since our models are calculated with 
$\alpha=0.25$, this would suggest that pair processes are not likely to be 
important.  Nevertheless, in the interest of consistency we have included 
pair interactions, though with some simplifications as noted below.  We 
confirm that pair effects are negligible for the branch of (low-$\mdot$) 
ADAFs studied in this paper.

In our calculations we assume that the pair fraction, defined as the ratio 
of the positron and proton number densities, $z=n_+/n_p$, is low.  In 
addition, although we do include particle-photon, and photon-photon 
interactions 
in our calculations, we assume that particle-particle pair production 
dominates.  This is valid since ADAFs produce very few photons with energy 
$\gsim m_e c^2$ needed to create pairs. These assumptions allow us to neglect 
the photon balance equation.  Similarly, we drop the advection term in the 
continuity equation for pairs.  This term acts as an additional sink term and 
reduces the equilibrium pair density; since we neglect it, our results 
correspond to an upper limit on $z$.  With these simplifications, the problem 
reduces to solving just the pair balance equation,
\be
\label{pairs}
(\dot{n}_+)_{\rm prod} = (\dot{n}_+)_{\rm ann},
\ee
where the term on the left is the rate of production of pairs and that on the 
right is the annihilation rate.  Appendix B gives detailed expressions (taken
from the literature) for the various pair production and pair annihilation 
processes in a thermal plasma.  We also show there how equation (\ref{pairs}) 
can be reduced to a quadratic equation in $z$ and solved iteratively.  

Our results indicate that pair production is not a significant process in 
ADAFs, at least in the parameter range explored in this paper (see \S3 for 
details); in all cases we find $z\lsim 10^{-5}$.  This result is not 
surprising.  Even though electrons in advection-dominated flows can reach 
temperatures on the order $\sim m_e c^2$, the highest temperatures are 
achieved only at 
relatively low $\mdot$, where the optical depth and the luminosity of the 
flow are very low.  At accretion rates near Eddington, the flow is much 
cooler with $T_e \lsim 0.2 m_e c^2$, and the bulk of the emission is 
significantly below the pair production threshold.  In other words, the 
compactness parameter $l = L_{\gamma} \sigma_T/R m_e c^3$, where $L_{\gamma}$ 
is the luminosity above $\sim 511\,{\rm keV}$ and $R \sim 10 R_{Schw}$ is 
the size of the emitting region, is always significantly below unity, 
ensuring that pair processes do not dominate the physics of the flow.

\subsection{Numerical Solution}

Given the system parameters $M,\ i,\ r_{out},\ \rtr,\ \alpha,\ \beta$ and  
$\delta$, the mass accretion rate $\mdot$, and neglecting pair production,
the calculation proceeds as follows.  We begin with a trial value for the 
advection parameter $f_{av}$, and obtain the global dynamical solution.  
This gives the following gas properties in the various radial shells of the 
ADAF and the corona: $v_j,\ \Omega_j,\ c_{s,j}^2,\ \rho_j,\ n_{e,j},\ Q_j^+$.  
Next we assign reasonable values to the electron temperature $T_{e,j}$ in 
all radial shells, and calculate the ion temperature $T_{i,j}$ via equation 
(\ref{pgas}), and also the quantities $Q_j^{ie}$ and $Q_j^{e,adv}$.  We 
now compute the spectrum of radiation emitted by each radial shell via 
synchrotron, bremsstrahlung and Compton scattering.  This calculation is 
done by the iterative scattering method of Poutanen \& Svensson (1996) as 
described in NBM.  It includes full coupling between the ADAF and the thin 
disk, namely irradiation of the ADAF by disk radiation, irradiation of the 
disk by the ADAF, reflection from the disk, generation of the iron K$\alpha$ 
line, etc.  Once the radiation transfer problem is solved, we obtain $Q_j^-$ 
corresponding to each shell and check whether the electron energy balance 
equation (\ref{eenergy}) is satisfied. If necessary, we iteratively modify the 
electron temperature profile $T_{e,j}$ until equation (\ref{eenergy}) is 
satisfied.  Once this is achieved, we compute $f_{av}$ via equation (\ref{fav}) 
and check if it agrees with our trial value.  If not, we choose a new value 
for $f_{av}$ and repeat the rest of the calculation described above.

To include pair processes we need to compute one additional quantity as a 
function of radius, namely the pair fraction $z_j$.   This is done by solving 
equation (\ref{pairs}) in each shell.   However, this is not trivial 
and requires an additional iteration loop in the calculation just described 
because the radiation solution is sensitive to $z_j$ and the pair production 
rate, $(\dot{n}_+)_{{\rm prod},j}$, depends on the radiation field.

In some cases, for instance when we want to determine the critical 
accretion rate $\mdot_{crit}$ (see \S3.3) or the coronal accretion rate 
$\mdot_{c,0}$ (see \S\S3.4, 3.5), we are interested in a solution with a 
given value of $f_{av}$.  The calculation then proceeds as described above, 
except that $\mdot$ or $\mdot_{c,0}$ is the adjustable parameter rather 
than $f_{av}$.  We start with a guess for $\mdot$  
or $\mdot_{c,0}$ and adjust it at each iteration to ensure that $f_{av}$ 
computed via equation (\ref{fav}) is equal to the desired value.  

The calculations presented here are a significant improvement over 
earlier versions.  In particular, by introducing the average advection 
parameter $f_{av}$, the dynamical solution is made consistent with the 
radiative transfer and cooling part of the calculation.  The only serious 
approximation is that we set $f (r)$ equal to a constant, $f_{av}$, whereas 
$f$ is in general expected to vary with radius (Esin 1997).  In order to 
compute a realistic $f(r)$, we would need to include a radial energy transport 
term in the electron energy equation (\ref{eenergy}).  Potential transport 
mechanisms include convection (cf. Narayan \& Yi 1994, 1995a; Igumenshchev,
Chen, \& Abramowicz 1996) and ion and electron conduction.  These transport
phenomena can cause a significant flow of energy from one radius to another, 
and they are not
easy to calculate from first principles.  Furthermore, radial energy 
transport may tend to smooth out variations in $f(r)$, driving the 
system toward a constant $f_{av}$. We therefore feel that it is reasonable 
to set $f (r) = f_{av}$ in determining the global dynamics of the flow.  
In effect, we treat the ADAF under a one-zone approximation for the purposes 
of the advection parameter $f$, while calculating the 
detailed variation with radius of all other gas properties.

Note that our models satisfy two separate energy equations: an energy
equation for the electrons given in equation (\ref{eenergy}), and an energy
equation for the total gas (ions plus electrons) which is satisfied by
the global flow solution (see Narayan et al. 1997; Chen et al. 1997).
Thus, we treat both the ions and the electrons self-consistently.
Nakamura et al. (1997) describe another approach where they write down
individual energy equations for the ions and the electrons.  The two
approaches are in principle equivalent.

\section{Modeling the Spectral States of Black Hole X-Ray Binaries}

In this section we investigate the spectral properties of the model
described in the previous section and demonstrate how the various
spectral states observed in BHXBs can be understood in the context of
this model.  The application of the model to the quiescent state of
BHXBs has been extensively discussed by NMY, NBM and Hameury et
al. (1997).  Here we summarize the main results of that work and use
the quiescent state as the starting point to explore the effect of
increasing the mass accretion rate.  A preliminary discussion of some
of the new results presented here may be found in Narayan (1996).

Figure 1 summarizes the broad features of the scenario we develop in
detail in the succeeding subsections.  The figure shows how the geometry 
of the flow changes as the mass accretion rate is varied, and identifies 
various well-defined stages in the evolution of the flow with the known 
spectral states of BHXBs.  At the lowest $\dot m$ ($\lsim10^{-2}$) we have 
the quiescent state, where the thin accretion disk is truncated at a fairly 
large transition radius and the accretion flow switches to an ADAF farther in 
(the bottom panel in Figure 1).  The physics which determines the exact value 
of the transition radius is not well understood, but it is likely to be 
related to the angular momentum of the material flowing in from the companion 
star.  In low-mass BHXBs, where mass is transferred via Roche lobe overflow, 
the circularization radius of the incoming stream is large ($r_{circ}\sim
10^4-10^5$).  Making the reasonable assumption that $r_{tr}$ is smaller than
$r_{circ}$ by a factor of a few, we expect in these systems $r_{tr}\sim10^4$.  
In high mass BHXBs, on the other hand, the angular momentum of the accreting 
material is much lower since mass transfer is driven by winds and the
gravitationally captured gas circularizes at a much smaller radius
(cf. Frank et al. 1992).  Thus, $r_{tr}$ may be as small as
$10^2$ or even smaller in these systems.

The other panels in Figure 1 show the effect of increasing $\dot m$ above 
its quiescent range.  For $\mdot$ up to a critical value $\dot m_{crit}\sim
0.08$ (the exact value depends on the value of $\alpha$ and $\beta$, \S3.2), 
the geometry remains essentially the same as in the quiescent state.  However, 
since the radiative efficiency of the flow increases rapidly with increasing 
$\dot m$ (Narayan \& Yi 1995b), the flow becomes quite luminous.  We identify 
such flows with the low state.  Once $\dot m$ exceeds $\dot m_{crit}$, the 
hot ADAF zone radiates too efficiently to remain advection-dominated.  As a 
result, the ADAF begins to shrink in size and the inner edge of the thin 
disk moves inward to smaller
radii.  We identify such flows, where the central ADAF is still
present but with a reduced size compared to the quiescent and low
state, with the intermediate state.  At still higher $\dot m$, the
central ADAF zone disappears altogether and the thin disk moves in
all the way to the marginally stable orbit.  A somewhat weak corona is
present above the disk.  We associate this configuration with the high state.  
Finally, at accretion rates close to Eddington we assume that the flow makes a
transition to a different state where the corona is much more massive
and active.  We tentatively identify this flow configuration with the very 
high state, although this is the weakest aspect of our scenario.

In the calculations presented below, unless otherwise stated, we use
the ``standard'' parameter set summarized in Table 1.  For those quantities that
can be derived from observations we adopt system parameters corresponding 
to the SXT Nova Muscae 1991 (see \S4.1).  For the parameter
$\delta$, we invariably choose a value of $10^{-3}$, but this quantity
plays no role in the calculations presented here and could equally
well be set to zero.  This still leaves two important parameters,
$\alpha$ and $\beta$.  We choose what we consider to be the most
natural values for these.  We assume that the magnetic field is in
equipartition with the gas pressure, which corresponds to $\beta=0.5$. The
assumption of equipartition fields is very common in many areas of high
energy astrophysics.  In particular, equipartition fields are quite plausible 
in accretion flows since the Balbus-Hawley (1991) instability, which 
presumably is the mechanism whereby the field grows, is known to shut off 
when $\beta\sim0.5$.  For the viscosity parameter $\alpha$ we follow
the prescription suggested by Hawley \& Balbus (1996, see also Hawley, 
Gammie \& Balbus 1995, 1996), viz. $\omega_{R \phi} \sim 0.5-0.6 p_{mag}$, 
where $\omega_{R \phi}$ is the shear stress.  For $\beta=0.5$ this gives a
value of $\alpha$ in the range $0.2-0.3$.  We choose a value in the
middle of the range, $\alpha=0.25$.

It should be emphasized that we have no adjustable parameters in the
calculations presented in this paper except for the mass accretion rate 
$\dot m$ (and to a very limited extent $r_{tr}$),  We could, in principle, 
optimize $\alpha$ and $\beta$ so as to obtain the best fit between the model 
and the Nova Muscae data discussed in \S4, but we feel that the data are not 
really good enough for such an exercise.  

\subsection{Quiescent State}

Between successive outbursts, transient BHXBs are generally found in
the quiescent state, where the observed luminosity is many orders of
magnitude below Eddington.  In the systems for which optical and X-ray
observations in quiescence exist (A0620-00, V404 Cyg, and GRO J1655-40), 
the data are explained quite well with the model 
shown in the bottom panel of Figure 1 (NMY; NBM; Hameury et al. 1997).  In
addition, the same model also explains observations of the
underluminous supermassive black hole at the center of our Galaxy, Sgr
A$^*$ (Narayan et al. 1995), as well as the supermassive black hole in
NGC 4258 (Lasota et al. 1996a).  These applications represent the most
important successes so far of the ADAF model.

On the basis of the above work, we define the quiescent state of
BHXBs to correspond to mass accretion rates $\mdot \lsim 10^{-2}$.
In Figure 2 we plot a sequence of spectra computed with our standard
parameter set. The blackbody-like optical/UV peak is produced by
self-absorbed synchrotron emission, while the peak at high energies
$\sim 100\, {\rm keV}$, visible especially clearly at very low
$\mdot$, is due to bremsstrahlung.  Inverse Compton scattering of
synchrotron photons by the hot ADAF electrons is responsible for the
rest of the spectrum and produces one or more bumps between the
synchrotron and bremsstrahlung peaks.

With increasing $\mdot$, two effects modify the shape of the spectrum.
Since gas pressure increases roughly linearly with $\mdot$, the
magnetic field grows as $\mdot^{1/2}$ (equipartition); consequently,
the synchrotron peak moves towards higher frequencies.  At the same
time, the photon spectral index in the 1-10 keV X-ray band steepens
from $\alpha_N \sim 1.7$ to $\sim 2.2$.  This is because at low
$\mdot$ high energy photons are produced primarily by bremsstrahlung
emission, whereas at higher $\mdot$ Comptonization dominates, which
results in a smoother but steeper spectrum.  Since the radiative
efficiency of the flow, defined as the ratio $\e_{-1} = L_{bol}/0.1\Mdot c^2$,  
is proportional to $\mdot$ (Narayan \& Yi
1995b), the overall normalization of the spectrum changes roughly as
$\mdot^2$.   Even for $\mdot \sim 0.01$, the efficiency is quite low 
($\e_{-1} \sim 0.02$) and these quiescent state models are very underluminous.

In the models shown here, the radiation from the outer thin disk is
negligibly small compared to the emission from the ADAF.  This would
not be true if the transition radius were closer to the black hole, as
we discuss at the end of \S3.2 (see also Figure 5).

\subsection{Low State}

The five heavy curves in Figure 3a show a sequence of model spectra
where the flow geometry is exactly the same as in the quiescent state;
now, however, $\mdot \ge 10^{-2}$ (Figure 1).  At these
relatively high accretion rates, Comptonization of synchrotron photons
by the hot gas in the ADAF constitutes the dominant cooling mechanism.
As $\mdot$ increases, the optical depth of the ADAF goes up causing a 
corresponding increase in the Compton $y$-parameter.  Consequently,
the high energy part of the spectrum becomes harder and smoother, and
the photon index in the X-ray band reverts from $\alpha_N \sim 2.2$
back to $\sim 1.5$.  The radiative efficiency for the highest $\mdot$
model is reasonably high and the total energy output is on the order of 
$10^{37}\,{\rm erg\,s^{-1}}$, which
corresponds to a few percent of the Eddington luminosity (for the
assumed $6 \msun$ black hole).  Most of the flux is emitted at around
100 keV and the spectrum falls off exponentially at higher energies.
The model spectra shown in Figure 3a reproduce well both the spectral
shapes and X-ray luminosities of BHXBs observed in the low state
(Tanaka \& Lewin 1995; Tanaka \& Shibazaki 1996).  Based on this 
correspondence, we identify these models with mass accretion rates 
$10^{-2} \lsim \mdot \lsim 10^{-1.1}$ with the low state.

The electron temperature profiles corresponding to our low state
models are shown in Figure 3b.  There is a clear anti-correlation
between the mass accretion rate and the electron temperature in the
inner part of the flow ($r \leq 10^2$).  This trend arises because, as
$\mdot$ increases, a larger fraction of the dissipated energy is
radiated away.  The flow thus becomes more cooling dominated and its
temperature decreases.  We note that OSSE
observations of the X-ray nova GRO J0422+32 in the low state showed an
inverse dependence between the high energy cutoff energy and the 100
keV flux (Kurfess 1996), as our low state models predict.  

The decrease of $T_e$ with increasing mass accretion rate is important 
in considering pair production in the ADAF.  At low $\mdot$ the radiative 
efficiency of the flow is small, and at high $\mdot$ the electron 
temperature is considerably below the pair creation threshold of 511 keV.
Consequently, only the particle-particle pair production processes are 
important, which are very inefficient for $T_e \lsim 10^11 K$.  We obtain
the equilibrium pair density of $z\sim 5\times 10^{-7}$ at $\mdot=10^{-2}$
and $z\sim 6\times 10^{-9}$ at $\mdot=\mdot_{crit}$.  Our values of $z$ are 
significantly lower than those obtained by Kusunose \& Mineshige (1996) for 
ADAFs.  The difference is most likely due to the fact that they considered
only cooling by bremsstrahlung and inverse Compton scattering, and consequently
obtained much higher electron temperatures.

Most of the emission in the low state is from the ADAF, and the outer thin 
disk is almost invisible.  When $\mdot$ is low, the optical/UV peak in the 
spectrum is mostly due to self-absorbed synchrotron emission and the thin 
disk contribution is hardly seen.  However, when $\mdot$ is high,
the thin disk blackbody component becomes more 
prominent and is visible as a peak in the optical.  The transition is
especially clear in the spectrum corresponding to $\mdot = 10^{-1.4}$,
where both the synchrotron peak and the shoulder corresponding to the disk 
component are evident.  This discussion pertains to models with a large 
transition radius, $r_{tr}\sim10^4$.  We discuss later the effect of a
smaller transition radius.

One of the new effects we have introduced in the present model is the
electron advection term in the energy balance equation (see \S2.2 and
Appendix A and the original discussion of Nakamura et al. 1997).  To
illustrate the effect of this term, we show by the thin lines in Figure 3a 
spectra computed without including electron advection.  For high
$\dot m$ the difference between spectra with and without electron
advection is very small.  However, the advective heating term depends
linearly on $\mdot$, while the ion-electron Coulomb exchange rate,
$Q^{ie}$, is proportional to $\mdot^2$.
Thus, with decreasing $\mdot$ we expect that advection will become
the dominant heating mechanism for the electrons.  This trend is
evident in Figure 3a.  For $\mdot=10^{-2}$, for example, the inclusion
of electron advection makes the system nearly twice as luminous as
when the term is not included.

Figure 3b makes a similar comparison of the electron temperature profiles.
The solid thin line shows $T_e (r)$ for $\mdot=10^{-2}$
when electron advection is not included.  This is to be compared with
the solid thick line which corresponds to the same model with electron
advection included.  We see that the addition of the electron advection term
increases the electron temperature in the inner parts of the flow by
acting as an extra heating term.  In the outer parts of the flow,
however, the electron temperature is reduced compared to its old
value.  Thus the energy output is increased in the inner parts of the
flow, and decreased in the outer parts, and the emission becomes more
centrally concentrated.
 
In Figures 4a and 4b we demonstrate how low state spectra vary as a function 
of other model parameters.  The effect of varying the inclination angle
$i$ is illustrated in Figure 4a for $\mdot=10^{-1.4}$.  When the
system is viewed almost face-on ($i=5^{\circ}$) it appears brighter by
a factor of $\sim 2$ compared to an edge-on system ($i=90^{\circ}$).
Both the disk and the ADAF contribute to this effect.  As $i$
increases, the projected area of the thin disk visible to the observer
decreases as $\cos{i}$, and the corresponding blackbody component in
the spectrum decreases proportionately.  The hot ADAF is not
completely spherical either, as the flow is flattened at the poles.
This leads to an anisotropy in the optical depth from the center to the edge
of the ADAF.  Thus, the hot Comptonized photons from the
center of the ADAF tend to escape preferentially along the rotation
axis rather than along the equator, making face-on systems brighter.
Clearly, this effect is important only for relatively large $\mdot$
when most photons are scattered at least once before escaping.  At
lower $\dot m$, the ADAF is optically thin and its emission is virtually 
independent of $i$ (except for relativistic Doppler factors, see 
Jaroszy$\acute{\rm n}$ski \& Kurpiewski 1997), and only the rather weak thin 
disk component varies with inclination.

The shape of the iron line is again a function of $i$.  In our model
the line width is simply proportional to $\sin{i}$, so that when the
system is viewed face-on the 6.4 keV feature is narrower and more
prominent.  In our model the equivalent width of the iron line has the
same value (a few eV) for all $i$, since we do not take into
account the dependence on the viewing angle.  In a more realistic
calculation, the equivalent width falls off sharply for $i\gsim
60^{\circ}$ (George \& Fabian 1991).

In Figure 4b we show how low state spectra depend on the values of $\alpha$ 
and $\beta$.  The three spectra shown here each correspond to the critical 
accretion rate, $\mdot_{crit}$, the maximum $\mdot$ for which the low state 
is possible (see \S3.3 for a more detailed discussion of $\mdot_{crit}$).
The solid and dotted curves correspond to $\alpha = 0.2\ {\rm and}\ 0.3$, 
respectively.  We see that an increase in $\alpha$ leads to an increase in 
the value of $\mdot_{crit}$ (the scaling is proportional to $\alpha^2$, see 
Narayan \& Yi 1995b), which increases the maximum 
luminosity that one can observe in the low state.  In addition, the larger 
$\alpha$ also causes the spectrum to be harder.  The dashed line in Figure 4b
demonstrates the effect of increasing $\beta$ from our standard value
of 0.5 to 0.95.  Since $(1-\beta)$ is the ratio of the magnetic
pressure to the total pressure in the ADAF, the change in $\beta$
causes the magnitude of the $B$-field to be reduced by a factor of
$\sqrt{10}$.  Consequently, the ADAF produces less synchrotron
emission and this causes the electron temperature to go up in order to
maintain the cooling rate.  The increase in $T_e$ is clearly seen in
Figure 4b: the spectrum corresponding to $\beta=0.95$ (indicated by
the dashed line) is harder and cuts off at a higher energy than the
equivalent $\beta=0.5$ model (dotted line).  The spectrum also has a
slightly harder slope.  

In those BHXBs where the donor star is more massive than the accreting black 
hole (e.g. Cyg X-1) the mass transfer is often driven by a stellar wind from 
the companion.  In this case the net angular momentum of the accreting material
is low, and the circularization radius and the transition radius are likely
to be much smaller than in low mass BHXBs (cf. Frank et al. 1992).  
In such systems, the emission from the thin disk in the quiescent and low 
states will be significantly enhanced.  To illustrate this effect we show in 
Figure 5 a sequence of quiescent and low state spectra computed for a model 
with $\rtr = 10^2$.  The disk blackbody emission now clearly dominates over
the synchrotron peak.  Since the maximum disk temperature varies as 
$\propto \rtr^{-3/4}$ (Frank et al. 1992), the blackbody peak
component is shifted into the far UV band for low $\mdot$ models, and into 
the soft X-ray band for higher $\mdot$.  The appearance of the 
iron K$\alpha$ line also changes with decreasing $\rtr$.  Its width is 
proportional to $r^{-1/2}$ and its equivalent width depends sensitively on 
the solid angle subtended by the thin disk as viewed from the hot flow
(see Figure 8 and the discussion in \S3.3).  As a result, in wind-fed systems
we expect both quantities to be significantly larger in the quiescent and low
states compared to Roche lobe overflow systems.

\subsection{Low State to High State Transition: Intermediate State}

The hot two-temperature advection-dominated solution on which our
models are based is known to exist only for accretion rates below a
critical value $\mdot_{crit}$.  For $\mdot>\mdot_{crit}$ the ratio of
the cooling rate to the heating rate becomes so large that no hot
equilibrium solution is possible.  The topology of the solution space
in the vicinity of $\mdot\sim\mdot_{crit}$ was described via a
local analysis by Narayan \& Yi (1995b) and Chen et al. (1995).

According to the local analysis, for $\mdot<\mdot_{crit}$, there are actually 
{\it two} hot solutions (Narayan \& Yi 1995b).  One of these is the ADAF 
solution which the present paper is based on, while the other is a hot 
two-temperature solution discovered originally by SLE.  The SLE solution is 
a cooling-dominated solution which is thermally unstable (Piran 1978, Wandel 
\& Liang 1991).  In terms of $f$, the ratio of advected energy to heat input, 
the ADAF solution corresponds to $f\to1$ (for $\mdot\ll\mdot_{crit}$) while 
the SLE solution corresponds to $f\to0$.  As $\mdot$ approaches
$\mdot_{crit}$ from below, the ADAF solution begins to radiate more
efficiently and its value of $f$ decreases.  At the same time, the SLE
solution becomes more advection-dominated and its value of $f$
increases.  At $\mdot=\mdot_{crit}$, the two solutions merge at an
intermediate value of $f\sim0.3$ and disappear.  For
$\mdot>\mdot_{crit}$ neither of the two hot solutions is present, and
the only solution available is the cool thin accretion disk.

We find that the global solutions described in this paper have an
exactly analogous behavior, except that the global advection parameter
$f_{av}$ which we defined in \S2.2 replaces the local parameter $f$
of the Narayan \& Yi (1995b) analysis.  To explore the topology of the
solutions, we fixed $f_{av}$ at various values between 0 and 1 and
solved for $\mdot$ by the methods described in \S2.5.  Figure 6a
shows the result, and reveals that $\mdot$ is not a monotonic function
of $f_{av}$, but reaches a maximum value, which we call
$\mdot_{crit}$, at $f_{av} \simeq 0.35$.  For $\mdot<\mdot_{crit}$,
there are two solutions for $f_{av}$.  The solution branch with $f_{av}>0.35$
corresponds to our global ADAF, while the branch with $f_{av}<0.35$
represents the global equivalent of the SLE solution.  In analogy with
the local analysis, we believe that only the ADAF branch represents a
viable flow.  The local instability of the SLE solution, which is known to 
be quite violent (Piran 1978), will likely carry over to the global SLE
branch (though technically this has not been demonstrated), making
this solution uninteresting for applications to real flows.

Figure 6a shows that no hot solution is possible for $\mdot >
\mdot_{crit}$.  The exact value of $\mdot_{crit}$ depends in general
on model parameters, e.g. it is proportional to $\alpha^2$ (Narayan \& Yi 
1995b, Abramowicz et al. 1995).  However,
we find that we almost always have $f_{av} \sim 0.35$ at the critical
solution.  Therefore, a quick and convenient way
of obtaining $\mdot_{crit}$ is to set $f_{av}$ equal to 0.35 and to
solve for $\mdot$.  This is the procedure we have used in the calculations
described below.

What happens when $\mdot>\mdot_{crit}$?  A qualitative answer to this
question is given by Esin (1997) who shows that in response to an
increase in $\mdot$ the outer regions of the ADAF zone switch from the
hot configuration to a thin disk, leaving the inner regions still in
the hot configuration.  In other words, the transition radius between
the thin disk and the ADAF zone moves in and adjusts self-consistently
to maintain an inner ADAF zone.  In the context of the present global
solutions, we can study this process by calculating $\mdot_{crit}$ as
a function of the transition radius.  The thick solid line in Figure 6b
shows the result, where each model was computed by fixing $r_{tr}$ to
the desired value, setting $f_{av}=0.35$, and determining the value of
$\mdot_{crit}$ corresponding to the particular $r_{tr}$.

Consider now a flow that initially has $\log(r_{tr})=3.9$ (our
standard value, see Table 1) and whose $\mdot$ increases above
$\mdot_{crit}(r_{tr})\sim 0.083$ for this $r_{tr}$.  The outer regions of
the ADAF zone will cool and $r_{tr}$ will decrease.  However, Figure 6b
shows that, as $r_{tr}$ decreases, the local value of $\mdot_{crit}$
becomes even smaller than it was at $r_{tr}=10^{3.9}$.  Therefore, the ADAF
continues to cool and shrink and to convert its outer parts into a
thin disk.  During this stage, the system configuration follows the 
short-dashed line (with the arrow pointing to the 
left).  This continues until $r_{tr}$ reaches a value of about
$10^{1.5}$.  At this point, $\mdot$ becomes equal to the local
$\mdot_{crit}(r_{tr})$, and a stable self-consistent solution is
possible. With further increases in $\mdot$, the system changes its
$r_{tr}$ continuously by tracking the $\mdot_{crit}(r_{tr})$ curve
until $\mdot\sim0.092$, when $r_{tr}$ reduces again discontinuously along 
the short-dashed line.  At this point, the inner ADAF zone shrinks 
and disappears altogether, and for all $\mdot>0.092$ the thin disk extends 
down to the marginally stable orbit.

Figure 7 shows a sequence of spectra corresponding to the transition
described above.  The curves are labeled by $r_{tr}$, and are computed
for the values of $\mdot$ given by the short-dashed line in Figure 6b.
As $r_{tr}$ changes from $10^{3.9}$ to about $10^{1.5}$, the hard end
of the spectrum is hardly affected because this radiation is produced
by the inner regions of the ADAF which are insensitive to what is
happening at larger radii.  The soft blackbody peak due to the outer
thin disk does change by a large amount and increases substantially in
luminosity.  However, this radiation is in the inaccessible EUV band.
Therefore, based on X-ray observations alone,  a system with $\rtr \sim 10^2$ 
looks almost identical to one with $\rtr \sim 10^4$, and would be classified 
as being in the low state.
As $r_{tr}$ decreases below $10^{1.5}$, however, the spectrum changes more
dramatically.  Now, the cooling effect of the radiation from the disk
Compton-scattering off the hot gas in the ADAF becomes more important
and the ADAF emission becomes softer.  At the same time the disk
emission moves into the soft X-ray band and becomes an important
component in the observed spectrum.  Indeed, the radiation in the
X-ray band makes a remarkable switch from a very hard spectrum to a
soft spectrum with an ultrasoft bump.  The final model with
$r_{tr}=10^{0.5}$ (indicated in Figure 7 by a dot-long-dashed line) is
reminiscent of the high state observed in several BHXBs (e.g. Tanaka \& 
Shibazaki 1996).  Therefore we identify the sequence of models discussed 
here with the famous low state to high state transition observed in several 
BHXRs.  In our model, the transition is primarily due to a change in $r_{tr}$ 
(Figure 1),  driven by a small change in $\mdot$.

Sudden changes from the low to the high state through a series of so called 
``intermediate state'' spectra have been observed in several black hole 
systems, e.g. Cyg X-1 (Belloni et al. 1996), Nova Muscae (Ebisawa et al. 1994), 
and GX 339-4 (Mendez \& van der Klis 1997).  Our model reproduces well several 
characteristics of the observations;  for example, the bolometric luminosity 
does not change very much during the transition.  
There are two reasons for this.   First, the change in $\mdot$ across the 
transition is quite small, $\sim10\%$.  Second, the radiative efficiencies 
at the two ends are not very different.  The low state, despite having an 
ADAF, has a fairly high radiative efficiency when $\mdot=\mdot_{crit}$: 
$\e_{-1} = 1-f_{av} \sim 0.65$ (since $f_{av}=0.35$ for the critical model).  
The high state model has an efficiency of $\sim 1$.  Thus, the 
total change in the efficiency is only $\sim 30\%$ and the change in 
the bolometric luminosity is no more than about 50\%.  In fact, the
observed change will be even less than this since the high state has most
of its emission in an ultrasoft component, a part of which is
not visible due to low-energy energy absorption by a detector window or 
the ISM.  Therefore, if one considers
say the $1-100$ keV band, the observed luminosity hardly changes at
all during the transition.  This feature has been noted by Zhang et
al. (1997) in a low-high transition observed in Cyg X-1.  Another very
interesting feature of the model is that during the transition the
spectrum pivots around a fixed luminosity at 10 keV (see Figure 7).
This again has been observed (e.g. E94, Zhang et al. 1997).

In addition to the broad features discussed above, other signatures in
the spectra may be used to follow the change in $r_{tr}$ during the
transition.  X-ray reflection from the disk increases in magnitude as
$r_{tr}$ decreases and this can be discerned in the shape of the hard
X-ray continuum.  Also, the strength and shape of the iron fluorescence
line changes significantly.  Figure 8 shows the variation of the line
equivalent width.  With decreasing $\rtr$, the fraction of the high
energy photons incident on the thin disk increases from $\sim 1\%$ to
$30\%$.  On the other hand, when $\rtr \lsim 10$ the incident spectrum
steepens, so that there are fewer photons available above the iron K-edge.  
The net result is that the equivalent width of the line increases by roughly 
an order of magnitude as $\rtr$ moves in from $10^{3.9}$ to $10^{0.5}$.  
The line width, however, also increases as $\rtr^{-1/2}$ due to the 
Keplerian rotation of the inner edge of the disk.  Therefore, while the 
equivalent width increases, the line also becomes physically much broader.

So far we have discussed a system that makes a transition from low to
high values of $\mdot$.  In SXTs, we usually consider the opposite case.  
In outburst, these systems quickly reach a nearly Eddington level of 
accretion and then undergo a steady decrease of $\mdot$ during the 
decline phase of the outburst.  When these systems approach and cross
$\mdot_{crit}$, they make a high to low state transition, moving along
the long-dashed lines with arrows pointing to the right in Figure 6b.
Because the $\mdot_{crit} (\rtr)$ curve is not monotonic, this track
is slightly different than in the previous case.  The difference could
in principle be observed as hysteresis in the light curves or spectra.
Systems like Cyg X-1 and GX 339-4 (cf. Grebenev et al. 1991) which
move back and forth through the high-low state transition may be good 
candidates to study in this connection.

The results in Figure 6b, however, depend
somewhat sensitively on our assumptions.  The fact that the high-low
transition occurs over a narrow range of $\mdot$ is probably a robust
result, as are the broad features of the spectra shown in Figure 7.
However, the exact shape of the $\mdot_{crit}(r_{tr})$ curve in
Figure 6b depends on details.  For instance, the dotted line in this
figure shows what happens if we assume that  half the energy required
to evaporate gas from the disk into the the corona comes from the energy 
budget of the thin disk; our standard assumption (represented by the solid 
line), on the other hand, is that all the evaporation energy is supplied by 
the hot flow (see \S2.2).  There is a fairly
significant change in the curve.  Furthermore, even if the shape of
$\mdot_{crit} (\rtr)$ is known accurately, the exact behavior of the
system during the transition will still depend on the various
timescales such as the time scale on which $\mdot$ changes, the
thermal time scale of the hot flow, and the evaporation time scale for
the thin disk.  In plotting the curves in Figure 6b we have assumed
that the latter two are much shorter than the first; however,
detailed numerical simulations are required to treat these effects
properly.

\subsection{High State}

For accretion rates higher than the values discussed in the previous
subsection ($\mdot \gsim 0.092$), a pure ADAF zone is not possible for any 
choice of $r_{tr}$.  Therefore, the only flow configuration allowed is a 
thin accretion disk that extends down to $r=3$, with a hot corona above it 
(Figure 1).  To compute spectra of 
such models, we need a prescription for the mass accretion rate in the 
corona.  As usual we assume that the corona behaves like an ADAF and we 
take the functional dependence of $\dot m_c$ to be of the form given in 
equation (\ref{mdotc}) with $r_{tr}=3$ (the marginally stable orbit).  This 
still leaves free the overall normalization constant $\mdot_{c,0}$.  Now, 
we expect $\mdot_{c,0}$  to be determined
primarily by a competition between two major processes: (1) the
evaporation of matter from the thin disk tends to boost
$\mdot_{c,0}$, and (2) radiative cooling of the hot gas causes
$\mdot_{c,0}$ to decrease.  In analogy with the discussion in the
previous subsection, we feel that a reasonable prescription is that
the corona has the maximum $\dot m$ that is allowed for the ADAF
solution.  Therefore, we adjust $\mdot_{c,0}$ until the advection
parameter $f_{av}=0.35$ in the coronal ADAF.  This provides a
self-consistent prescription for the corona that does not involve any
additional parameters.

Figure 9 shows the values of $\mdot_{c,0}$ that we obtain as a
function of the total mass accretion rate $\mdot$ using this
prescription.  We see that $\mdot_{c,0}$ decreases quite steeply with
increasing $\mdot$.  This behavior is easy to understand.
Comptonization of cool photons from the thin disk is the dominant
cooling mechanism in the coronal gas.  As $\mdot$ increases, the disk
flux irradiating the corona increases.  This causes the corona to cool more
efficiently, and so the corona saturates at a smaller value of $\mdot_{c,0}$.

A sequence of spectra corresponding to these disk plus corona models is
shown in Figure 10, the curves being labeled by the total mass
accretion rate $\mdot$.  We see that the emission is completely
dominated by the blackbody disk component, which peaks in the soft
X-ray band.  In addition, there is a weak high energy tail which is
produced by Compton scattering of disk photons in the hot corona.  The
temperature of the soft peak ($\sim 1$ keV) as well as the
overall normalization of the spectra (which correspond to X-ray
luminosities of around $10\%$ or more of Eddington) agree with
observations of the high state in several black hole systems (e.g. Cyg
X-1, Nova Muscae and GX 339-4).  This is, however, not a new
result since it has been generally accepted for many years that the high
state corresponds to a standard thin accretion disk (e.g. Tanaka \&
Shibazaki 1996).

One new result of our model for the high state is that we find the mass 
accretion rate in the corona decreases with increasing $\mdot$.  As a result, 
we predict that the more luminous systems should have less high energy 
emission.  This is is consistent with observations on Nova Muscae 1991 (E94, 
see \S4).

It should be mentioned that at the values of $\mdot$ considered here
the inner regions of a standard thin accretion disk are subject to
thermal and viscous instabilities (Frank et al. 1992).  Since BHXBs
are seen in the high state (e.g. Nova Muscae, see \S4) it is clear
that the instabilities do not pose a serious threat to real systems.
How do the systems survive?  It has been suggested that perhaps the
viscous stress in a thin disk is proportional to $\alpha$ times the
gas pressure, rather than $\alpha$ times the total pressure
(e.g. Sakimoto \& Coroniti 1981).  Such a prescription renders the
disk stable to both the thermal and viscous instability.  Magnetic
fields do not alter the stability provided the magnetic pressure
scales with the gas pressure, as we have assumed.  Incidentally, radiation
pressure is small in the ADAF zone of all the models described in this
paper; our ADAFs are therefore stable (Kato, Abramowicz \& Chen 1996).

\subsection{Very High State}

Some BHXBs, e.g. Nova Muscae 1991 (E94) and GX 339-4 (Miyamoto et al. 1991), 
have been observed to reach luminosities close to the Eddington 
luminosity, which is higher then the luminosities typically seen in 
the high state.  In this ``very high state,'' the spectra are found
to be significantly harder than in the high state.  A substantial
fraction of the flux is emitted in a hard X-ray tail extending to well
over 100 keV, with $\alpha_N \sim 2.5$.

The model we have described so far cannot explain the very high state.
As we showed in \S3.4, with increasing $\mdot$ the coronal emission
actually decreases in the high state so that by the time we approach
the Eddington mass accretion rate there is practically no hard
component in the spectrum.  Clearly we need a new idea to explain the
very high state.

One ad hoc idea which has been quite popular for models of active
galactic nuclei is to invoke enhanced energy dissipation in the
corona, as suggested by Haardt \& Maraschi (1991).  A possible
mechanism could be coronal reconnection of magnetic field loops
that are buoyantly ejected from the thin disk (Field \& Rogers 1993).
We have attempted to incorporate this possibility in our
models by introducing a new parameter $\eta$, defined as the
fraction of the disk viscous energy that is dissipated directly in the
corona; the models described so far in the paper correspond to $\eta
=0$.  By setting $\eta>0$ we can increase the coronal emission in our
model by a fairly large factor.  Not only does more of the energy go
into the corona, a larger $\eta$ also results in a larger fraction of
the mass being accreted via the corona.  This is because the disk now
produces less soft photons and so the the hot gas in the corona
undergoes less Compton cooling.  The coronal density thus
increases above the value obtained for $\eta=0$.

Figure 11 illustrates the effect of varying $\eta$.  We show spectra
corresponding to two values of $\mdot$, in each case for models with
$\eta = 0.1,\ 0.3$ and $0.5$.  When we increase $\eta$ we 
transfer energy from the thin disk into the corona, and we see that
this results in an anti-correlation between the magnitude of the
ultra-soft blackbody disk component and the hard X-ray emission
between 10 and 100 keV.  Note, however, that the cutoff energy for the
hard tail is relatively low.  These models are just not hot enough to
produce significant emission at 100 keV, whereas observations indicate
a substantial flux even at a few hundred keV (Gilfanov et al. 1993).

The modeling of the very high state is
the most speculative part of this paper.  The idea we have explored
here, based on Haardt \& Maraschi (1991), is just one of a number of
possible scenarios.  One reason for the uncertainty is that the
luminosities observed in the very high state are generally comparable to
the Eddington luminosity (e.g., see the
discussion of Nova Muscae in \S4), so that radiation pressure is
likely to be important.  We do not include radiation pressure in our
models.  Moreover, if the gravitational energy of the thin disk is
indeed dissipated in the corona via magnetic field reconnection, the
resulting particle distribution could be significantly non-thermal,
which would modify the shape of the emerging spectrum.

\section{Outburst of Nova Muscae 1991}

Having developed an understanding of the various spectral states in
BHXBs, and the relationship of the states to one another, we now
compare the predictions of the model against observations of the soft
X-ray transient Nova Muscae 1991.  We discuss the binary parameters of
Nova Muscae in \S4.1 and describe the spectral data in \S4.2.  We then
use our models to generate synthetic light curves and compare them
against the observations in \S4.3.  Finally, we compare theoretical
and observed spectra in \S4.4.

\subsection{Optical Observations in Quiescence: Model Parameters}

After Nova Muscae returned to quiescence following its outburst in the fall of
1991, it was extensively studied by two optical observing teams and
this work has provided a detailed and confirmed picture of the system.
The binary has an orbital period of P = 10.4 hours and a mass function
of f(M) = 3.01 $\pm$ 0.15~M$_{\odot}$ (Orosz et al.  1996; see also
Casares et al. 1997).  The mass ratio is q = M$_{2}$/M$_{1}$ = 0.133
$\pm$ 0.019 (Orosz et al. 1994), a result that has been confirmed
using an independent technique (Casares et al. 1997).  Two
measurements of the inclination angle are in agreement and give a
value of $i \approx 60^{\rm o}$ (Orosz et al. 1996; Shahbaz et
al. 1997).  The limits on $i$ given by Orosz et al. are $54^{\rm o} <
i < 65^{\rm o}$.  The corresponding mass and mass limits for the
primary are $M_{1}$ = 6.0~M$_{\odot}$ and $5.0 < M_{1} <
7.5~M_{\odot}$ (Orosz et al. 1996; see also Casares et al. 1997).

We estimated the distance to Nova Muscae as follows. The apparent
magnitude of the K3-K5 secondary corrected for reddening and for the
light contribution of the accretion disk is V$_{0} \approx 20.0$
(Orosz et al. 1996). Using the dynamical data summarized above,
Eggleton's (1983) formula for the radius of the Roche lobe, and
Kepler's Law we computed the mean radius of the Roche-lobe-filling
secondary: R$_{2}$ = 1.03~R$_{\odot}$.  Finally, using the radius and
apparent magnitude of the secondary, and the visual absolute flux of a
K4 dwarf (Popper 1980) we find d = 5.7 $\pm$ 1.0 kpc.  About half the
estimated uncertainty is due to the uncertainty in $M_{1}$ given
above; the remainder is due to uncertainties in the spectral type of
the secondary and in the fraction of the total light that is
contributed by the accretion disk. A virtue of this method of
determining the distance is that it makes no assumption whatsoever
about the uncertain evolutionary state of the secondary star.  A
somewhat greater distance is obtained by this method if one adopts the
best-fit values of the mass function, f(M) = 3.35~M$_{\odot}$, and
orbital inclination, $i = 54^{\rm o}$, obtained by Casares et
al. (1997) and Shahbaz et al. (1997), respectively: d = 6.2 kpc.  On
the other hand, Shahbaz et al.  (1997) argue from evolutionary
considerations that the secondary star must be less massive than a
main-sequence K3-K5 star, from which they conclude that d $<$ 4.0 kpc.
For the purposes of this study, we adopt d = 5.0 kpc.

In summary, as inputs to our model calculations we adopt $i = 60^{\rm
o}$, $M_{1}$ = 6.0~M$_{\odot}$, and d = 5.0 kpc.  In addition, from a
study of the H$_{\alpha}$ emission line profile we adopt v$_{\rm in}$
= 2000 km~s$^{-1}$, which implies r$ _{tr} = 8400$,
i.e. $\log r_{tr}=3.9$ (NMY; Orosz et al. 1994).  We take the outer
edge of the thin disk to be at 80\% of the Roche lobe radius of the black
hole, which gives $\log r_{out}\sim 4.9$.

\subsection{X-ray Observations in Outburst}

The outburst of Nova Muscae 1991 was first detected on January 8 by
the Ginga ASM (Kitamoto et al. 1992).  It was subsequently observed in
detail by the ASM and by the Ginga LAC for about 8 months (E94).  Here
we give an abbreviated account of its spectral evolution based on  
the works just cited (see also Brandt et al. 1992; Lund
1993; Goldwurm et al.  1992; Gilfanov et al. 1993).

Figure 12a shows two lightcurves of Nova Muscae, one corresponding to
a standard X-ray band (2--12 keV, solid circles) and the other to a
soft $\gamma$-ray band (20--100 keV, open circles).  These light curves 
were computed by straightforward integration of the 51 Ginga Lac spectra 
described below, augmented with SIGMA data where available.  The following 
is a brief description of the lightcurves based on the detailed 
accounts given by Kitamoto et al. 1992, E94 and Figure 12a.  

Seven days after its discovery, Nova Muscae reached its maximum
intensity of 8 Crab (1-6 keV).  During this rising phase, the spectrum
was initially dominated by a hard power-law component with a photon
index $\alpha_{N} \sim 2.5$; however, as the total X-ray flux
increased, the hard flux ($\gsim$ 8 keV) gradually decreased.  From
about the time of maximum and for about four months thereafter, the
ultrasoft blackbody component was dominant.  For about 80 days
following the maximum, this soft component decayed exponentially with
a characteristic time scale of 31.2 days.  Meanwhile, the hard flux
decayed by more than two orders of magnitude.  However, at about 80 days after
the outburst the hard component began to rise in intensity, and a second 
broad maximum was observed in the soft component, which approximately 
doubled in intensity.  Thereafter the soft component decayed more rapidly 
with an e-folding time of 21.9 days (Kitamoto et al. 1992).
About 60 days after the outburst, the hard power-law component almost
disappeared completely for a month.  Then at about 130 days the
spectrum underwent a high-low state transition during which the
blackbody component softened and the power-law tail brightened
substantially and became much harder ($\alpha_{N} \sim 1.6$) than at
any previous time during the outburst.  This same hard spectrum was
observed during the remaining eight observations by the Ginga LAC,
which spanned the period from 157 to 238 days after the outburst (E94).

Spectral fitting parameters characterizing 51 spectra, which were
derived from Ginga LAC observations, are summarized in Table 2 of E94.
Each set of parameters was obtained by fitting the 1.2--37 keV Ginga
LAC pulse-height data to a spectral model composed of a soft disk
blackbody component plus a hard power-law component.  Ebisawa et
al. used a total of seven spectral parameters: the blackbody
temperature and normalization, the power-law photon index and
normalization, the column density N$_{H}$, and two parameters
describing a broad absorption feature above the Fe-K edge.  Using the
values of these spectral parameters given by Ebisawa et al., we
computed six representative spectra of Nova Muscae over the interval
1-40 keV.  Specifically, we used the four blackbody and power-law
parameters mentioned above.  We set N$_{\rm H}$ = 0 to approximate the
intrinsic source spectrum, and we ignored the Fe-K absorption feature.

These six best-fit Ginga Lac spectra are shown in Figure 14 as heavy
solid lines extending from 1-40 keV (1-10 keV for Figure 14b).  The
pairs of lighter solid lines, which flank each heavy line, give upper 
and lower limits on the spectra.  The upper (lower) limit was obtained 
by varying all four spectral parameters simultaneously by 1~$\sigma$ 
in order to maximize (minimize) the sum of the blackbody and power-law 
fluxes at all energies.  These limits should be conservative, except 
possibly in the case of the spectrum for day 79 for which the fit to 
the data is unsatisfactory (see Table 2 in E94). The dotted
extension of the heavy solid line in Figure 14a and Figures 14c-f is
an extrapolation of the Ginga spectrum to 300 keV. The rationale for
this extrapolation is the good agreement during the first 30 days of
the outburst between the photon index determined by Ginga (E94) and the 
index determined by SIGMA (Goldwurm et al. 1992).  In Figure 14b, the 
heavy line extends only from 1-10 keV because Ebisawa et al. were unable 
to determine the photon index of the hard component during this period 
and they fixed the photon index at 2.6.  In the
10-40 keV range, this spectrum is represented in Figure 14b by a
dotted line which does not extend to higher energy.  The triple of UV
data points in Figures 14a-c are interpolated in time from the
compilation of fluxes given by Shrader and Gonzalez-Riestra (1993).

\subsection{Modeling the X-ray and $\gamma$-ray Lightcurves}

As shown in Figure 12 and described in \S4.2, and in E94, the bolometric
luminosity of Nova Muscae reached a maximum around day 7 and then declined 
over the next several months.  The decay of the 2--12 keV X-ray flux
was relatively smooth and monotonic, whereas the 20--100 keV $\gamma$-ray 
flux went through quite large variations.  This behavior is readily 
interpreted in terms of the sequence of spectral states described in \S3.  
{\em Very High State}: Before day $\sim 60$, the X-ray and $\gamma$-ray fluxes
were intense; $\mdot$ was near or even somewhat greater than the Eddington
rate for our adopted distance of 5 kpc.  {\em High State}: From day 70 to 
day 120, the $\gamma$-ray flux was extremely faint while the X-ray flux 
remained fairly intense.  {\em Intermediate State}: Around day 130 the source
spectrum switched dramatically from soft to hard and the $\gamma$-ray flux 
became much larger than the soft flux.  This was clearly a high--low state
transition.  Following this, from day 130 to day 200, both the X-ray and 
$\gamma$-ray fluxes remained fairly constant.  {\em Low State}:  From 
day 200 onward, the source maintained a very hard spectrum.  On day 238, 
the last Ginga LAC observation, the source was quite faint relative to the 
peak of the outburst.  By this time Nova Muscae was well into decline and 
approaching the quiescent state.

Thus, there is a qualitative relationship between the spectral states 
described by our model (\S3) and the behavior of the X-ray and $\gamma$-ray 
light curves.  We now seek a quantitative relationship between the model and 
the lightcurve data by specifying a mapping between the models described in 
\S3 and the time of observation.  Ignoring the very high state, our models
are specified by only two parameters: (1) the mass accretion rate
$\mdot$, which determines which of the four states the system is in,
and (2) the transition radius $r_{tr}$, which follows the track shown
by the long-dashed line in Figure 6b.  In fact, $r_{tr}$ is an
independent parameter only in the intermediate state.  In all other
states it has a specific value, either $r_{tr}=3$ (high
state), or $r_{tr}=10^{3.9}$ (low and quiescent state).

The data in Figure 12a indicate that the luminosity in the 2--12 keV band, 
which dominates the X-ray emission in the first four months following the 
maximum, declines roughly exponentially with time (see \S4.2).  
The theoretical work of Cannizzo (1993) also
suggests that the mass accretion rate during decline following a dwarf nova
instability will have an exponential dependence under plausible
conditions.  We therefore assume an exponential scaling of $\mdot$ with time.
In the intermediate state, we expect
that it takes a finite period of time for the transition radius to
move out from $r_{tr}=3$ (corresponding to the high state) to
$r_{tr}=10^{3.9}$ (low state).  Unfortunately we have very little
understanding of the time dependence of $\rtr$, because of large
uncertainties in the mechanism responsible for disk evaporation.  In
analogy with the $\mdot$ scaling we make the rather ad hoc assumption
that $\rtr$ also has an exponential time dependence.  We then obtain
the following very simple mapping between our model parameters and the
time of observation:
\be
\label{ltcurve}
{\rm Time} - 7\,{\rm days} = - t_{\mdot} \ln\left(\frac{\mdot}{\mdot_0}\right) +t_{tr}
\ln\left(\frac{\rtr}{3}\right) .  
\ee 
This prescription involves three parameters: (1) The mass accretion rate 
at maximum luminosity (day 7), $\mdot_0$, (2) the e-folding decay time of 
the mass accretion rate, $t_{\mdot}$, and (3) the e-folding time of the 
transition radius, $t_{tr}$ (the last parameter is needed only in the
intermediate state).

As explained in \S3, we are confident that our model captures the
essential physics of four of the five spectral states, viz. the high
state, the intermediate state, the low state and the quiescent state,
but we are less comfortable with our proposal for the very high state
as it involves a somewhat ad hoc assumption about the disk corona.
Therefore, we concentrate first on the portion of the lightcurves
from day 70 (where we place the transition between the very high and high
states) onward.

By inspection of the light curves it is straightforward to estimate
the three parameters in equation (4-1).  First, during the period from
day 200 to day 238, the source was in the low state, which in our
model is dominated by the ADAF.  Since the model predicts how the
luminosity of the ADAF varies as a function of $\mdot$, the ratio of
the fluxes on day 200 and day 238 immediately gives the change in
$\mdot$ between these two points.  From this we estimate that $t_{\mdot}\sim35$
days.  Next, we see that the source made the high-low transition
around day 130.  In our model this happens when the mass accretion rate has
a value $\mdot\sim 0.092$ (see Figure 6b).  This fixes the normalization
of the mass accretion rate: $\mdot_0\sim3$.  Around day 130 the
source first entered the intermediate state when the thin accretion
disk began to shrink back and the central region was replaced by an
ADAF.  With time, the ADAF grew.  According to the model, the high-low
transition occurs during the period when $r_{tr}$ increases from $10^{0.5}$ to
$\sim10^{1.5}$ (see Figure 7).  Beyond this point, the X-ray spectrum
should remain basically unchanged while the transition radius continues
to grow to its maximum value ($r_{tr}=10^{3.9}$ in our model).
Indeed, the data show such behavior; following the high-low transition
on day 130, there is a pronounced plateau in both the X-ray and
$\gamma$-ray lightcurves extending up to day 200.  If we assume that
the ADAF reached its maximum size on day 200, then we immediately
obtain an estimate of the third parameter: $t_{tr}\sim8$ days.

Now that the three mapping constants are known, the model is completely 
specified; Figure 12b shows the variations of $\mdot$ and $r_{tr}$ with 
time for the particular constants derived above.  We are now in a position 
to calculate theoretical light curves.  For each model described in 
\S\S3.1--3.4 we use eq (4-1) to calculate the corresponding time coordinate 
in the light curve.  Then, from the calculated model spectra we compute 
the luminosities in the 2--12 keV and 20--100 keV bands.  The heavy and thin 
solid lines in Figure 12a show the theoretical X-ray and $\gamma$-ray 
lightcurves that result from this calculation.

The overall agreement between the theoretical lightcurves and the data
is quite good, and the model reproduces most of the key features in
the data rather well.  Recall that we used only the following three
pieces of information to fit the mapping constants: (1) The start time
of the intermediate state (day 130), (2) the end time of the
intermediate state (day 200), and (3) the change in total flux between
day 200 and day 238.  Several other features of the lightcurves which were
not fitted are successfully predicted by the model:

\noindent
(1) Days 200--238: The model has the correct ratio of the X-ray to 
$\gamma$-ray flux
in the low state.  That is, the model predicts the correct spectral
index on day 200 and also has the correct variation of spectral index
between day 200 and day 238 (see also Figure 12d and the discussion
of the spectral index time variation below).

\noindent
(2) Days 130--200: The model predicts the correct flux in the intermediate 
plateau in both the X-ray and $\gamma$-ray lightcurves (see Figure 12a).  
Thus, the prediction
that the intermediate state occurs at $\mdot\sim0.08$ seems to be
verified by the observations.  Actually, the predicted fluxes are somewhat 
high, but this could be easily corrected by a slight change in
the assumed black hole mass or the distance to Nova Muscae, or by
reducing $\alpha$ slightly (since flux$\,\propto\mdot_{crit}\propto\alpha^2$).

\noindent
(3) Days 70--130: The model does a surprisingly good job in the high state.  
In particular, the very
different slopes of the X-ray and $\gamma$-ray lightcurves are
predicted correctly.  The X-ray flux is nearly proportional to $\mdot$
(except for a modest bolometric correction) because the high
state spectrum is dominated by the ultrasoft component.  It is pleasing
that although the slope of the X-ray lightcurve here was not used to derive 
any of the parameters (recall that the $\mdot$ decay time scale of 35 days
was obtained by considering the data beyond day 200), nevertheless the
model fits the observations quite well.  However, the overall
normalization of the X-ray flux is about a
factor of two less than the observations.  The agreement in the
$\gamma$-ray lightcurve is more impressive.  As explained in \S3.4, the
model predicts that the coronal emission in the high state should
increase with decreasing $\mdot$ and this prediction is well
confirmed by the data.

We now consider the first 70 days of the outburst when Nova Muscae was
in the very high state.  As described in \S3.5, to reproduce the very
high state we need to invoke an extra parameter, $\eta$, which
is the fraction of the disk gravitational energy that is
dissipated in the corona.  Since the exact mechanism for such energy
transfer is not understood, the value of $\eta$, as well as its dependence
on $\mdot$, are not known.  A conservative assumption is that $\eta = 0$
below some value of $\mdot$ and changes to a constant non-zero value
for higher accretion rates.  The solid and dashed curves in Figure 12a
correspond to two such models where we have set $\eta = 0.3$ and 0.1,
respectively, and where we have assumed that the transition between
the very high and high states occurred around day 70.  (Thus there are
two additional parameters in our model of the very high state: $\eta$
and the day of the transition.)  Both models reproduce the observed
$\gamma$-ray lightcurves during the first 70 days reasonably well.
However, they predict an abrupt change in the $\gamma$-ray flux on day
70 (because of the abrupt change in $\eta$), whereas the data show a
gradual transition.

Perhaps a more physical model, but one that involves too much freedom
in fitting the data, is one that allows $\eta$ to vary with $\mdot$ in a
smooth fashion.  The dotted line in Figure 12a shows a theoretical
lightcurve where we have adjusted $\eta$ to achieve a good agreement with 
the data during the entire period of the very high state except near day 10; 
note that we still set $\eta = 0$ once the system goes into the high state.
Figure 12c shows our three models for $\eta$ as a function of time.
Clearly the variable $\eta$ model gives a better fit to most of the observed 
flux points in Figure 12a, but it is not clear if there is any physical
significance to the derived run of $\eta(t)$.

As a further test of the models we show in Figure 12d the predicted 
variation of the photon index of the power-law component of the spectrum
(calculated between 10 and 20 keV) 
compared with the photon indices measured by E94.  The gap in the data 
between day 50 and day 130 corresponds to the time during which 
Nova Muscae was too faint at high energies for E94 to determine the 
photon index.  Figure 12d shows that the model is in very
good agreement with the measured indices in the intermediate and low
states (days 130 to 238).  In the very high state, the $\eta=0.3$
model gives a reasonable fit, while the $\eta = 0.1$ model (not
surprisingly) predicts a power-law tail that is too steep.  Oddly
enough, the variable $\eta$ model, which gives the best fit to the
broad band fluxes, does not fit the observed photon indices very well.

Finally, Figure 13a shows the bolometric (solid line), UV (dashed line) 
and optical (dotted line) lightcurves of Nova Muscae during the outburst, 
computed from the model.  Figure 13b gives the radiative efficiency normalized 
to $10^{-1}$. All three curves in Figure 13a show a small step near day 70 
which is associated with the transition from the very high state to the 
high state.  In the very high state considerable energy is required 
to evaporate gas from the disk to maintain the massive corona; thus 
the radiative efficiency $\e_{-1}$ of this state is reduced relative to the 
high state, as demonstrated in Figure 13b. Therefore, for a given value 
of $\mdot$, the very high state is somewhat less luminous than the high
state. The X-ray flux in Nova Muscae did show a secondary maximum between 
days 70 and 80 (see Figure 12a), and indeed such features are common in many 
SXTs during decline (Tanaka \& Lewin 1995). 

Another feature in the bolometric lightcurve in Figure 13a is the plateau 
corresponding to the intermediate state 
between day 120 and 200.  This is the period during which $\mdot$ does not 
change very much but $r_{tr}$ changes by a large factor.  As we explained 
in \S3.3, the bolometric luminosity changes by only about 50\% during this 
entire transition, which lasted about 80 days in the case of Nova Muscae.
During the same period, there is a noticeable rise in the optical and UV 
flux from the system.  This emission is produced in the outer parts of the 
thin disk.  As the transition radius increases, the outer thin disk experiences
more irradiation by the ADAF and this causes the disk temperature and 
UV/optical luminosity to increase considerably.  A similar 
increase was seen in the optical lightcurve of A0620-00 during its 1975
outburst (see van Paradijs \& McClintock 1995 and references therein).
  
Finally, once the low state is reached, the bolometric 
luminosity drops very rapidly because by this point the flow is
dominated by the ADAF.  As Narayan \& Yi (1995b) showed, a characteristic 
feature of an ADAF is a steep decrease of radiative efficiency with decreasing
$\mdot$ (Figure 13b), which causes the luminosity to fall rapidly.  A
sudden drop in the luminosity at late time was seen in Nova Muscae 
(Figure 12a), and also in
other soft X-ray transients (Tanaka \& Shibazaki 1996 and references 
therein).  By comparison, the optical and UV fluxes decrease much more slowly 
until about day 300.  During this time the system is in the low state when
the optical and UV emission is produced by self-absorbed synchrotron.  As 
is clear from Figure 3a, the flux in this part of the spectrum decreases
more slowly with $\mdot$ than the X-ray flux.  However, after day 300, the 
system is the quiescent state, where the total flux  is dominated by the
synchrotron emission.  When, following the decline in $\mdot$, the 
synchrotron peak moves outside the UV band, the UV flux falls off more 
rapidly than the bolometric luminosity.  The corresponding rapid decline
occurs somewhat later in the optical light curve.

To summarize, we feel that our simple three-parameter ($\mdot_0,\ t_{\mdot},\ 
t_{tr}$) model reproduces the main features of the observations of Nova 
Muscae 1991 extremely well.  By varying $\mdot$ and $\rtr$ in a systematic 
way we are able to follow the system through more than three orders of
magnitude in the observed flux.  There is only one minor problem, namely
the predicted X-ray flux is lower than the data in the
high state.  This could probably be fixed by using a better model for the
thin disk, e.g. including the effects of electron scattering in the upper 
layers of the disk (Shimura \& Takahara 1995). 

\subsection{Modeling the Spectral Evolution of the System}

We computed model spectra that correspond in time with the six observed 
spectra  discussed earlier (\S4.2; Figure 14).  
Our model spectra are shown as dashed lines in Figure 14.
Note that there are no adjustable parameters here other than
the three mapping constants which were fitted in the previous section.
Note, in particular, that we have not normalized any of the predicted
spectra.  (If we allowed ourselves a normalization constant in
each panel we would obviously get a better fit.)

The comparison shows that the model gives a good description of the 
spectral evolution of the system.  This is not
surprising, of course, since we have already shown that the model fits
the light curves well.  The purpose of this section is, in fact, not
to demonstrate the goodness of fit but rather to use the spectral
comparison to identify problem areas where further work is needed.

First, it is clear that the model seriously underpredicts the
optical/UV flux (by a factor of 2 to 3) in the cases where data are
available (panels a--c).  This problem is most likely related to our
simplified treatment of the thin disk.  In computing the irradiation
of the disk by the ADAF/corona, we assume that the disk is flat and
infinitely thin (see NBM).  However, a real disk is expected to be
flared on the outside, so that the disk will intercept a larger
fraction of the irradiating flux than we allow in the model.  It is
straightforward to include disk flaring and it is clear that this will
lead to an increase in the predicted optical/UV flux (e.g. Vrtilek et al. 
1990).  Whether or not the problem will be completely eliminated remains 
to be seen.

The second problem is that the model spectrum in the very high state
(panel a) cuts off at quite a low energy $\sim 50$ keV.  This is not
consistent with observations which indicate that the power-law
component of the spectrum extends well over 100 keV (Gilfanov et al. 1993),
a result that is also confirmed by observations of V404 Cyg, GRO J0422+32, 
and GS2000+25 in the very high state (Sunyaev et al. 1993).
This problem convinces us that the model of the very high 
state presented here is probably not correct.

In the high state (panel b), there is an apparent discrepancy between
the high energy slopes of the predicted and observed spectra.  However, 
the slope shown by the dotted line in this panel was assigned arbitrarily by
E94 since they were unable to measure it from their data.  Therefore, 
it is not clear that the discrepancy is real.  There is also a small problem 
with the normalization of the soft component in the intermediate state (panel 
c); the problem is in fact slightly worse at the end of the high state (days 
100 to 125).  In general, the predicted soft X-ray flux in the high 
state is too low by a factor of $\sim$2.  This is the same problem which 
we mentioned already in \S4.3 in connection with the X-ray lightcurve.  
We are unsure at this point whether or not this indicates a serious difficulty
with the model.

The remaining panels in Figure 14 all show quite good agreement between
the model and the data.  In the low state spectra shown in panels d--f, the 
apparent soft components in the observed spectra are quite uncertain, and the 
agreement between the model and the data is in fact good.  The calculated 
low state spectra cutoff at around 100 keV. The Nova Muscae data do not rule 
out such a cutoff (the dotted lines are just 
extrapolations of the Ginga data, since there were no observations at these 
energies available after day 30).  In fact, observations of other SXTs in
the low state seem to indicate an exponential cutoff near 100 keV,
e.g. GRO J0422+32 (Kurfess 1996).  We therefore believe that this
feature of the model is correct.

\section{Discussion and Conclusions}

One of the major goals of high energy astrophysics is the development
of a consistent predictive theory of accretion flows around black
holes.  Ideally, such a theory should start with a few basic
parameters, such as the black hole mass $M$, its spin parameter $a$,
the inclination of the system $i$, the outer radius of the accretion
flow $r_{out}$, and the Eddington-scaled mass accretion rate $\mdot$,
and predict the spectrum and variability of the system.

The thin accretion disk model of Shakura \& Sunyaev (1973) and Novikov
\& Thorne (1973) (see Frank et al. 1992 for a modern review)
was an important step in developing such a theory.  Unfortunately, 
the model predicts a very soft blackbody-like spectrum with
$kT\sim 2 (M/M_\odot)^{-1/4}\mdot^{1/4}$ keV.  While this spectrum fits
observations of some accreting black holes, it does not explain many
other systems that emit much of their radiation in hard X-rays and
$\gamma$-rays, with photon energies up to 100 keV or more.  Indeed,
among the five spectral states that have been identified in BHXBs,
namely the quiescent, low, intermediate, high and very high states,
the thin accretion disk model explains only the high state.  The other
four states all involve hot optically thin gas with electron
temperatures $\gsim10^9$ K.  Until recently there were no good
dynamical models of accretion flows with these properties (see Liang 1997
for a review).

The situation has changed with the discovery of a new class of hot, optically 
thin, advection-dominated accretion flows (ADAFs) (Narayan \& Yi 1994, 1995b, 
Abramowicz et al. 1995, Chen et al. 1995), which have electron temperatures 
in the range needed to explain the observations, and which do not suffer 
from the strong thermal and viscous instabilities that plagued earlier 
attempts to develop hot accretion models (e.g. SLE, Piran 1978, Wandel \& 
Liang 1991).  ADAF models 
with low (Eddington-scaled) mass accretion rates, $\mdot\sim10^{-3}-10^{-2}$, 
have been successfully used to explain the quiescent state of BHXBs (NMY, 
NBM, Lasota et al. 1996b, Hameury et al. 1997).  The geometry of the flow is
shown schematically in the lowest panel of Figure 1, and consists of an
ADAF that extends from the black hole horizon to a transition radius
$r_{tr}\sim10^4$ (in Schwarzschild units) and a thin accretion disk
that extends from $r_{tr}$ to an outer radius $r_{out}$.  A similar
model (either with or without the outer disk) also works for low
luminosity galactic nuclei (Narayan et al. 1995; Fabian \& Rees 1995;
Lasota et al. 1996a; Reynolds et al. 1996; Mahadevan 1997).

A preliminary study of the spectral states of BHXBs using the ADAF model 
was made by Narayan (1996).  The present paper demonstrates that when the 
ADAF model of the quiescent state is extended to higher mass accretion rates 
it reproduces the low state, intermediate state, and high state of BHXBs and 
naturally explains the transition from the low state to the high state.  
Thus, the model presented here provides a unification of the quiescent, low,
intermediate and high states and represents the first consistent and
predictive theory that succeeds in encompassing a large fraction of
the phenomenology of accreting black holes (but not all of the
phenomenology since the very high state still remains to be
explained).

The basic scenario that emerges from the present study is shown 
schematically in Figure 1 and is discussed in detail in \S3.  We propose
that the low state is similar to the quiescent state in its flow geometry, 
but with a higher mass accretion rate.  This state survives until $\mdot$ 
reaches a critical value which we calculate to be $\mdot_{crit}\sim0.083$ 
(for values of $\alpha$ and $\beta$ given in Table 1).  Once $\mdot$ 
crosses $\mdot_{crit}$, the ADAF 
zone begins to shrink and the thin disk moves in by a corresponding
amount.  By the time $\mdot$ reaches about 0.092 the disk moves all the
way down to the marginally stable orbit and there is no central ADAF.
An ADAF still persists, however, in the form of a corona above the
disk.  We identify the transition states, which correspond to mass
accretion rates in the range $0.083\lsim\mdot\lsim0.092$, with the
intermediate state.  However, until the transition radius decreases 
to $\rtr \sim 10^{1.5}$, the X-ray spectrum of the system in the 
intermediate state is roughly the same as in the low state 
(see Figure 7); the spectrum begins to soften only for $\rtr<10^{1.5}$.
Because of this,  our theoretical definition of the intermediate state
is somewhat different from the one adopted by the observational community;
we assume that the system is in the intermediate state whenever $\rtr$ is 
smaller than its quiescent value, whether or not the X-ray spectrum softens.
Finally, accretion flows with $\mdot>0.09$ correspond
to the high state.  The X-ray luminosities and spectra calculated with
the above scenario are in good agreement with observations of BHXBs.

An important point which should be emphasized is that the scenario shown
in Figure 1 is an inevitable consequence of the ADAF model and is not
put in by hand or adjusted in any way.  This is because of two major
features that set this model apart from previous studies of accreting
black holes.  First, the present model achieves a fairly high level of
self-consistency in its combined treatment of dynamics and radiation
processes (\S2.5).  Specifically, the model (1) satisfies mass and momentum 
conservation at each radius by making use of global flow solutions (\S2.2, 
and Narayan et al. 1997c, Chen et al. 1997), (2) solves for the
electron and ion temperature at each radius of the accreting
two-temperature gas by solving the energy
equation (\S\S2.2, 2.3), and (3) incorporates a fairly sophisticated
treatment of radiative processes (\S2.4, and NBM).  Some loose ends do
remain and there is need for further improvement as we discuss below.
Nevertheless, we emphasize that a very high level of self-consistency has 
been achieved here.  One consequence of the self-consistency is that we do not
have the freedom to adjust the gas density or temperature or radiation
as we please.  (Recall that coronal models, to pick an example, do not 
include any dynamics and often treat the optical depth of the corona as a 
free parameter.)

Secondly, the model is essentially parameter-free.  Given the
binary parameters $M$, $i$ and $r_{out}$, the model makes a
unique prediction of the flow configuration of the gas, and the spectrum 
of the radiation, as a function of $\mdot$.  The model does  involve a 
second parameter, the transition radius $r_{tr}$, but in principle this 
is a function of $\mdot$ (Figure 6).  At low $\mdot$, $r_{tr}$ has a 
relatively large value determined by the nature of the mass transfer from
the companion star (see the discussion in \S2).  When
$\mdot=\mdot_{crit}$, $r_{tr}$ begins to decrease; this decrease
is calculated self-consistently in the model.  Finally, at high
$\mdot$ there is no transition radius since the thin disk comes down
to the marginally stable orbit.  The model is thus quite
well-constrained and does not allow much room for adjustment.  The
model does involve two important microscopic parameters, $\alpha$ and
$\beta$, which parameterize the viscosity and the magnetic field
strength (plus a third parameter $\delta$ which does not influence any
of the results presented here).  We have not treated $\alpha$ and $\beta$ as 
adjustable constants but rather assigned to them what we consider their most 
natural values ($\beta=0.5,\ \alpha\approx(1-\beta)/2=0.25$).  We anticipate 
that with improved data in the future it may be
possible to fit $\alpha$ and $\beta$.  Hopefully, they are universal
constants which once fitted on one source can be applied to any
other source.

The main results obtained in this paper are the following:

\noindent
1. The luminosities and spectra that we calculate in the low state agree 
well with observations published in the literature.  For instance, when 
$\mdot$ approaches $\mdot_{crit}$, we obtain luminosities of a few percent 
of Eddington and hard spectra with photon indices $\sim 1.4-1.5$, which is 
consistent with the data (e.g. Nowak 1995).  One interesting result is that 
we find the maximum $\mdot$ in the low state to be $\mdot_{crit}\sim 1.3
\alpha^2$ whereas the preliminary work of Narayan (1996) gave a scaling 
$\mdot_{crit}\sim0.3\alpha^2$.  Also, we find the luminosity of the 
critical model to be $L_{crit}/L_{Edd} \sim 0.4 \alpha^2$ instead of
$\sim 0.03\alpha^2$.  As a result of the new scalings, we can now 
explain the luminosities observed in the low state with a
reasonable value of $\alpha\sim0.25$ instead of $\alpha\sim1$ required
with the old scaling.  The change appears to be related to 
improved global dynamics and better treatment of Compton-scattering in this
paper (and in NBM) compared to Narayan (1996).

\noindent
2. A major success of the model is the fact that it provides a natural
explanation for one of the most dramatic features of BHXBs, namely
the low-high state transition.  As mentioned above, the ADAF which
dominates the physics of the low state can exist only for 
$\mdot<\mdot_{crit}$.  Once $\mdot$ crosses $\mdot_{crit}$, most of the
gas in the hot flow settles into a thin accretion disk, with a weak
optically thin corona above it.  The spectrum therefore switches quite
rapidly from the hard power-law spectrum characteristic of the low
state to a soft spectrum which is dominated by the modified blackbody
disk component.  The intermediate spectral stages in this transition
are shown in Figure 7, and these correspond quite well with the
intermediate state which has been observed between the low and the
high state (e.g. E94, Belloni et al. 1996, Mendez \& 
van der Klis 1997).  Since the low-high transition occurs over a
relatively narrow range of $\mdot$, the total luminosity of the system
stays nearly constant.  Furthermore, the X-ray spectrum displays an 
interesting pivoting behavior around $\sim 10$ keV.  These features have
been seen in Nova Muscae (E94, \S4) and Cyg X-1 (Zhang et al. 1997), and are 
convincingly explained in the model.

\noindent
3. The model predicts that, in the high state, as $\mdot$ increases
the mass in the corona decreases (Figure 9) and correspondingly the
fraction of the flux in the hard tail goes down.  This is confirmed by
observations of Nova Muscae 1991 (Figure 13a).  If the model is correct,
this should be a universal feature of BHXBs in the high state, but we
are not aware if the anti-correlation between the ultrasoft bump and the 
hard tail has been seen in other systems in the high state.

In \S4, we present a quantitative comparison between the theoretical
predictions of the model and the extensive observations of the outburst 
of Nova Muscae 1991 carried out by E94.  The system parameters are 
well-known for this binary (\S4.1, Table 1).  Using a very simple model for the 
mapping between the mass accretion rate $\mdot$, the transition radius 
$r_{tr}$ and the time of observation (equation \ref{ltcurve}), we present 
in \S\S4.3, 4.4 a detailed comparison of the models
and the data.  We use the theoretical model to predict the spectrum of
Nova Muscae as a function of time and we compare the predictions
against the data.  The results are very encouraging (Figures 12--14).
Not only is there a near-perfect qualitative agreement with the main
features of the X-ray and $\gamma$-ray lightcurves, it even appears
that the model is close to being a quantitatively predictive
theory.  Considering the level of self-consistency we have achieved 
and the relatively few parameters used in the comparison we find
the results quite encouraging.

We also present in this paper a model of the very high state.  The
model is based on an idea pioneered by Haardt \& Maraschi (1991),
namely that a part of the viscous energy released in the thin accretion
disk may be input directly into the corona.  We incorporate this
possibility via an additional parameter $\eta$ (\S3.5) and we compute
model spectra as a function of $\mdot$ and $\eta$ (Figure 11).  The
results are not unreasonable.  However, in contrast to the quiescent,
low, intermediate and high states, where our model is robust and
physically consistent, the proposal we make for the very high state is
ad hoc.  In fact, we include the very high state in this paper mainly
for reasons of completeness, particularly for the comparisons in \S4.

The modeling techniques we have used in this paper involve a number of
improvements over previous versions of the model which were developed for 
applications to quiescent SXTs (NMY, NBM, Hameury et al. 1997).  (1) By means 
of the parameter
$f_{av}$ (equation \ref{fav}) we allow for the back reaction of radiative 
losses on the global dynamical solution of the ADAF.  (2) Following the work 
of Nakamura et al. (1997) we have included energy advection in the thermal 
balance equation of the electrons (equation \ref{eenergy} and Appendix A).
This allows us to compute a more accurate temperature profile for the
electrons.  (3) We have significantly enhanced the manner in which we model 
the coupling between the thin disk and the ADAF, taking care to include 
Compton reflection and the iron fluorescence line (\S2.3).  (4) We have
also included pair processes in the ADAF and the corona (\S2.4 and 
Appendix B), though we agree with the conclusion of Bj$\ddot{\rm o}$rnsson 
et al. (1996), and Kusunose \& Mineshige (1996) that pairs are not important.  

Despite these enhancements to the model, several further improvements
are possible.  One long standing uncertainty in the model is the
process by which material in the thin disk evaporates into the
corona.  A mechanism suggested by Meyer \& Meyer-Hofmeister (1994)  
is surface heating via electron conduction from the
corona, while Honma (1996) suggested turbulent heat transport from the
inner regions of the ADAF.  Whatever the mechanism, one would like to
be able to compute the evaporation rate from basic physical principles
rather than fix $\rtr$ empirically.

Two simple changes would substantially improve our modeling of the thin disk.
Instead of assuming that the disk emission is a pure blackbody, we could allow 
for graybody effects associated with electron scattering in the upper layers
of the disk.  This will have the effect of shifting the disk spectrum to 
higher energies (e.g. Shimura \& Takahara 1995).  Secondly, we need to include 
vertical flaring so as to improve our model of the X-ray irradiation of the
disk.  We believe this will significantly
reduce the discrepancy in the predicted optical/UV flux of Nova Muscae
discussed in \S4.4. 

The models presented here are still based on Newtonian physics, and an
obvious next step is to incorporate fully relativistic global flow
solutions in Kerr geometry.  Such solutions have been calculated by
Abramowicz et al. (1996), Peitz \& Appl (1997) and Gammie \& Popham
(1997).  Apart from being more consistent, these solutions will allow
us the opportunity to study the role of the spin parameter
$a$ of the black hole.  The effect of this parameter on spectra of
ADAFs is at present completely unexplored.  It is also necessary to
include in the radiative transfer relativistic effects such as
gravitational redshift, ray bending and Doppler boosts due to the
motion of the gas.  Titarchuk et al. (1996, 1997) have shown that the
bulk motion of the gas as it flows into the black hole can have an
important effect on the high energy end of the spectrum.  This effect
is not included in our present models. 

So far we have not addressed the question of source variability, but
the model makes several predictions in this regard.  Though ADAFs are
to some degree rotationally supported, the radial velocity reaches a
significant fraction of free-fall.  Using the self-similar solution of
Narayan \& Yi (1994) the viscous timescale can be estimated to be $t_v
\sim R/v_r \sim 0.025 (M/10 \msun) (0.25/\alpha) (r/10)^{3/2}$ s.  The
dynamical timescale is even shorter, $t_d \sim R/c_s \sim 0.003 (M/10
\msun) (r/10)^{3/2}$ s. Thus, ADAFs can in principle produce very
rapid variability, as shown by Manmoto et al. (1996).  In comparison, the 
characteristic timescales associated with a thin accretion disk are 
significantly longer, especially when $r_{tr}$ is large.  We thus expect 
short time scale variability to be strongly correlated with the power-law 
component of the emission, as seems to be observed (Nowak 1995).

Interactions between the hot inner ADAF and the cool outer thin disk
at or near the transition radius can be an additional source of
variability on longer timescales (Mendez \& van der Klis 1997).  In
this case, the characteristic time will be a multiple of the Keplerian
rotation period at $r_{tr}$, $t_K = 2 \pi R/v_K \sim 1 (M/10 \msun)
(r/100)^{3/2}$ s, and it is conceivable that the variations might 
be quasi-periodic.  The model thus makes a very specific prediction.
During the low-high state transition, any characteristic frequency
in the variability spectrum (e.g. QPOs) should be low when the spectrum 
is hard ($r_{tr}$ is large) and high when the spectrum is soft 
($r_{tr}$ is small).  The detection of $\sim 0.1$ Hz
QPO in the power spectra of BHXBs in the low state and $\sim 10$ Hz
QPO in the high state (Nowak 1995, Cui et al. 1997) is certainly very
suggestive.  If this behavior is confirmed as a general trend in
BHXBs, then it would be a strong confirmation of the model.

The basic paradigm of our model, namely that the thin disk is limited to a 
large radius in the low state and quiescent state but comes close to the 
black hole in the high state and very high state,  
can also be tested through direct determination of $\rtr$.  One method is by
fitting the ultrasoft component in the spectrum to a multicolor blackbody 
(technically graybody).  This was attempted in the case of Nova 
Muscae by E94.  However, during the high-low state transition, which is when
$\rtr$ is expected to vary, the disk emission peaks 
at energies below $1\,{\rm keV}$, making the results rather uncertain.  A more
indirect, but equally promising way of detecting the change in $\rtr$ during 
this transition is through observations of Compton
reflection of hard X-rays off the
surface of the disk, and by measuring the strength and spectral
width of the iron fluorescence line.  The physical width of the line
is directly correlated with the rotational velocity at the inner edge
of the disk which is $\propto r_{tr}^{1/2}$, while the equivalent
width contains information about the solid angle subtended by the thin
disk as viewed from the radiating zones of the ADAF.  We have already
included Compton reflection in the model, and we compute some of the
properties of the iron line.  There is a clear qualitative prediction
according to the model: both Compton reflection and the iron line
equivalent width should become stronger as a BHXB proceeds from the
low state to the high state.  In order to make direct quantitative
comparisons with the data we will need to use a more realistic iron
line profile shape than we have done so far and take into account the 
3-D shape of the accretion disk.

We conclude by clarifying a point of terminology.  An ADAF is defined
as an accretion flow where energy advection dominates over cooling.
However, in the models with $\mdot>\mdot_{crit}$ described in this
paper only 35\% of the heat energy is advected (corresponding to
$f_{av}=0.35$) and the flows are not ADAFs in the strict definition of
the term.  These flows nevertheless belong to the same solution branch
which at lower values of $\mdot$ are truly advection-dominated with
$f_{av}\to 1$.  For this reason, we feel it is appropriate to
label them as ADAFs.

\acknowledgments

We thank K. Ebisawa, S. Kitamoto and J. Orosz for providing data and for 
helpful discussions.  This work was supported in part by NSF grant 
AST 9423209 and NASA grant NAG 5-2837.  A.~A.~E. was supported by a National 
Science Foundation Graduate Research Fellowship, and partial support for 
J.~E.~M. was provided by the Smithsonian Institution Scholarly Studies Program.

\vfill\eject
\begin{appendix}
\section{Energy Balance Equation for Electrons}

Electrons in the accretion flow must satisfy the general energy conservation 
requirement 
\be
\label{elen}
\rho T_e v \frac{d s_e}{d R} = q^{ie} + \delta q^+ - q^-,
\ee
where $s_e$ is the entropy of the electrons per unit total gas mass, 
$q^{ie}$ is the Coulomb energy transfer rate from ions to electrons per 
unit volume, $\delta q^+$ is the fraction of the viscous energy dissipation 
that goes directly into the electrons, and $q^-$ is the cooling rate per 
unit volume.  In our previous work we set the left hand side of this 
equation to zero and required that the heating and cooling rates balance at
all radii.  However, Nakamura et al. (1997) have made the important point that 
energy advection by electrons may dominate under some circumstances.  We 
therefore consider the full equation in this work.

We know that 
\be
T_e d s_e = d u_e + P_e d \left(\frac{1}{\rho}\right),
\ee
where $u_e$ is the internal energy of the electrons per unit mass and 
$P_e$ is the electron pressure.  The quantities on the right side of this
equation are  in general easy to
compute.  However, our case is complicated by the fact that we need to consider
a mixture of gas and magnetic fields.  If gas pressure contributes 
a constant fraction $\beta$ to the total pressure in the flow, $P_{tot}$ can
be written as
\be
P_{tot} = \frac{\rho k T_i}{\mu_i m_u} + \frac{\rho k T_e}{\mu_e m_u} +
\frac{B^2}{24 \pi} = \frac{\rho k T_i}{\beta \mu_i m_u} + 
\frac{\rho k T_e}{\beta \mu_e m_u}.
\ee
It seems natural then to denote the two resulting terms as the effective ion 
and electron pressures, so that $P_e = \rho k T_e/(\beta \mu_e m_u)$.

The internal energy of the gas is a sum of the ion and electron internal 
energies plus the magnetic energy density:
\be
\label{inten}
u = \frac{3}{2} \frac{k T_i}{\mu_i m_u} + a(T_e) \frac{k T_e}{\mu_e m_u} + 
\frac{B^2}{8 \pi \rho} = \frac{6-3\beta}{2 \beta} \frac{k T_i}{\mu_i m_u} +
\left[\frac{3 (1-\beta)}{\beta} + a(T_e)\right] \frac{k T_e}{\mu_e m_u},
\ee
where the coefficient $a(T_e)$ varies from $3/2$ in the case of a
non-relativistic electron gas, to $3$ for fully relativistic electrons.  
The general expression for $a$ as a function of the dimensionless electron
temperature $\theta_e = k T_e/m_e c^2$ was derived by 
Chandrasekhar (1939, equation [236]) to be
\be
a(\theta_e) = \frac{1}{\theta_e} \left(\frac{3 K_3 (1/\theta_e) + 
K_1 (1/\theta_e)}{4 K_2 (1/\theta_e)} - 1\right).
\ee
Note that the ion thermal energy in these flows never exceeds 10\%
of the ion rest mass, so that the corresponding coefficient for the ions is
always $\sim 3/2$.

As we have done for the pressure, the right hand side of 
equation (\ref{inten}) is naturally divided into two terms, the ion 
and electron internal energies:
\be
u_i = \frac{6-3\beta}{2 \beta} \frac{k T_i}{\mu_i m_u},\ \ \ {\rm and}\ \ \ 
u_e= \left[\frac{3 (1-\beta)}{\beta}+ a(T_e)\right] \frac{k T_e}{\mu_e m_u}.
\ee
In this interpretation, $u_e$ and $u_i$ are again ``effective'' quantities, 
which include contributions from the particles as well as the associated 
magnetic field.  Note that the contribution 
of the magnetic field to the internal energy of each particle 
species is proportional to the contribution of these particles to the total 
pressure, a natural choice in our model where the ratio of the magnetic
to gas pressure is fixed.

Having defined $P_e$ and $u_e$, we can now compute $d s_e/d R$.  Then
equation (\ref{elen}) becomes
\be
\frac{\rho v k}{\mu_e m_u} \left[\frac{3 (1-\beta)}{\beta} + a(T_e) + 
T_e \frac{d a}{d T_e}\right] \frac{d T_e}{d R} + \frac{\rho k T_e}{\beta
\mu_e m_u} \left[\frac{d v}{d R} + \frac{2 v}{R}\right] = q^{ie} +
\delta q^+ - q^-,
\ee
where we have used mass conservation to replace the radial derivative 
of gas density, $d \rho/d R$, by terms that depend on the radial velocity $v$
and its derivative $d v/d R$.  Finally, to obtain the electron energy 
equation for each shell of the ADAF, we multiply both sides of the 
equation above by the appropriate shell volume.

\section{Electron-Positron Pair Production and Annihilation}

Local pair equilibrium requires that the pair creation rate is balanced by 
pair 
annihilation, as expressed in equation (\ref{pairs}).  The pair annihilation
rate per unit volume for a thermal distribution of electrons and positrons 
was derived by Svensson (1982a) 
\be
(\dot{n}_+)_{\rm ann} = \frac{\pi c r_e^2 n_+ n_-}{1+2 \theta_e^2/\ln{(1.12
\theta_e +1.3)}},
\ee
where $n_+$ and $n_-$ are the positron and electron number densities 
respectively (in the main body of the paper the electron number
density is denoted by $n_e$).

Pair production in relativistic plasmas occurs through several 
interactions.  In this paper we consider electron-electron ($e e$), 
photon-electron ($\gamma e$), photon-proton ($\gamma p$), and photon-photon 
($\gamma \gamma$), collisions (note that ``electron'' here means either 
electron or positron).  Thus
\be
(\dot{n}_+)_{\rm prod} = (\dot{n}_+)_{e e} + (\dot{n}_+)_{\gamma e} +
(\dot{n}_+)_{\gamma p} + (\dot{n}_+)_{\gamma \gamma}.
\ee  
We ignore electron-proton interactions, which were shown by 
Svensson (1982b, hereafter S82b) to be less efficient than $e e$ collisions 
by a factor of $\sim 10$.  Though Svensson's result was derived for a 
one-temperature plasma where
$\theta_p = \theta_e$, it is probably still valid in our case, since the
protons in the ADAF are not relativistic.

The electron-electron pair production rate per unit volume in the 
non-relativistic and extreme relativistic limits is (White \& 
Lightman 1989) 
\be
(\dot{n}_+)_{e e} = c r_e^2 (n_+ + n_-)^2 \left\{ \begin{array} {ll}
        2 \times 10^{-4}\ \theta_e^{3/2} \exp{(-2/\theta_e)}\ (1 + 0.015 
	\theta_e), & \theta_e \ll 1, \\
        (112/27 \pi) \ \alpha_f^2\ (\ln{\theta_e})^3\ (1+0.058/\theta_e)^{-1},
        & \theta_e \gg 1,
        \end{array}\right.
\ee
where $\alpha_f= 1/137.04$ is the fine structure constant.  To join the two 
limiting expressions smoothly at $\theta_e \sim 1$, we use $(\dot{n}_+)_{e e} 
= (\dot{n}_+)_{e e}^{\theta_e \ll 1} \times e^{-\theta_e} + 
(\dot{n}_+)_{e e}^{\theta_e \gg 1} \times e^{-1/\theta_e^3}$.

The pair production rate via photon-electron collisions is given by the 
double integral (S82b)
\be
(\dot{n}_+)_{\gamma e} = \frac{c (n_+ + n_-)}{2 K_2 (1/\theta_e)} 
\int_0^{\infty} {d x \frac{n_{\gamma} (x)}{x^2} \int_4^{\infty} {d y\, y\,
\sigma_{\gamma e} (y) \exp{[-(x/y+y/x)/2 \theta_e]}}},
\ee
where $n_{\gamma} (x)$ is the number density of photons with energy $x = h \nu/
m_e c^2$.
Because of the exponential factor, the greatest contribution to 
the second integral comes from photons with energies $x\sim y > 4$.  
Using the numerical fit for the cross section $\sigma_{\gamma e} (y)$ given 
in Stepney \& Guilbert (1983) we numerically
evaluate the two integrals at each point in the accretion flow. 

Since the proton temperature in ADAFs is always below $10^{12}\,{\rm K}$,
protons are never relativistic ($\theta_p \lsim 0.09$) and the photon-proton
pair production rate can be simply written as (S82b)
\be
(\dot{n}_+)_{\gamma p} = c n_p \int_2^{\infty} {d x\, n_{\gamma} (x)\, 
\sigma_{\gamma p} (x)},
\ee
where we again adopt the numerical fit to the cross section 
$\sigma_{\gamma p} (x)$ from Stepney \& Guilbert (1983), and integrate 
over the photons with energies above $2 m_e c^2$.

Finally, the photon-photon pair production rate is given by S82b
\be
(\dot{n}_+)_{\gamma \gamma} = \frac{c}{2} \int_0^{\infty}{d x\, n_{\gamma} (x)
\int_{1/x}^{\infty}{d y\, n_{\gamma} (y) \langle \sigma_{\gamma \gamma} 
(x y, \phi)\rangle_{\phi}}},
\ee
where $\langle \sigma_{\gamma \gamma} (x y, \phi)\rangle_{\phi}$ is the
interaction cross section averaged over the photon-photon interaction angle
$\phi$, taken from Gould \& Schreder (1966).  The factor of $1/2$ takes care
of double counting of interacting photon pairs.  Note that only the pairs with
$x y \geq 1$ contribute to pair production.

We now to rewrite the above rate equations in terms of the pair fraction
$z = n_+/n_p$.  For simplicity, we assume that the accreting gas contains 
only ionized hydrogen, so that $n_p = \rho/m_p$.  Then we can write
$n_+ = z n_p$ and $n_- = n_p + n_+ = (z+1) n_p$.  With these definitions, 
equation (\ref{pairs}) becomes 
\be
n_p z (1+z) g_{\rm ann} = n_p (1+2 z)^2 g_{e e} + n_p (1+2 z) g_{\gamma e} + 
n_p g_{\gamma p} + g_{\gamma \gamma},
\ee
where the coefficients $g_{\rm ann}$, $g_{e e}$, $g_{\gamma e}$, $g_{\gamma p}$,
and $g_{\gamma \gamma}$ contain all the detailed physics described above.
This new equation is quadratic in $z$ and can be readily solved at each 
point in the advection flow, given the gas density, electron temperature and
the radiation field.  Because of the dependence on the latter, the final 
solution has to be obtained iteratively, since newly created electrons and 
positrons contribute to the radiative processes, and thus help create 
more pairs.

\end{appendix}

\vfill\eject
\references
\def\refpar{\hangindent=3em\hangafter=1}
\def\reference{\refpar\noindent}
\def\apj{ApJ}
\def\apjs{ApJS}
\def\mnras{MNRAS}
\def\aa{A\&A}
\def\aas{A\&A Suppl. Ser.}
\def\aj{AJ}
\def\araa{ARA\&A}
\def\nat{Nature}
\def\pasj{PASJ}

\reference Abramowicz, M. A., Chen, X., Grantham, M., Lasota, J.-P. 1996, 
\apj, 471, 762

\reference Abramowicz, M. A., Chen, X., Kato, S., Lasota, J. P., \& Regev, O.
1995, \apj, 438, L37

\reference Abramowicz, M. A., Czerny, B., Lasota, J. P., \& Szuszkiewicz, E.
1988, \apj, 332, 646

\reference Balbus, S. A. \& Hawley, J. F. 1991, \apj, 376, 214

\reference Begelman, M. C. 1978, \mnras, 184, 53

\reference Begelman, M. C. \& Chiueh, T. 1988,\apj, 332, 872

\reference Belloni, T., Mendez, M., van der Klis, M., Hasinger, G., 
Lewin, W. H. G., van Paradijs, J. 1996, \apj, 472, L107 

\reference Bj$\ddot{\rm o}$rnsson, G., Abramowicz, M. A., Chen, X., Lasota, 
J.-P. 1996, \apj, 467, 99

\reference Brandt, S., Castro-Tirado, A. J., Lund, N., Dremin, V., 
Lapshov, I., Sunyaev, R. A. 1992, \aa, 254, L39

\reference Cannizzo, J. K. 1993, in Accretion Disks in Compact Stellar 
Systems, ed. J. C. Wheeler, Singapore: World Scientific Publishing, 6

\reference Casares, J., Martin, E. L., Charles, P. A., Molaro, P., 
Rebolo, R. 1997, New Astr., 1, 299

\reference Chandrasekhar, S. 1939, in Introduction to the Study of Stellar 
Structure

\reference Chen, X. 1995, \mnras, 275, 641

\reference Chen, X., Abramowicz, M. A., Lasota, J. P. 1997, \apj, 476, 61

\reference Chen, X., Abramowicz, M. A., Lasota, J. P., Narayan, R., Yi, I.
1995, \apj, 443, L61

\reference Clayton, D. D. 1983, Principles of Stellar Evolution and 
Nucleosynthesis, U. of Chicago Press

\reference Cui, W., Heindl, W. A., Rothschild, R. E., Zhang, S. N., 
Jahoda, K., Focke, W. 1997, \apj, 474, L57

\reference Ebisawa, K. et al. 1994, \pasj, 46, 375 (E94)

\reference Ebisawa, K., Ueda, Y., Inoue, H., Tanaka, Y., White, N. E. 1996, 
\apj, 467, 419

\reference Eggleton, P. P. 1983, \apj, 268, 368

\reference Esin, A. A. 1997, \apj, 482, in press (astro-ph/9701039)

\reference Fabian, A. C. \& Rees, M. J. 1995, \mnras, 277, 55

\reference Field, G. B. \& Rogers, R. D. 1993, \apj, 403, 94 

\reference Frank, J., King, A., \& Raine, D. 1992, Accretion Power in
Astrophysics (Cambridge, UK: Cambridge University press)

\reference Gammie, C. \& Popham, R. 1997, submitted to \apj

\reference George, I. M. \& Fabian, A. C. 1991, \mnras, 249, 352

\reference Gierli$\acute{\rm n}$ski, M., Zdziarski, A. A., Done, C., 
Johnson, W. N., Ebisawa, K., Ueda, Y., Haardt, F., Phlips, B. F. 1997, 
\mnras, in press (astro-ph/9610156)

\reference Gilfanov, M. et al. 1993, \aas, 97, 303

\reference Goldwurm, A. et al. 1992, \apj, 389, L79

\reference Gould, R. J. \& Schreder, G. P. 1966, Phys. Rev., 155, 1404

\reference Grebenev, S. A. et al. 1991, 17, 413

\reference Haardt, F. \& Maraschi, L. 1991, \apj, 380, L51

\reference Hameury, J.-M., Lasota, J.-P., McClintock, J. E., \& Narayan, R.
1997, submitted to \apj

\reference Hawley, J. F. \& Balbus, S. A. 1996, in Physics of Accretion Disks,
eds. S. Kato et al., Gordon and Breach Sci. Publ., 273

\reference Hawley, J. F., Gammie, C. F., Balbus, S. A. 1995, \apj, 440, 742

\reference Hawley, J. F., Gammie, C. F., Balbus, S. A. 1996, \apj, 444, 690

\reference Honma, F. 1996, \pasj, 48, 77

\reference Huang, M. \& Wheeler, J. C. 1989, \apj, 343, 229

\reference Igumenshchev, I. V., Chen, X., Abramowicz, M. A. 1996, \mnras, 
278, 236

\reference Jaroszy$\acute{\rm n}$ski, M. \&  Kurpiewski, A. 1997, \aa, 
in press (astro-ph/9705044)

\reference Kato, S., Abramowicz, M. \& Chen, X. 1996, \pasj, 48, 67

\reference Katz, J. I. 1977, \apj, 215, 265

\reference Kitamoto, S., Tsunemi, H., Miyamoto, S., \& Hayashida, K. 1992, 
\apj, 394, 609

\reference Kurfess, J. D. 1996, \aas, 120, 5

\reference Kusunose, M. \& Mineshige, S. 1996, \apj, 468, 330

\reference Lasota, J. P., Abramowicz, M. A., Chen, X., Krolik, J., Narayan, 
R., \& Yi, I. 1996a, \apj, 462, 142L

\reference Lasota, J. P., Narayan, R., \& Yi, I. 1996b, \aa, 314, 813

\reference Liang, E. 1997, Phys. Rep., in press 

\reference Lightman, A. P. \& White, T. R. 1988, \apj, 335, 57

\reference Lund, N. 1993, \aas, 97, 289

\reference Mahadevan, R. 1997, \apj, 477, 585

\reference Manmoto, T., Takeuchi, M., Mineshige, S., Matsumoto, R., Negoro, H.
1996, \apj, 464, L135

\reference McClintock, J. E., Horne, K., \& Remillard, R. A. 1995, \apj, 
442, 358 

\reference Melia, F. \& Misra, R. 1993, \apj, 411, 797

\reference Mendez, M. \& van der Klis, M. 1997, \apj, 479, 926

\reference Meyer, F. \& Meyer-Hofmeister, E. 1994, \aa, 288, 175

\reference Mineshige, S. 1996, \pasj, 48, 93

\reference Mineshige, S. \& Wheeler, J. C. 1989, \apj, 343, 241 

\reference Miyamoto, S., Kimura, K., Kitamoto, S., Dotani, T., \& Ebisawa, K. 
1991, \apj, 383, 784

\reference Morrison, R. \& McCammon, D. 1983, \apj, 270, 119

\reference Nakamura, K. E., Kusunose, M., Matsumoto, R., \& Kato, S. 1997, 
to appear in \pasj

\reference Nandra, K. \& Pounds, K. A. 1994, \mnras, 268, 405

\reference Narayan, R. 1997, in Proc. IAU Colloq. 163 on Accretion 
Phenomena \& Related Outflows, ASP Conf. Series, eds. D. T. Wickramasinghe
et al., in press (astro-ph/9611113)

\reference Narayan, R. 1996, \apj, 462, 136

\reference Narayan, R., Barret, D., \& McClintock, J. E. 1997a, \apj, in 
press (NBM)

\reference Narayan, R., Garcia, M. R. \& McClintock, J. E. 1997b, \apj, 
478, 79

\reference Narayan, R., Kato, S. \& Honma, F. 1997c, \apj, 476, 49

\reference Narayan, R., McClintock, J. E., \& Yi, I. 1996, \apj, 457, 821 
(NMY)

\reference Narayan, R. \& Yi, I. 1994, \apj, 428, L13

\reference Narayan, R. \& Yi, I. 1995a, \apj, 444, 231

\reference Narayan, R. \& Yi, I. 1995b, \apj, 452, 710

\reference Narayan, R., Yi, I., \& Mahadevan, R. 1995, \nat, 374, 623

\reference Novikov, I. D. \& Thorne, K. S. 1973, in Black Holes, ed. 
DeWitt, C. and B. (Gordon \& Breach, NY), 343

\reference Nowak, M. A. 1995, \pasp, 107, 1207

\reference Orosz, J. A., Bailyn, C. D., Remillard, R. A., McClintock, J. E., 
\& Foltz, C. B. 1994, \apj, 436, 848

\reference Orosz, J. A., Bailyn, C. D., McClintock, J. E., \& Remillard, R. A.
1996, \apj, 468, 380

\reference Orosz, J. A., Remillard, R. A., Bailyn, C. D., \& McClintock, J. E. 
1997, \apj, 478, 83

\reference Peitz, J. \& Appl, S. 1997, \mnras, in press (astro-ph/9612205) 

\reference Piran, T. 1978, \apj, 221, 652 

\reference Popper, D. M. 1980, \araa, 18, 115

\reference Poutanen, J. \& Svensson, R. 1996, \apj, 470, 249

\reference Rees, M. J., Phinney, E. S., Begelman, M. C., Blandford, R. D. 
1982, \nat, 275, 17

\reference Reynolds, C. S., Di Matteo, T., Fabian, A. C., Hwang, U., 
Canizares, C. R. 1996, \mnras, 283, L111

\reference Sakimoto, P. J. \& Coroniti, F. V. 1981, \apj, 247, 19

\reference Shahbaz, T., Naylor, T. \&  Charles, P. A. 1997, \mnras, 285, 
607

\reference Shakura, N. I. \& Sunyaev, R. A. 1973, \aa, 24, 337

\reference Shapiro, S. L., Lightman, A. P., \& Eardley, D. M. 1976, \apj, 204,
187 (SLE)

\reference Shimura, T. \& Takahara, F. 1995, \apj, 445, 780

\reference Shrader, C. R. \& Gonzalez-Riestra, R. 1993, \aa,  276, 373

\reference Stepney, S., \& Guilbert, P. W. 1983, \mnras, 204, 1269

\reference Sunyaev, R. A. \& Titarchuk, L. G. 1980, \aa, 86, 121

\reference Sunyaev, R. A. et al. 1993, \aa, 280, L1

\reference Svensson, R. 1982a, \apj, 258, 321

\reference Svensson, R. 1982b, \apj, 258, 335 (S82b)

\reference Svensson, R. \& Zdziarski, A. A 1994, \apj, 436, 599

\reference Tanaka, Y. et al. 1995, \nat, 375, 659

\reference Tanaka, Y. \& Lewin, W. H. G. 1995, X-ray Binaries, eds. W. H. G. 
Lewin et al., Cambridge Univ. Press., 126 

\reference Tanaka, Y. \& Shibazaki, N. 1996, \araa, 34, 607

\reference Titarchuk, L. G., Mastichiadis, A. \& Kylafis, N. D. 1996, \aas, 
120, C171

\reference Titarchuk, L. G., Mastichiadis, A. \& Kylafis, N. D. 1997, \apj, in 
press (astro-ph/9702092)

\reference van der Klis, M. 1994, \apjs, 92, 511

\reference van Paradijs, J. \& McClintock, J. E. 1995, X-ray Binaries, eds. 
W. H. G. Lewin et al., Cambridge Univ. Press., 58

\reference Vrtilek, S. D., Raymond, J. C., Garcia, M. R., Verbunt, F., 
Hasinger, G., Kurster, M. 1990, \aa, 235, 162

\reference Wandel, A. \& Liang, E. 1991, \apj, 380, 84

\reference White, T. R. \& Lightman, A. P. 1989, 340, 1024 

\reference White, T. R. Lightman, A. P. \& Zdziarski, A. A. 1988, \apj, 
331, 939

\reference Zhang, S. N., Cui, W., Harmon, B. A., Paciesas, W. S., 
Remillard, R. E., van Paradijs, J. 1997, \apj, 477, L95

\clearpage
\include{aesintable}

\clearpage

\begin{figure}
\includegraphics{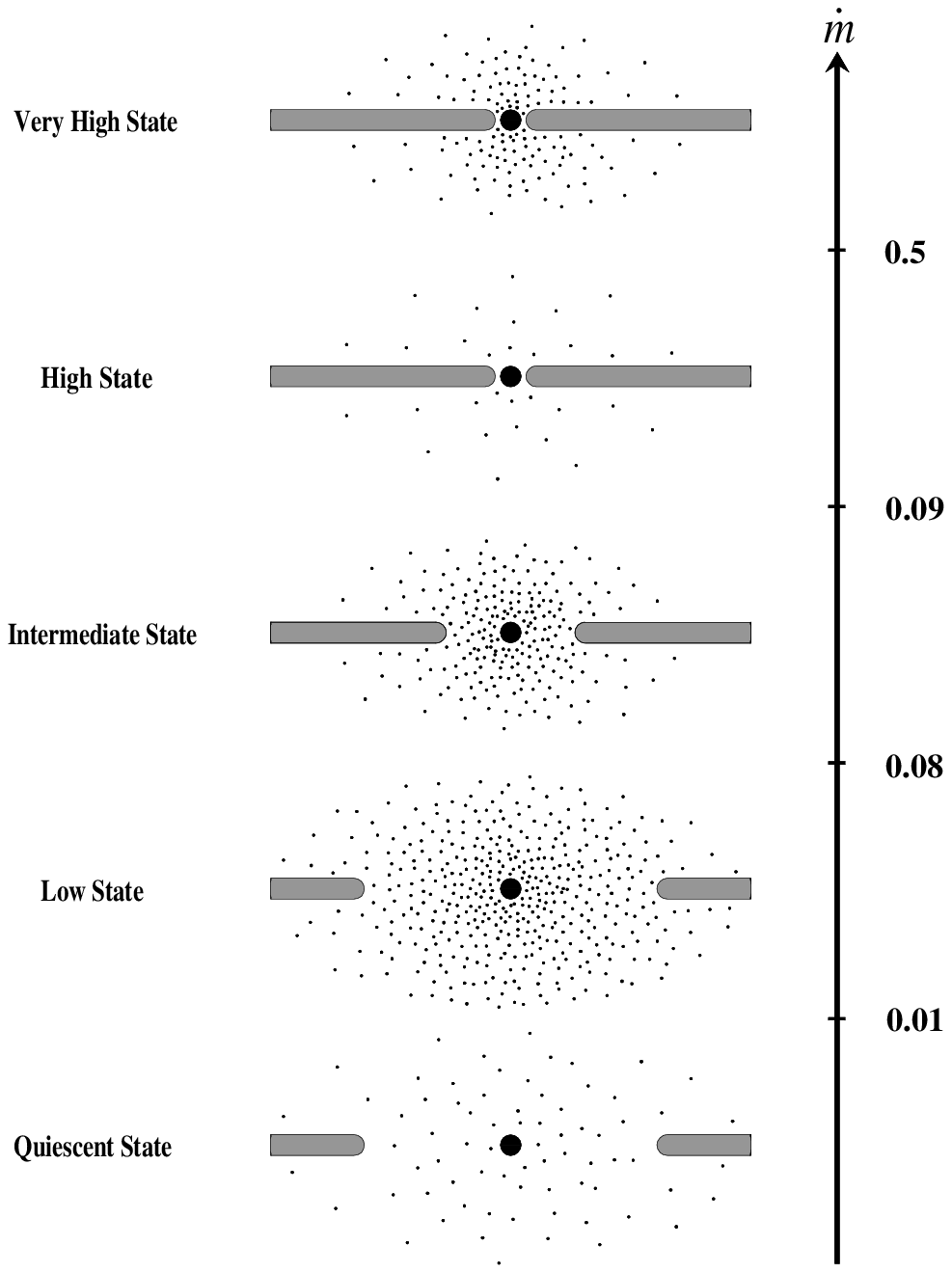}
\vskip 5.7in
\caption{The configuration of the accretion flow in different spectral 
states shown schematically as a function of the total mass accretion rate
$\mdot$.  The ADAF is indicated by dots and the thin disk by the horizontal
bars.  The lowest horizontal panel shows the quiescent state which corresponds
to a low mass accretion rate (and therefore, a low ADAF density) and a large 
transition radius.  The next panel shows the low state, where the mass 
accretion rate is larger than in the quiescent state, but still below the 
critical value $\mdot_{crit}\sim 0.08$.  In the intermediate state (the 
middle panel), $\mdot > \mdot_{crit}$ and the transition radius is 
smaller than in the quiescent/low state.  In the high state, the thin disk 
extends down to the last stable orbit and the ADAF is confined to a 
low-density corona above the thin disk.  Finally, in the very high state, we
make the tentative proposal that the corona has a substantially larger  
$\mdot$ than in the high state.}
\end{figure}

\begin{figure}
\includegraphics{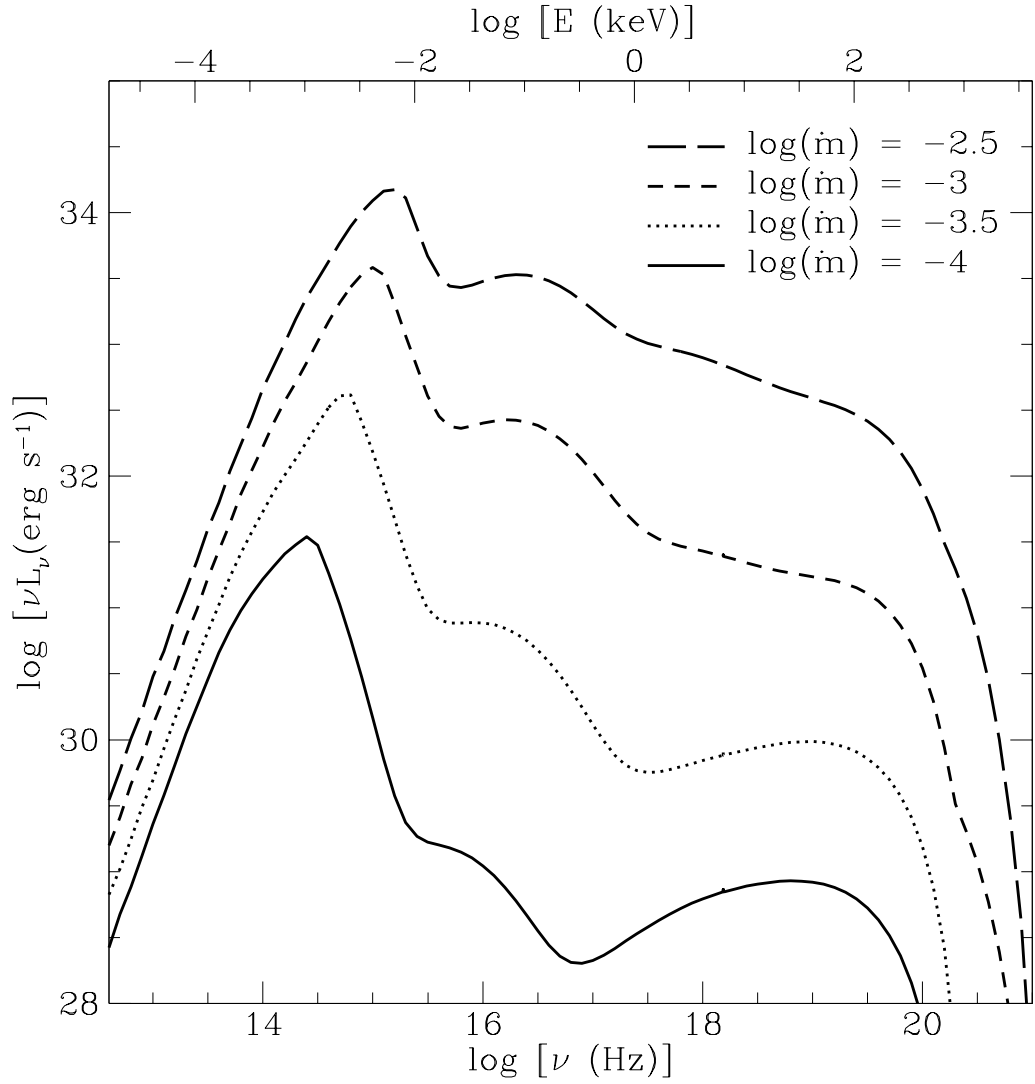}
\vskip 5.7in
\caption{Quiescent state spectra for models with $M = 6 \msun$, 
$\log{(\rtr)} = 3.9$, $i=60^{\circ}$, $\alpha = 0.25$, and  $\beta=0.5$.  The
values of $\mdot$ are indicated on the plot.}
\end{figure}
\clearpage

\begin{figure}
\includegraphics{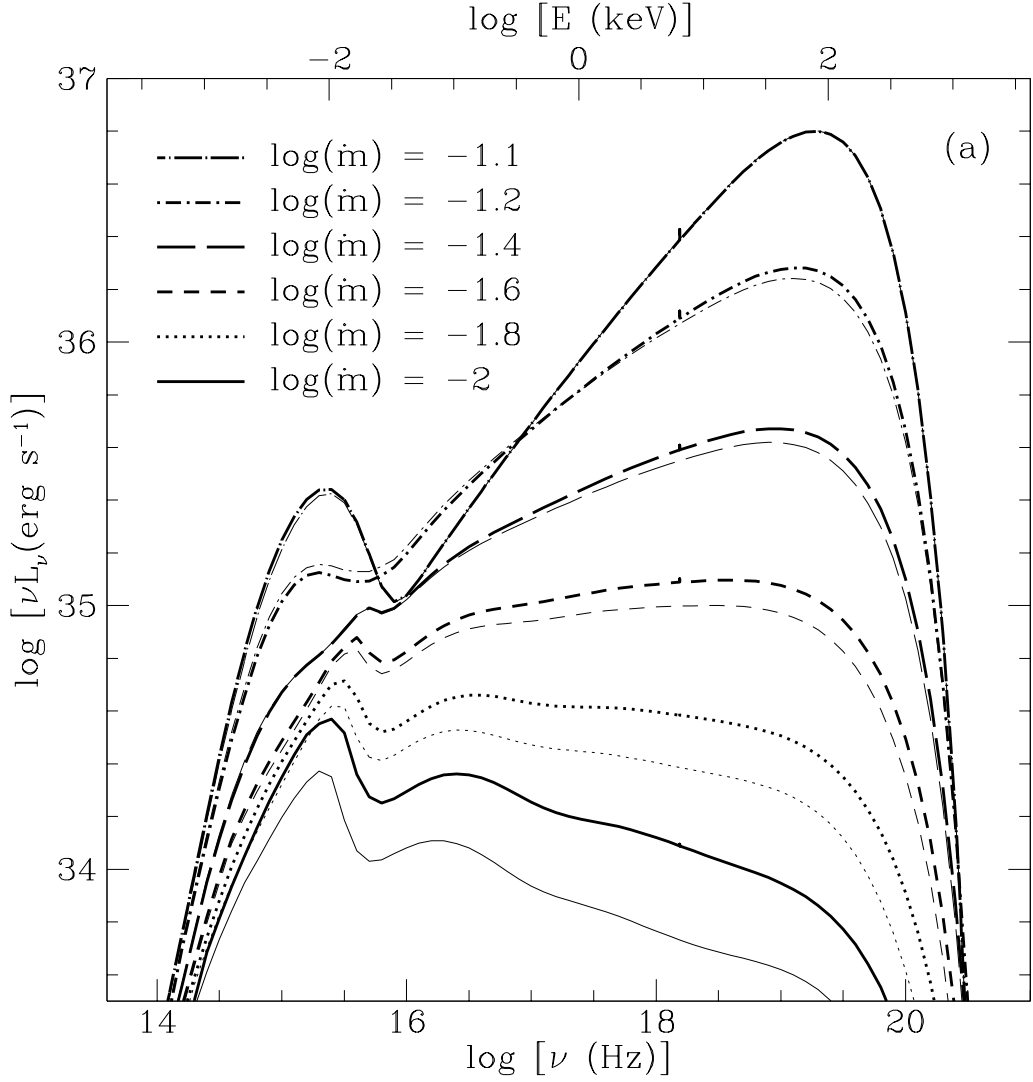}
\vskip 5.7in
\caption{(a) Low state spectra for models with $M = 6 \msun$, $\log{(\rtr)} = 
3.9$, $i=60^{\circ}$, $\alpha = 0.25$, $\beta=0.5$.  The values of $\mdot$ are 
indicated on the plot.  The largest $\mdot$ is equal to the critical 
accretion rate $\mdot_{crit}$. Heavy lines show spectra computed with the 
electron advection term included; for comparison, thin lines show 
spectra of models where this term is omitted.}  
\end{figure}

\setcounter{figure}{2}
\begin{figure}
\includegraphics{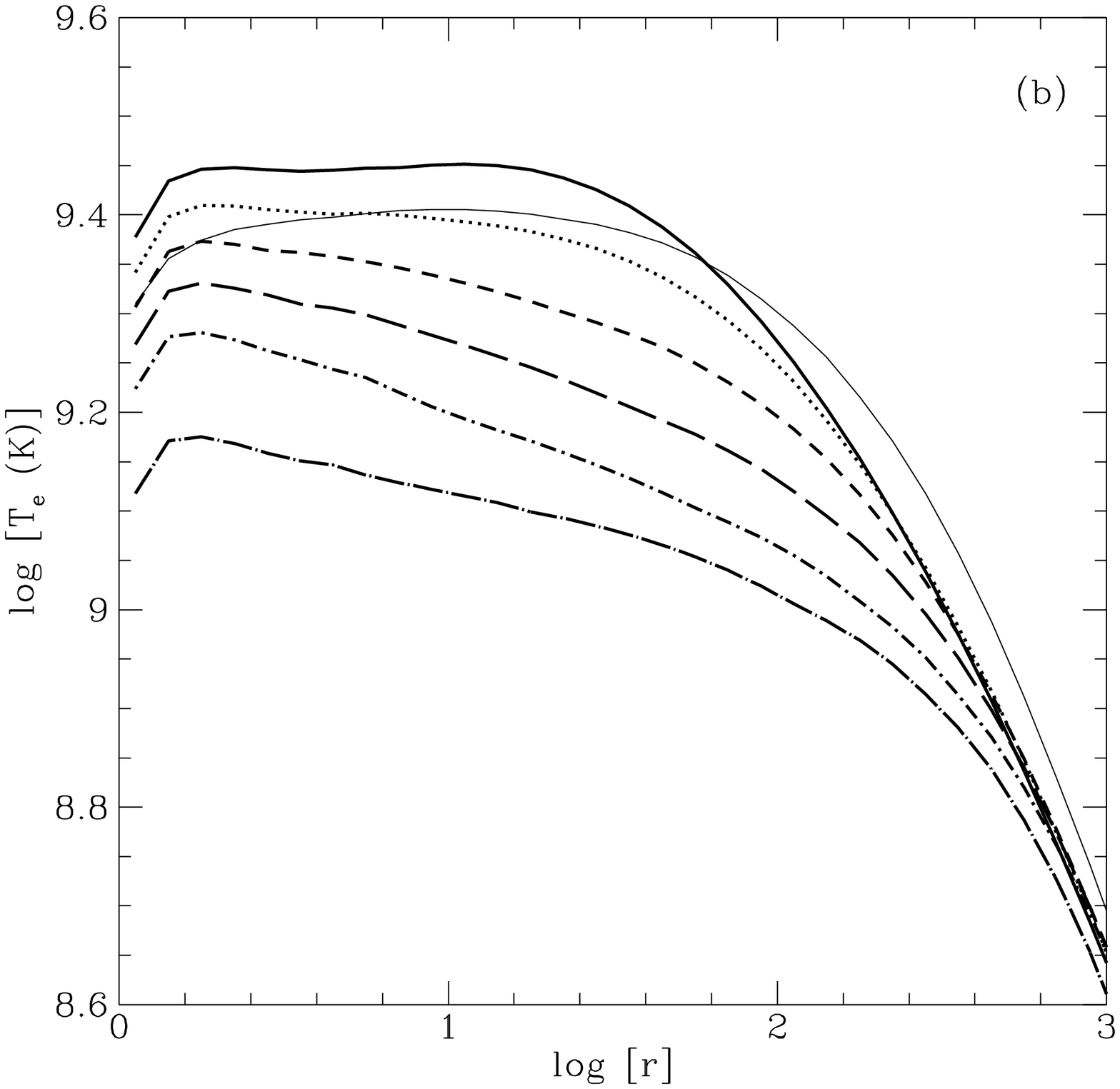}
\vskip 5.7in
\caption{(b) Variation of electron temperature $T_e$ with radius $r$ for the 
models shown in (a).  Note that the electron temperature decreases with 
increasing $\mdot$.  For the model with $\log{\mdot} = -2$ (solid line), the
temperature profile is computed both with (heavy line) and without 
(thin line) the electron advection term.}
\end{figure}

\begin{figure}
\includegraphics{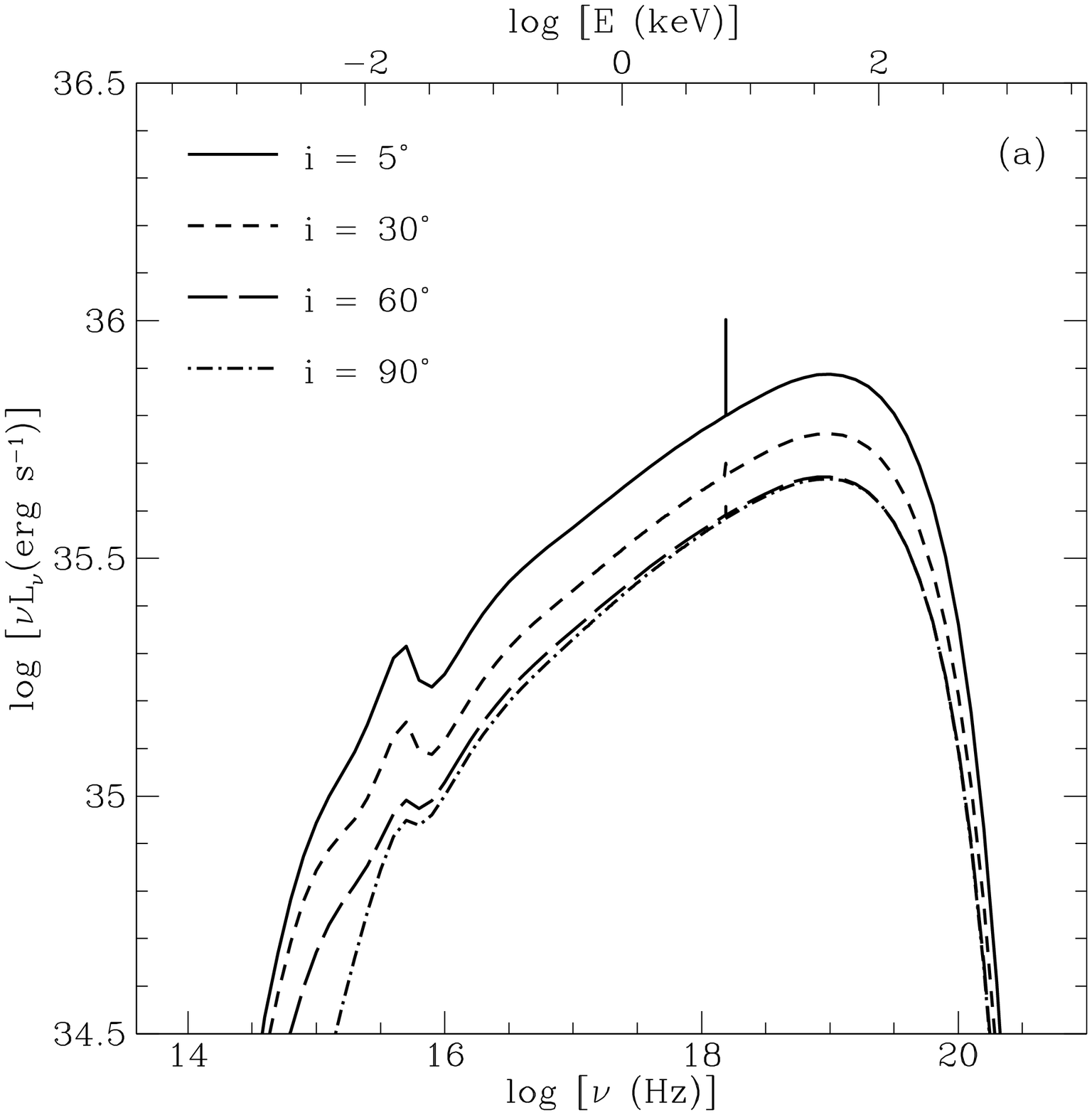}
\vskip 5.7in
\caption{(a) Spectra of a low state model with $M = 6 \msun$, 
$\log{(\rtr)} = 3.9$, $i=60^{\circ}$, $\alpha = 0.25$, $\beta=0.5$, and 
$\log{\mdot} = -1.4$ viewed at different inclination angles.
The feature at $6.4\,{\rm keV}$ is the iron K$\alpha$ line.  Its width is
proportional to $\sin{i}$, so the line is more prominent in face-on systems,
though the equivalent width is independent of $i$.}
\end{figure}

\setcounter{figure}{3}
\begin{figure}
\includegraphics{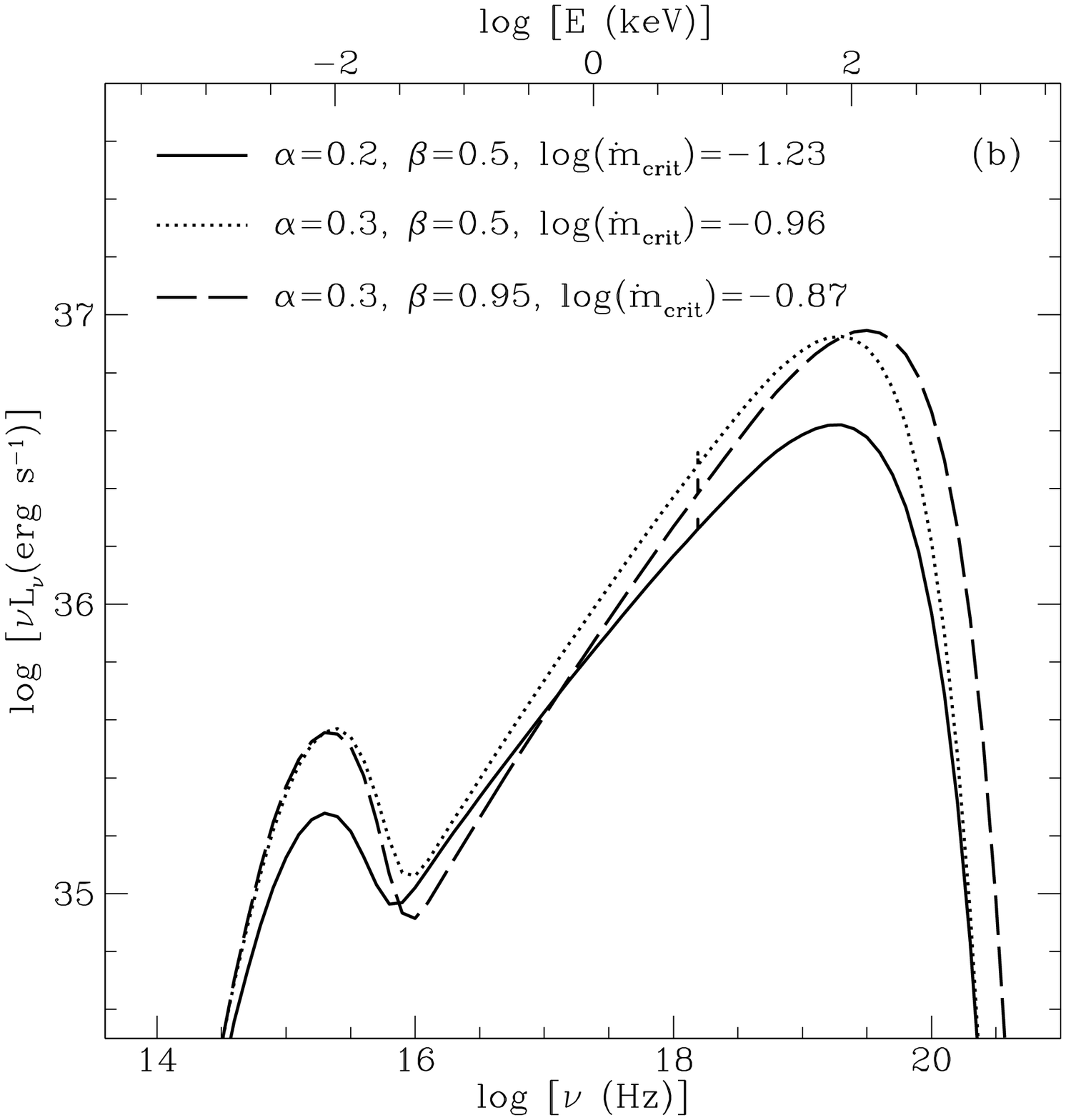}
\vskip 5.7in
\caption{(b) Low state spectra for a model with $M = 6 \msun$, 
$\log{(\rtr)} = 3.9$, $i=60^{\circ}$, and values of $\alpha$, $\beta$, and 
$\mdot$ as shown on the figure.  Each spectrum corresponds to a model with
 the critical mass accretion rate $\mdot_{crit}$.}
\end{figure}

\begin{figure}
\includegraphics{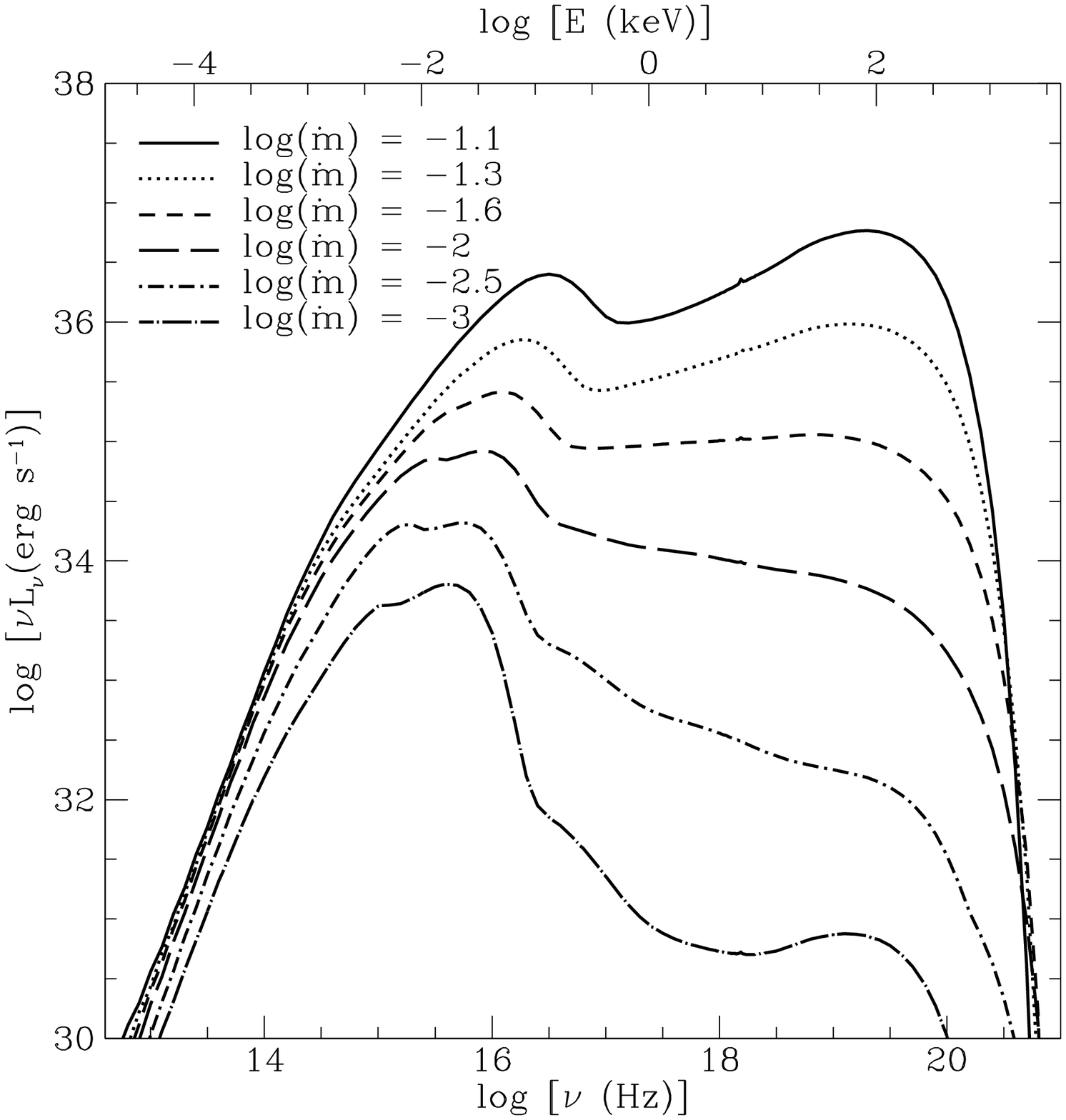}
\vskip 5.7in
\caption{Low state and quiescent state spectra for a wind-fed system with
$M = 6 \msun$, $\log{(\rtr)} = 2$, $i=60^{\circ}$, $\alpha = 0.25$, 
$\beta=0.5$ and values of the accretion rate as shown on the figure.
The spectra in the X-ray band are the same as for a system with a larger
$\rtr$ (compare with Figure 2 and Figure 3).  However in the optical and UV
bands the thin disk emission now dominates over the synchrotron emission, 
and the soft peak moves from the optical into the UV and soft X-ray band.}
\end{figure}

\begin{figure}
\includegraphics{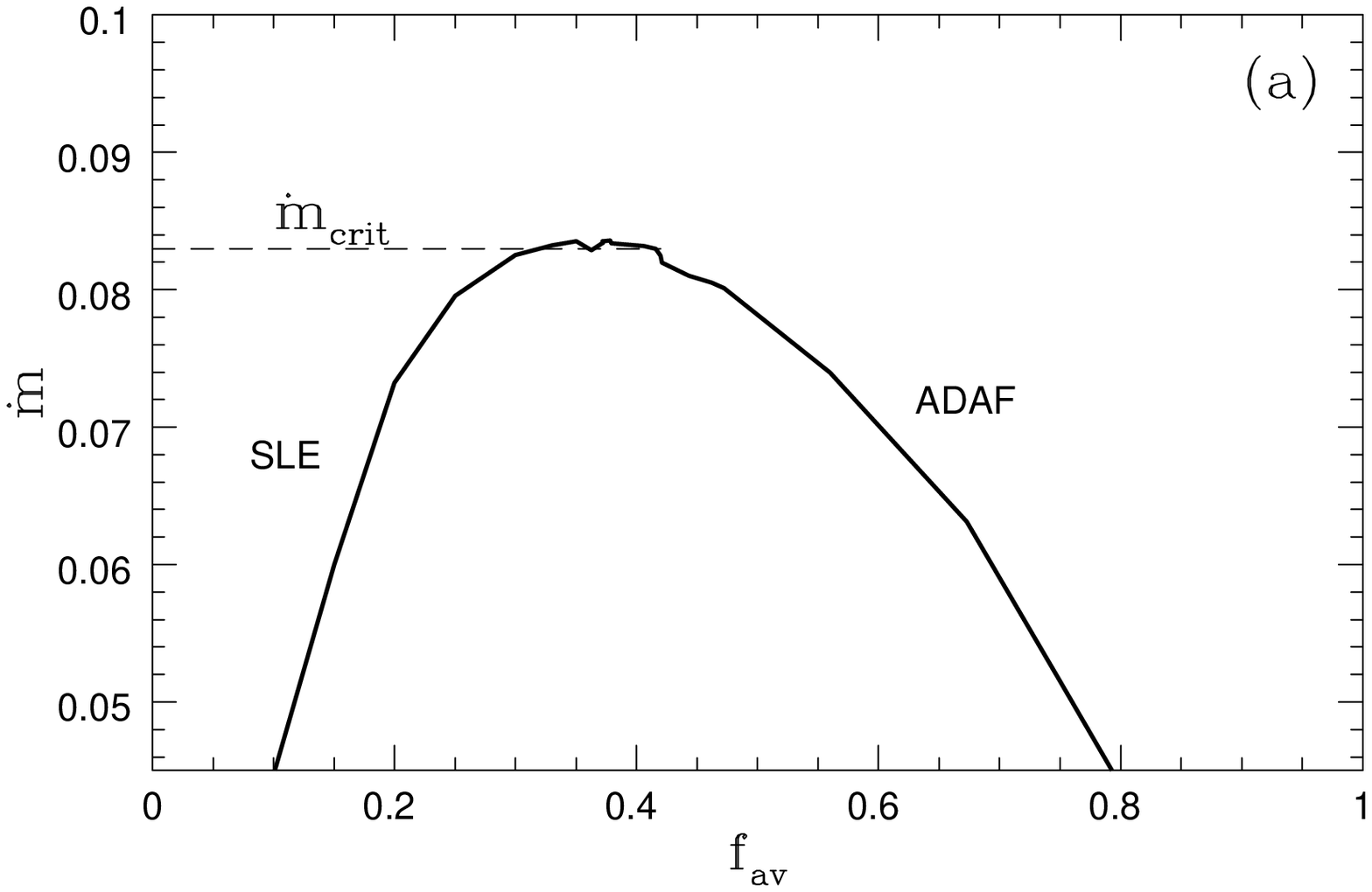}
\includegraphics{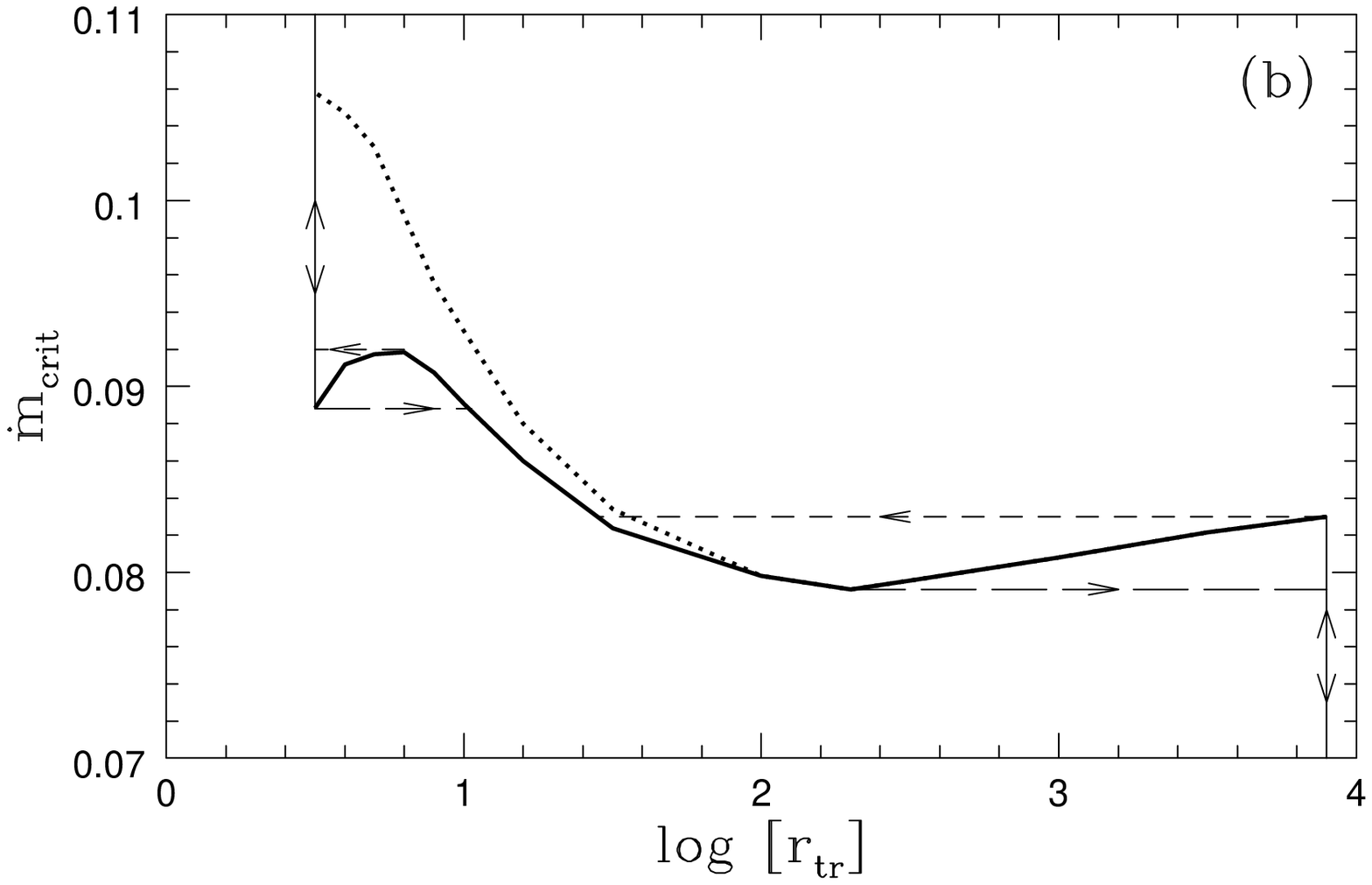}
\vskip 5.7in
\caption{(a) $\mdot$ vs. $f_{av}$ for a model with $M = 6 \msun$
$\alpha=0.25$, $\beta=0.5$, and $\log{(\rtr)} = 3.9$.  The maximum allowed
mass accretion rate, $\mdot_{crit}$, corresponds to $f_{av}\sim 0.35$.  
The models with $f_{av} < 0.35$ correspond to the thermally unstable
SLE solutions, while the models with $f_{av} > 0.35$ are the 
advection-dominated solutions used in this paper. 
(b) The heavy solid line shows $\mdot_{crit}$ as a function of $\rtr$. Thin 
lines indicate possible tracks followed by a system during the 
rise phase (short-dashed line) and decay phase (long-dashed line) of an
outburst.  The dotted line shows $\mdot_{crit} (\rtr)$ computed under the 
assumption that half of the energy needed for evaporation of the thin disk
material into the corona comes from the energy budget of the thin disk.}
\end{figure}

\begin{figure}
\includegraphics{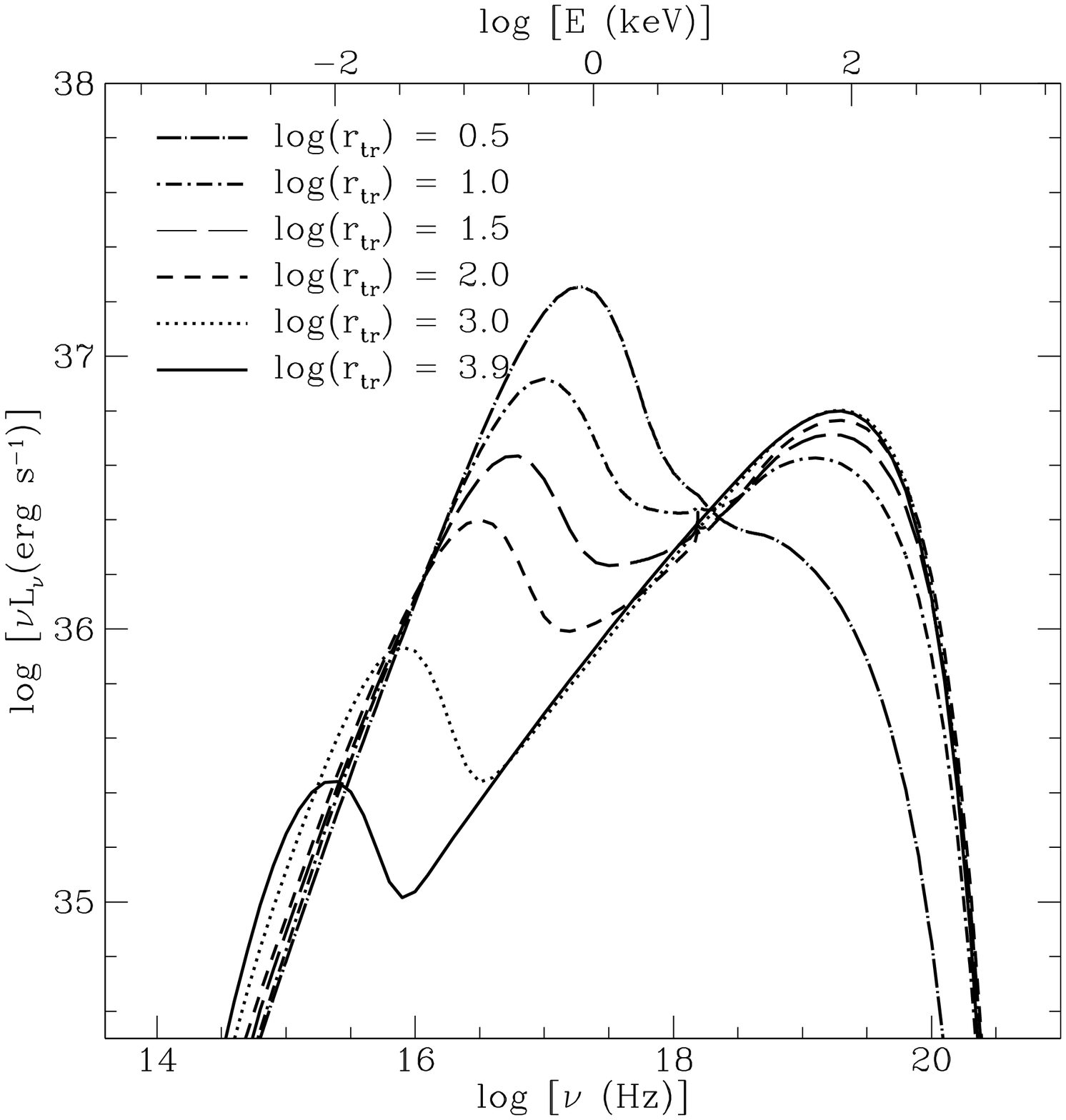}
\vskip 5.7in
\caption{Spectra of intermediate state models with $M = 6 \msun$,
$i=60^{\circ}$, $\alpha=0.25$, $\beta=0.5$.  Values of the transition radius 
are indicated on the figure; the accretion rate is taken to be the critical 
rate, $\mdot_{crit}(\rtr)$, shown in Figure 6b.  Note that the spectra pivot 
around 10 keV.  Note also the appearance of a reflection bump centered near 
30 keV in the spectra of models with $\rtr \le 10^2$, and the broadening of 
the iron K$\alpha$ line at $6.4\,{\rm keV}$.}
\end{figure}

\begin{figure}
\includegraphics{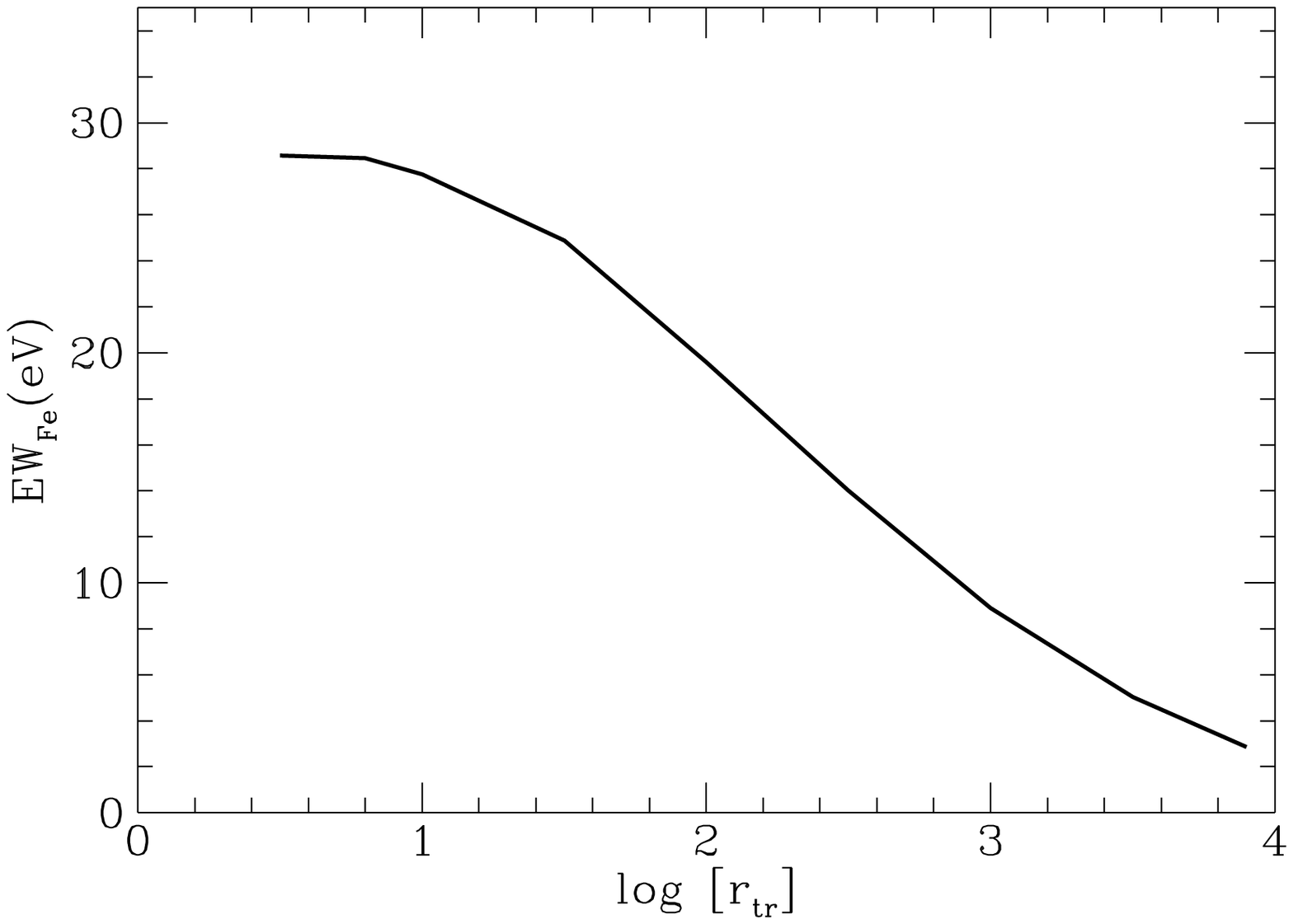}
\vskip 5.7in
\caption{Variation of the iron K$\alpha$ line equivalent width as a function 
of the transition radius $\rtr$ in the intermediate state models shown in
Figure 7.}
\end{figure}

\begin{figure}
\includegraphics{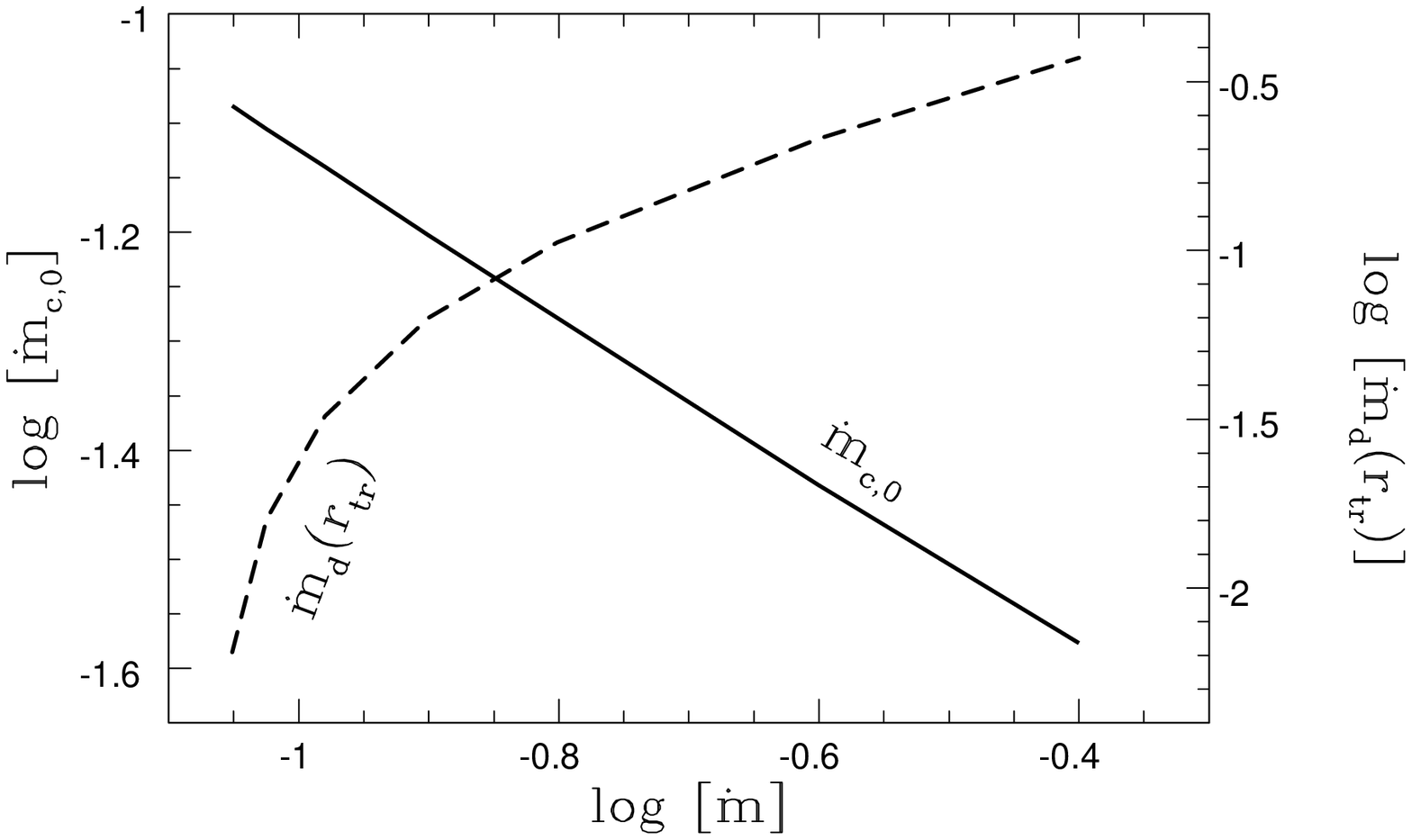}
\vskip 5.7in
\caption{The solid line corresponds to the critical accretion rate in the 
corona, $\mdot_{c,0}$, as a function of the total accretion rate, $\mdot$.
The dashed line is the corresponding  accretion rate in the thin disk at the 
last stable orbit, $\mdot_d (r=3) = \mdot-\mdot_{c,0}$.  Note that 
$\mdot_{c,0}$ decreases with increasing $\mdot$.  Thus the coronal emission 
becomes weaker with increasing total luminosity.}
\end{figure}

\begin{figure}
\includegraphics{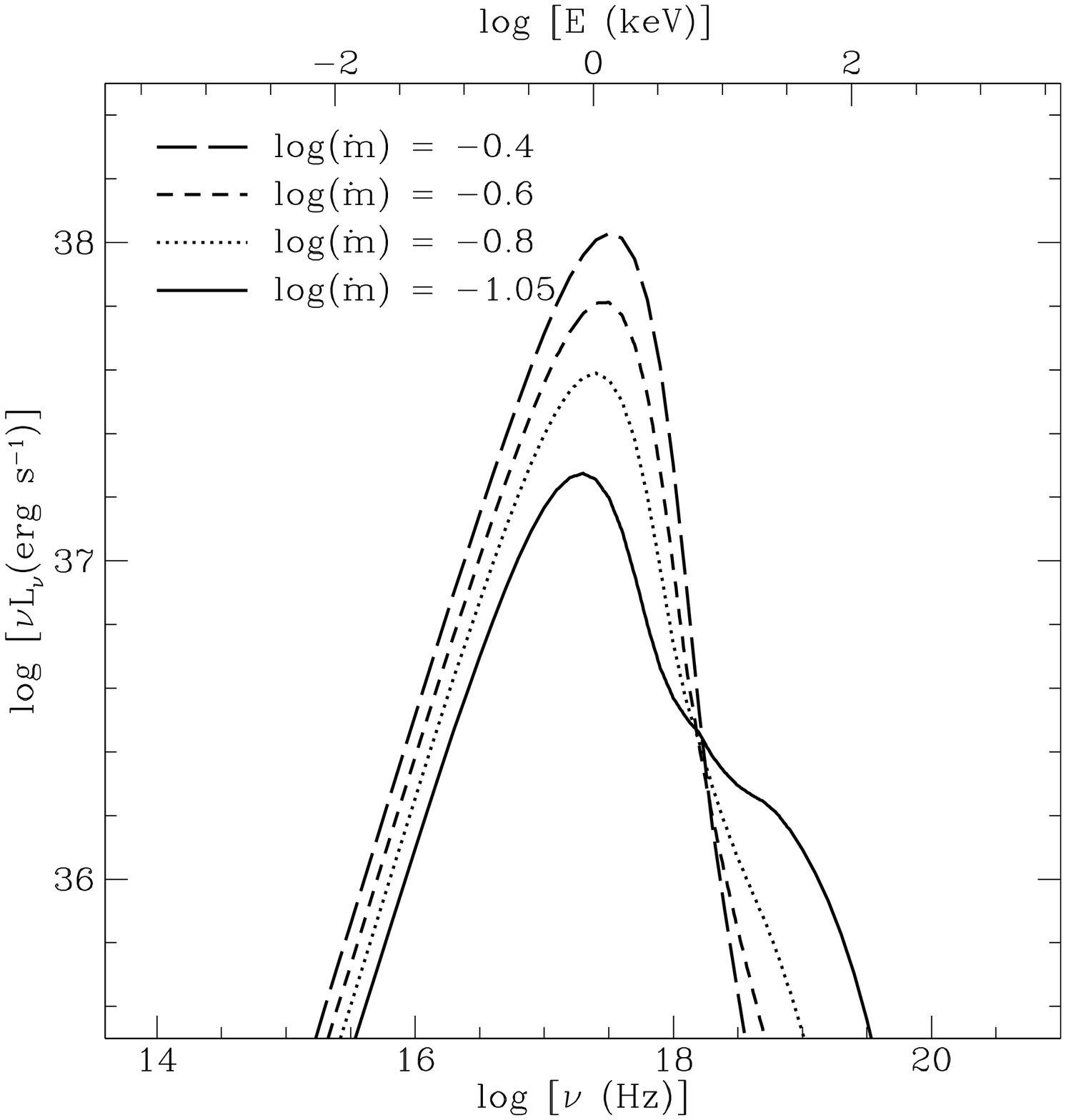}
\vskip 5.7in
\caption{Spectra of high state models with $M = 6 \msun$, $i=60^{\circ}$,
$\rtr = 3$, $\alpha=0.25$, $\beta=0.5$, and $\mdot$ as indicated on the 
figure.  At higher $\mdot$, the corona contains less gas (see Figure 9), and 
consequently, the amount of high energy emission decreases.}
\end{figure}

\begin{figure}
\includegraphics{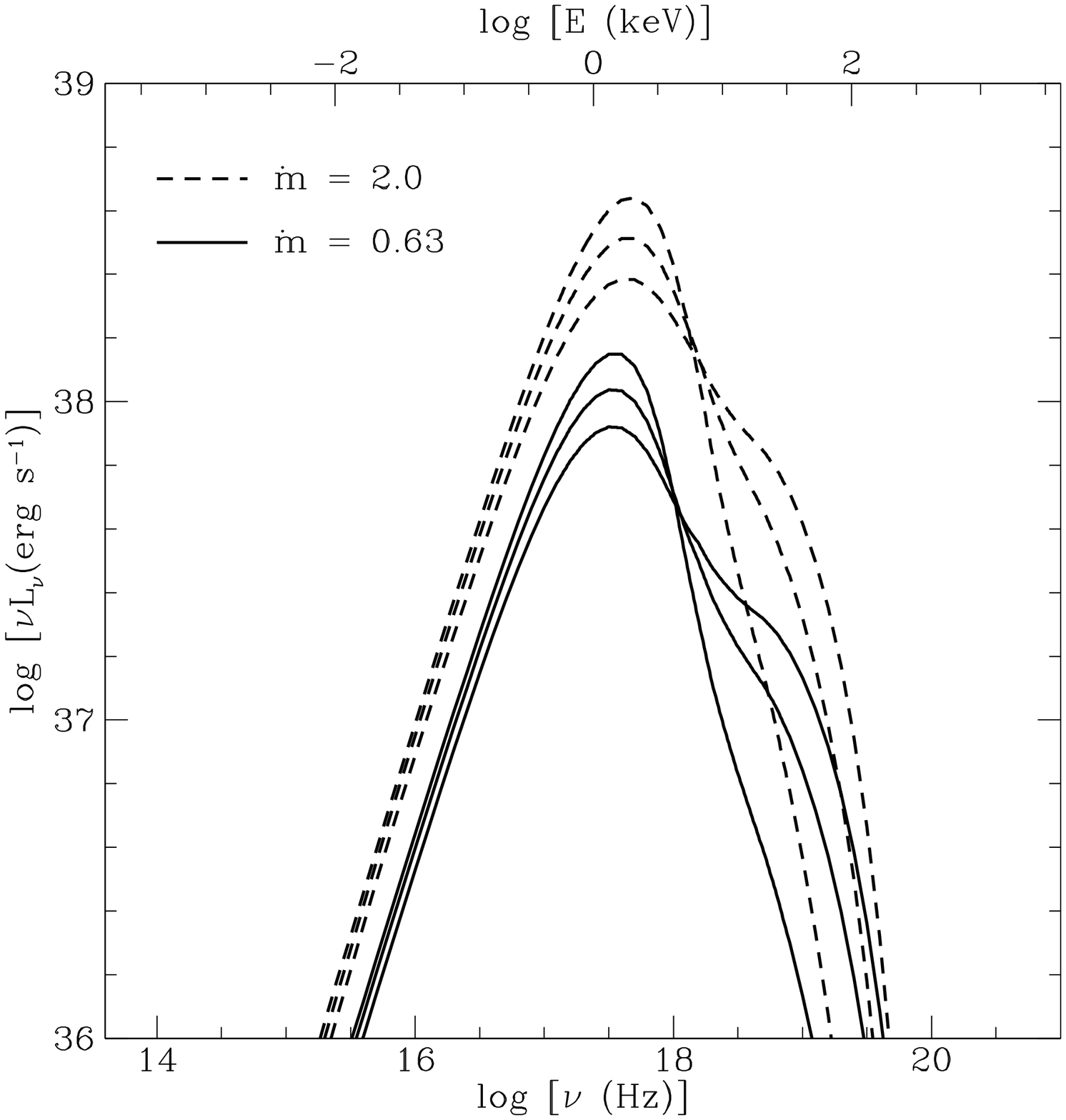}
\vskip 5.7in
\caption{Spectra of very high state models with $M = 6 \msun$, 
$i=60^{\circ}$, $\rtr = 3$, $\alpha=0.25$, $\beta=0.5$. For each 
value of $\mdot$ three spectra are shown with $\eta = 0.1,\ 0.3,\ 0.5$ in 
order of increasing flux at 10 keV.  Note that with increasing $\eta$
the disk blackbody emission at $\sim 2$ keV decreases, but the high energy 
tail becomes more prominent.}
\end{figure}
\clearpage

\begin{figure}
\includegraphics{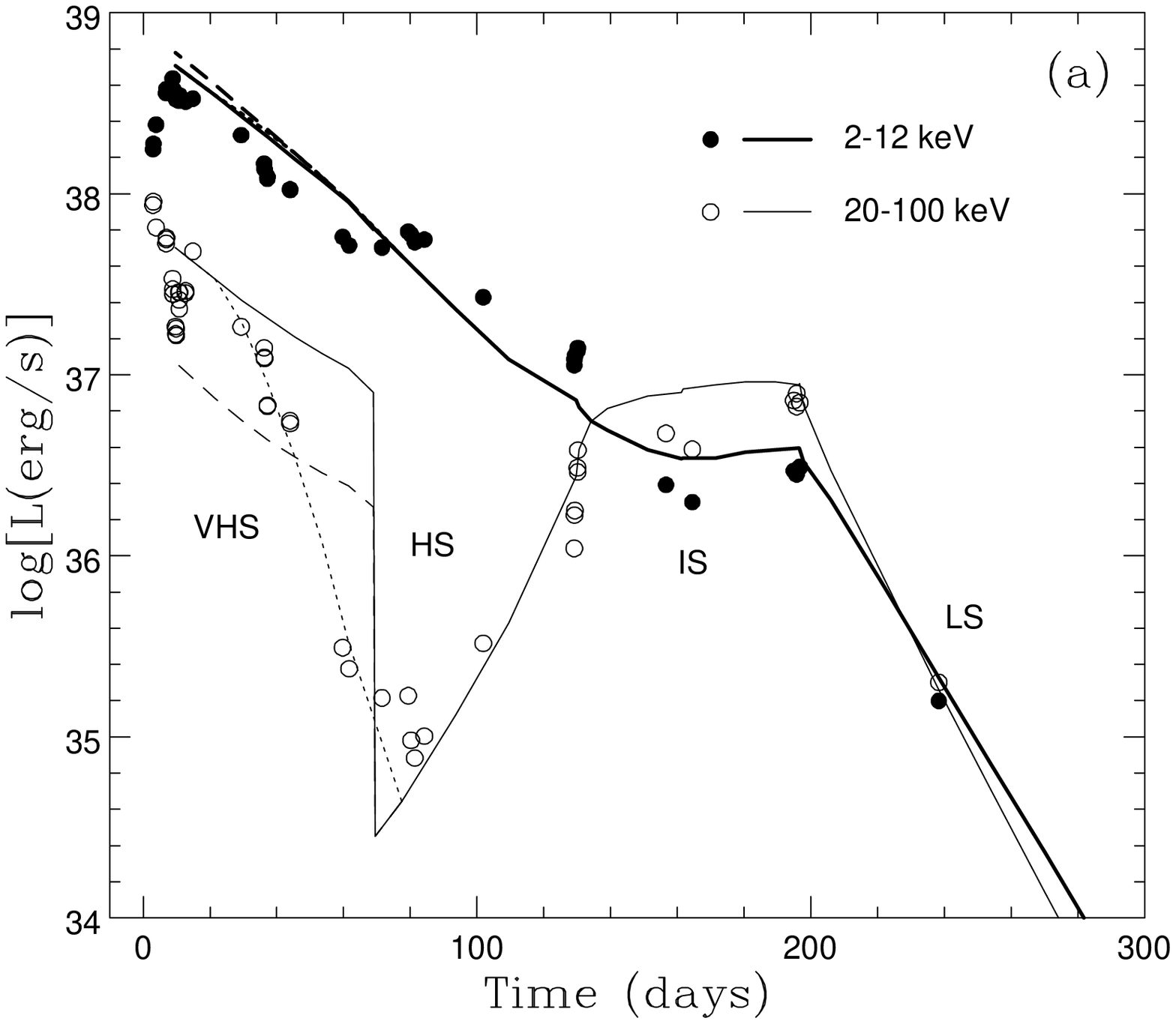}
\vskip 5.7in
\caption{(a) X-ray and $\gamma$-ray light curves of Nova Muscae 1991.  
Filled and open circles are data from E94 corresponding to the 2--12 keV and 
20--100 keV bands, respectively.  The heavy and thin lines are the 
corresponding model predictions.  The symbols VHS, HS, IS, LS correspond to
the very high state, high state, intermediate state and low state 
respectively.  In the very high state, the solid and dashed lines 
correspond to models with $\eta = 0.3$ and $0.1$, while the dotted line 
corresponds to a model with variable $\eta$.}
\end{figure}

\setcounter{figure}{11}
\begin{figure}
\includegraphics{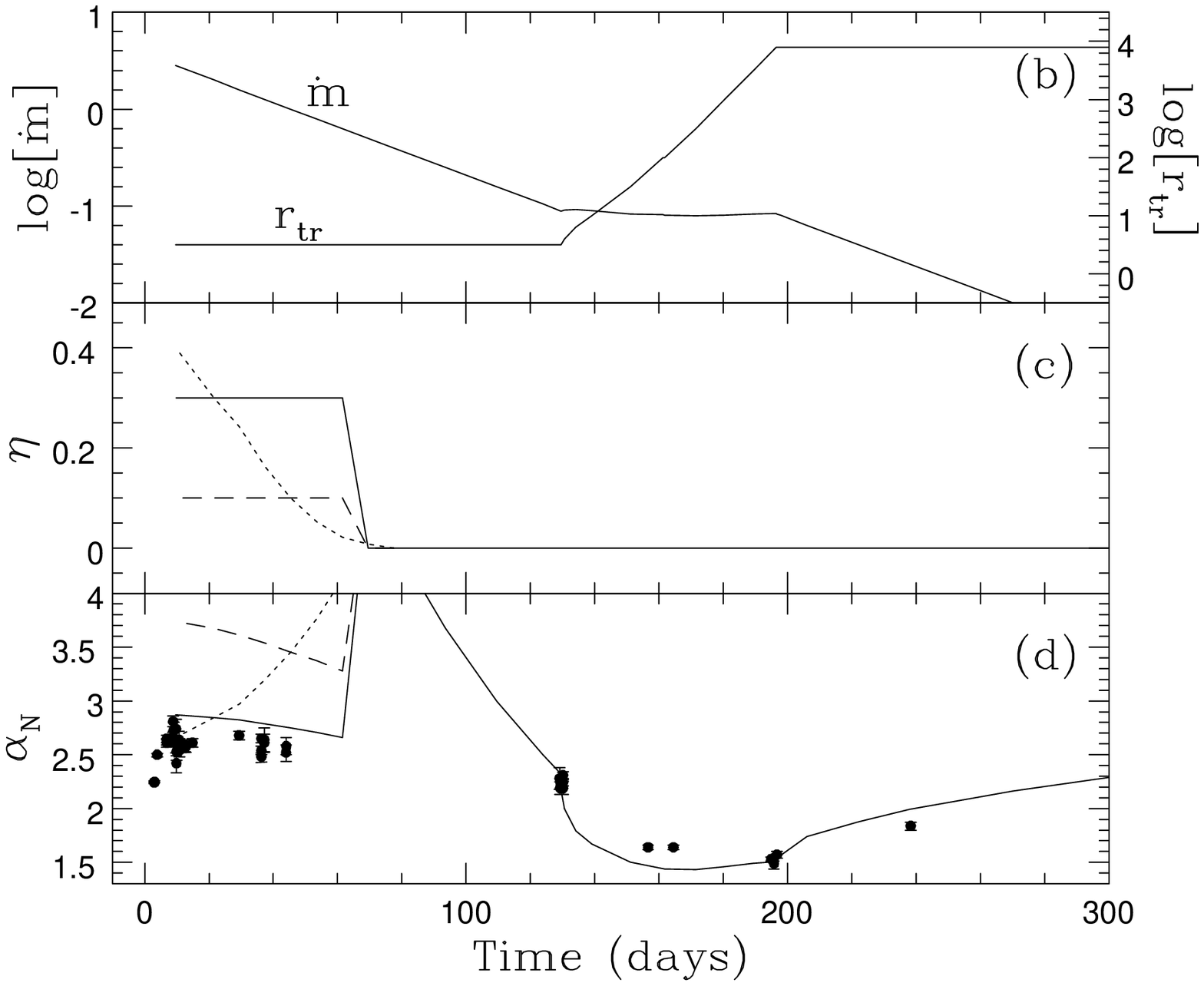}
\vskip 5.7in
\caption{(b) Variation of $\mdot$ and $\rtr$ with time during the Nova Muscae 
outburst, according to the model described by equation (4-1).
(c) Time variation of $\eta$ for the three models of the very high state: 
$\eta = 0.3$ (solid line), $\eta=0.1$ (dashed line), variable $\eta$
(dotted line).
(d) Time variation of the photon-index $\alpha_N$ of the hard power-law
component of the spectrum.  Filled points with error bars are the data from 
E94.  The lines correspond to model results, with the spectral index 
computed in the range $10-20\,{\rm keV}$ (the line types are as in 
panel c).}
\end{figure}

\begin{figure}
\includegraphics{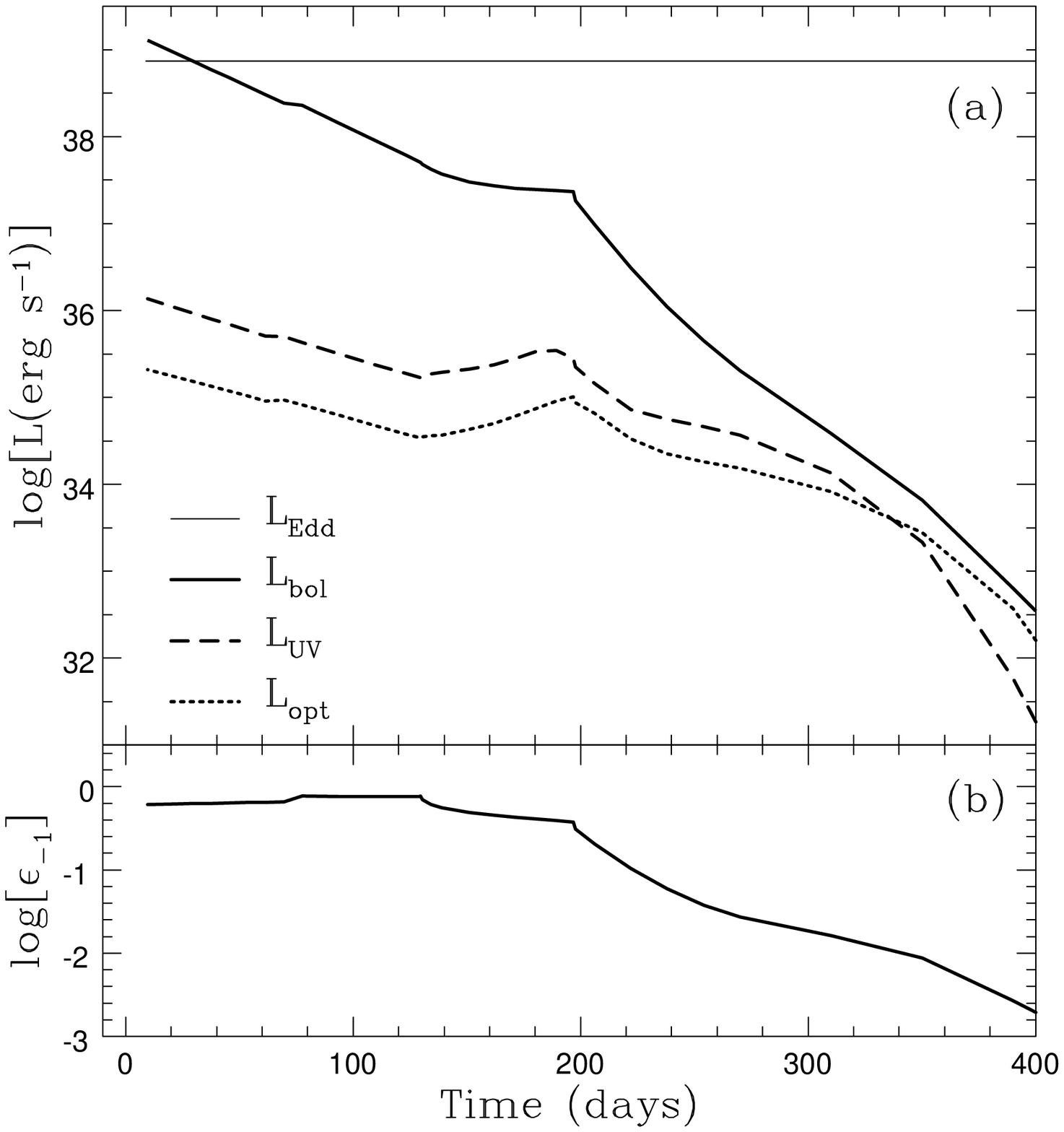}
\vskip 5.7in
\caption{(a) The bolometric, optical ($10^{14.5}-10^{15}$ Hz) and UV 
($10^{15}-10^{15.5}$ Hz) light curves of Nova Muscae, computed for the 
model with $\eta=0.3$ in the very high state.  The relationship between Time, 
$\mdot$ and $\rtr$ is given by equation (4-1) and plotted in Figure 
13.  The horizontal line at the top is the Eddington luminosity for a $6\msun$ 
accreting object. 
(b) Variation of the radiative efficiency of the accretion flow,
$\e_{-1} = L_{bol}/(0.1 \Mdot c^2)$, with time during the outburst.}
\end{figure}

\begin{figure}
\includegraphics{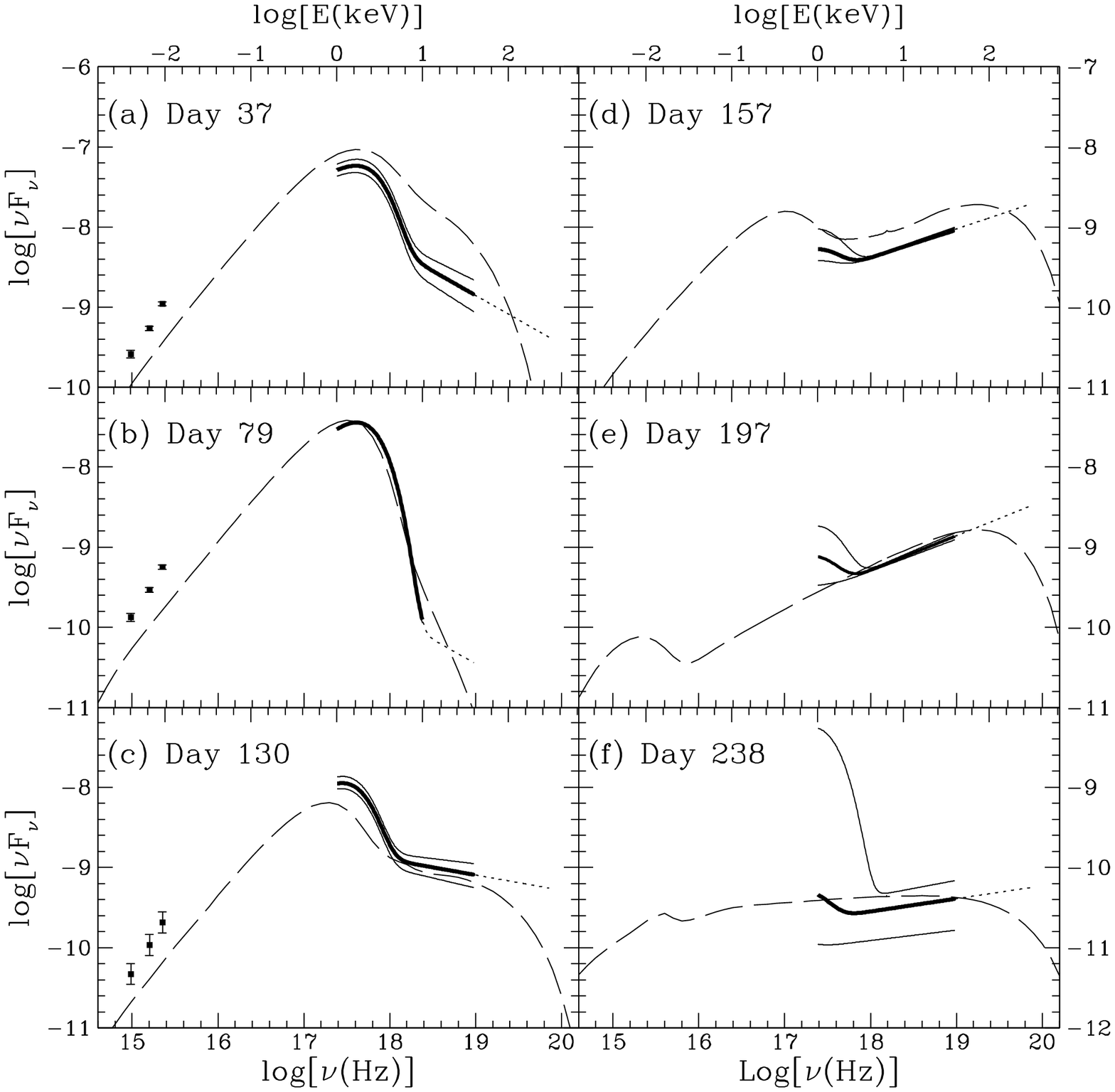}
\vskip 5.7in
\caption{Spectra of Nova Muscae on six different days during the outburst, 
which correspond to the following entries in Table 2 of E94: 26, 33, 43, 44, 
49, and 51.  The heavy solid lines show the observed spectra reconstructed 
from the spectral fit parameters published by E94. The thin solid lines are 
1-$\sigma$ upper and lower limits. The dotted lines in panels (a), (c)-(f) 
are extrapolations.  The dotted line in panel (b) indicates that the 
power-law index of the hard end of the spectrum is highly uncertain.  The 
UV data points are from Shrader \& Gonzalez-Riestra (1993).  The dashed 
lines are model spectra corresponding to the values of $\mdot$ and $\rtr$ 
shown in Figure 12b for each date.  The model spectrum in panel (a) 
corresponds to the very high state with $\eta = 0.3$.  Panel (b) corresponds 
to the high state, panels (c) and (d) to the intermediate state, and 
panels (e) and (f) to the low state.}
\end{figure}

\end{document}

%% file: aesintable.tex
\begin{deluxetable}{ccccccc} 
\tablecolumns{7}
\tablewidth{0pc}
\tablecaption{Standard parameter values}
\tablehead{\colhead{$M\ (\msun)$} & \colhead{$i$} & \colhead{$\log{r_{out}}$} & 
\colhead{$\log{\rtr}$\tablenotemark{a}} & \colhead{$\alpha$} & \colhead{$\beta$} &
\colhead{$\delta$}}
\startdata
6 & $60^{\circ}$ & $4.9$ & $3.9$ & 0.25 & 0.5 & 0.001 \nl
\enddata
\tablenotetext{a}{in quiescence}
\end{deluxetable}